\documentclass[fleqn,usenatbib]{mnras}

\usepackage{newtxtext,newtxmath}

\usepackage[T1]{fontenc}

\DeclareRobustCommand{\VAN}[3]{#2}
\let\VANthebibliography\thebibliography
\def\thebibliography{\DeclareRobustCommand{\VAN}[3]{##3}\VANthebibliography}

\usepackage{float}
\usepackage{epsfig}
\usepackage{graphicx}
\usepackage{graphics}
\usepackage{latexsym}
\usepackage{hyperref}
\usepackage{wasysym}
\usepackage{threeparttable}

\def\gsim{\;\lower4pt\hbox{${\buildrel\displaystyle >\over\sim}$}\;}
\def\lsim{\;\lower4pt\hbox{${\buildrel\displaystyle <\over\sim}$}\;}
\def\grls{\;\lower4pt\hbox{${\buildrel\displaystyle >\over <}$}\;}
\def\jaavso{JAAVSO}

\title[SN 2019ein: a Sub-Chandrasekhar Explosion]{SN 2019ein: A Type Ia Supernova Likely Originated from a Sub-Chandrasekhar-Mass Explosion}

\author[Xi et al.]{
Gaobo Xi,$^{1}$
Xiaofeng Wang,$^{1,2}$\thanks{E-mail: wang\_xf@mail.tsinghua.edu.cn}
Wenxiong Li,$^{3}$
Jun Mo,$^{1}$
Jujia Zhang,$^{4,5,6}$
Jialian Liu,$^{1}$
Zhihao Chen,$^{1}$
\newauthor
Alexei V. Filippenko,$^{7}$
Weikang Zheng,$^{7}$
Thomas G. Brink,$^{7}$
Xinghan Zhang,$^{1}$
Hanna Sai,$^{1}$
\newauthor
Shuhrat A. Ehgamberdiev,$^{8,9}$
Davron Mirzaqulov,$^{8}$
and Jicheng Zhang$^{10}$
\\
$^{1}$Physics Department and Tsinghua Center for Astrophysics, Tsinghua University, Beijing 100084, China\\
$^{2}$Beijing Planetarium, Beijing Academy of Science and Technology, Beijing 100044, China\\
$^{3}$The School of Physics and Astronomy, Tel Aviv University, Tel Aviv 69978, Israel\\
$^{4}$Yunnan Observatories, Chinese Academy of Sciences, Kunming 650216, China\\
$^{5}$Key Laboratory for the Structure and Evolution of Celestial Objects, Chinese Academy of Sciences, Kunming 650216, China\\
$^{6}$Center for Astronomical Mega-Science, Chinese Academy of Sciences, 20A Datun Road, Chaoyang District, Beijing 100012, China\\
$^{7}$Department of Astronomy, University of California, Berkeley, CA 94720-3411, USA\\
$^{8}$Ulugh Beg Astronomical Institute, Uzbekistan Academy of Sciences, Tashkent 100052, Uzbekistan\\
$^{9}$National University of Uzbekistan, Tashkent 100174, Uzbekistan\\
$^{10}$Department of Astronomy, Beijing Normal University, Beijing 100875, China
}

\date{Accepted XXX. Received YYY; in original form ZZZ}

\pubyear{2022}

\begin{document}
\label{firstpage}
\pagerange{\pageref{firstpage}--\pageref{lastpage}}
\maketitle

\begin{abstract}
We present extensive optical photometric and spectroscopic observations for the nearby Type Ia supernova (SN~Ia) 2019ein, spanning the phases from $\sim 3$ days to $\sim 330$ days after the explosion. This SN~Ia is characterized by extremely fast expansion at early times, with initial velocities of Si~II and Ca~II being above $\sim 25,000$--30,000~km~s$^{-1}$. After experiencing an unusually rapid velocity decay, the ejecta velocity dropped to $\sim 13,000$~km~s$^{-1}$ around maximum light. Photometrically, SN~2019ein has a moderate post-peak decline rate ($\Delta m_{15}(B) = 1.35 \pm 0.01$~mag), while being fainter than normal SNe~Ia by about 40\% (with $M^{\rm max}_{B} \approx -18.71 \pm 0.15$~mag). The nickel mass synthesized in the explosion is estimated to be 0.27--0.31~M$_{\odot}$ from the bolometric light curve. Given such a low nickel mass and a relatively high photospheric velocity, we propose that SN~2019ein likely had a sub-Chandrasekhar-mass white dwarf (WD) progenitor, $M_{\rm WD} \lesssim 1.22$~M$_{\odot}$. In this case, the explosion could have been triggered by a double-detonation mechanism, for which 1- and 2-dimensional models with WD mass $M_{\rm WD} \approx 1$~M$_\odot$ and a helium shell of 0.01~M$_{\odot}$ can reasonably produce the observed bolometric light curve and spectra. The predicted asymmetry as a result of double detonation is also favored by the redshifted Fe~II and Ni~II lines observed in the nebular-phase spectrum. Possible diversity in origin of high velocity SNe Ia is also discussed.

\end{abstract}



\section{INTRODUCTION}
\label{secintro}

Type Ia supernovae (SNe~Ia) are among the most luminous stellar explosions in the Universe. Their peak luminosities are relatively uniform and can be further standardized to have dispersion of only $\sim 0.09$~mag through empirical relations such as that of \citet{1993ApJ...413L.105P} and color-parameter relations \citep{2007A&A...466...11G,2007ApJ...659..122J,2005ApJ...620L..87W}, making them unrivalled standard candles at cosmological distances. Observations of high-redshift SNe~Ia have revealed the accelerating expansion of the Universe \citep{1998AJ....116.1009R, 1998ApJ...507...46S, 1999ApJ...517..565P} and hence the existence of dark energy, while the Hubble-flow sample of SNe~Ia enabled precise measurements of the local Hubble constant when combined with the calibrations of Cepheid variables \citep{2019ApJ...876...85R,2021arXiv211204510R}.

Although it is commonly accepted that SNe~Ia are thermonuclear explosions of white dwarfs (WDs) in binary systems \citep{1960ApJ...132..565H,2012ApJ...744L..17B}, the exact nature of their progenitors (such as properties of the companion stars) are still under debate \citep{2011NatCo...2..350H}. The companion star of the exploding WD in SNe~Ia could either be a nondegenerate main-sequence star  (single-degenerate model; \citealt{1973ApJ...186.1007W}) or another degenerate WD (double-degenerate model; \citealt{1984ApJS...54..335I}). Both of these models face challenges \citep[e.g.,][]{2014ARA&A..52..107M,2018PhR...736....1L}.

Moreover, several explosion mechanisms have been proposed for SNe~Ia. One is the delayed-detonation model (DDT; \citealt{1984ApJ...286..644N}; \citealt{1991A&A...245..114K}), in which a slow burning process (subsonic deflagration) initially occurs near the center of a carbon-oxygen (CO) WD when its mass approaches the Chandrasekhar limit ($M_{\rm Ch}$) by accreting material from its companion, and the deflagration wave then transitions to supersonic detonation under certain critical conditions. Another popular model is the sub-Chandrasekhar double-detonation scenario \citep{1982ApJ...253..798N, 1982ApJ...257..780N, 1990ApJ...354L..53L, 1994ApJ...423..371W, 1996ApJ...472L..81H}, which involves an initial helium detonation on the surface of the C+O WD and the resulting secondary detonation in its inner core.  

Observationally, most SNe~Ia ($\sim 70$\%) display relatively homogeneous observed properties and are called ``Branch-normal'' \citep{1993AJ....106.2383B}. On the other hand, there is an increasing number of peculiar subclasses of SNe~Ia \citep{2014Ap&SS.351....1P,2017hsn..book..317T}, including luminous super-Chandrasekhar \citep{2006Natur.443..308H} or SN~1991T-like events with slow decline rates \citep{1992AJ....103.1632P, 1992ApJ...384L..15F}, subluminous SN~1991bg-like events with fast decline rates \citep{1992AJ....104.1543F,1993AJ....105..301L}, and subluminous SN~2002es-like events with normal decline rates \citep{2012ApJ...751..142G,2015ApJ...799...52W}.

Based on the observed diversity of their spectral properties, some subclassifications were proposed for SNe~Ia. \cite{2005ApJ...623.1011B} suggested that SNe~Ia could be classified into high-velocity-gradient (HVG) and low-velocity-gradient (LVG) subgroups in terms of the velocity gradient of Si~II $\lambda$6355 measured within 10 days after the maximum light. \cite{2006PASP..118..560B} divided SNe~Ia into four subtypes according to the strength of Si~II absorption features at maximum light. Also, based on the Si~II $\lambda$6355 velocities at maximum light ($v_{\mathrm{max}}$(Si)), \cite{2009ApJ...699L.139W} divide Branch-normal SNe~Ia into two categories: high velocity (HV; $v_{\mathrm{max}}{\rm (Si)} > 11,800~\mathrm{km~s^{-1}}$) and normal velocity (NV; $v_{\mathrm{max}}{\rm (Si)} \le 11,800~\mathrm{km~s^{-1}}$). The HV group is found to be redder and have a lower extinction ratio $R_V = A_V/(A_B-A_V)$. \cite{2013Sci...340..170W} further found that $v_{\mathrm{max}}$(Si)  exhibits a double-component distribution, with the HV group being more concentrated in the inner regions of the host galaxies than the NV one, suggesting metal-rich stellar environments for the HV SNe~Ia. This idea is supported by analysis from both \cite{2015MNRAS.446..354P} and \cite{2020ApJ...895L...5P} based on a rolling-search sample of SNe~Ia. More recently, \cite{2019ApJ...882..120W} found that the HV group shows an excess of blue flux during the early nebular phase, while the NV group does not. This could be explained by light-scattering effects if there is abundant circumstellar material (CSM) around the HV objects.

Among normal SNe~Ia, the subclass of HV~SNe~Ia could originate from a different explosion mechanism than the others. \cite{2019ApJ...873...84P} examined a set of numerical models of sub-Chandrasekhar double-detonation explosions, finding that the double-detonation model can explain the observed behavior for some low-luminosity and part of the HV~SNe~Ia. In comparison with SNe~Ia that have more homogeneous velocities and luminosities, SNe~Ia predicted from double-detonation explosions are on average fainter and could have ejecta velocities ranging from $\sim 9500$~km~s$^{-1}$ to $\sim 16,000$~km~s$^{-1}$ around maximum light. The large range of velocity is attributed to different progenitor WD masses.

HV~SNe~Ia could also be explained by viewing-angle effects of asymmetric explosions. \cite{2010Natur.466...82M} found that SNe~Ia with higher $v_{\mathrm{max}}$(Si) preferentially display redshifted [Fe~II] features in the nebular phase, implying an asymmetric explosion scenario in which the outer ejecta are moving toward us while the inner core region is moving away. They found that, with a geometric effect, a single scheme of asymmetric delayed-detonation model can explain both HV and NV~SNe~Ia. \cite{2018MNRAS.477.3567M} support the idea of an asymmetric explosion, but the actual explosion mechanism is still debated. \cite{2019ApJ...878L..38T} also studied a multidimensional full-star simulation of a double-detonation model. The model velocity is relatively high when viewed from the He-detonation pole, and gradually reduced to the normal-velocity range when viewed from another side. A recent study of \cite{2021ApJ...906...99L} further supports the idea that HV and a portion of NV~SNe could be explained by a double-detonation model with a geometric effect. However, a single explosion mechanism with a geometric effect is facing challenges in explaining the different distributions of HV and NV~SNe~Ia among host environment and metallicity \citep{2013Sci...340..170W,2015MNRAS.446..354P}. Also, both the delayed-detonation and double-detonation models have difficulties in reproducing some observed features of HV~SNe~Ia. The physical origin of the HV objects is still an open question, and further subclassification (and multiple channels) in the current HV group may be needed to account for the observed diversity.

SN~2019ein is an SN~Ia showing extremely high ejecta velocity at early times, with measured velocities for Si~II and Ca~II absorption being (respectively) $\sim 23,900$ and $\sim 29,000$~km~s$^{-1}$ about 14 days before $B$-band maximum light \citep{2020ApJ...897..159P}. \cite{2020ApJ...893..143K} and \cite{2020ApJ...897..159P} found that the observed properties of SN~2019ein are largely consistent with those of a delayed-detonation model, except that the early velocity is even higher than predicted by the model. Also, the ejecta are expected to be asymmetric, but the measured continuum polarization of SN~2019ein is low (0.0--0.3\%; \citealt{2022MNRAS.509.4058P}), disfavoring significant global asphericity.

In this paper, we present our observations and analysis of this SN. The photometry and spectroscopy extend from early times to the late nebular phase, enabling us to further examine possible mechanisms of SN~2019ein and HV~SNe~Ia. Section~\ref{secobs} describes the observations and data reduction. Photometric and spectroscopic results are presented in Section~\ref{secphoto} and Section~\ref{secspec}, respectively. Possible explosion mechanisms and progenitor types are discussed in Section~\ref{secdis}. Section~\ref{secsum} provides our conclusions.

\section{OBSERVATIONS}
\label{secobs}

\begin{figure*}
\center
\includegraphics[angle=0,width=1\textwidth]{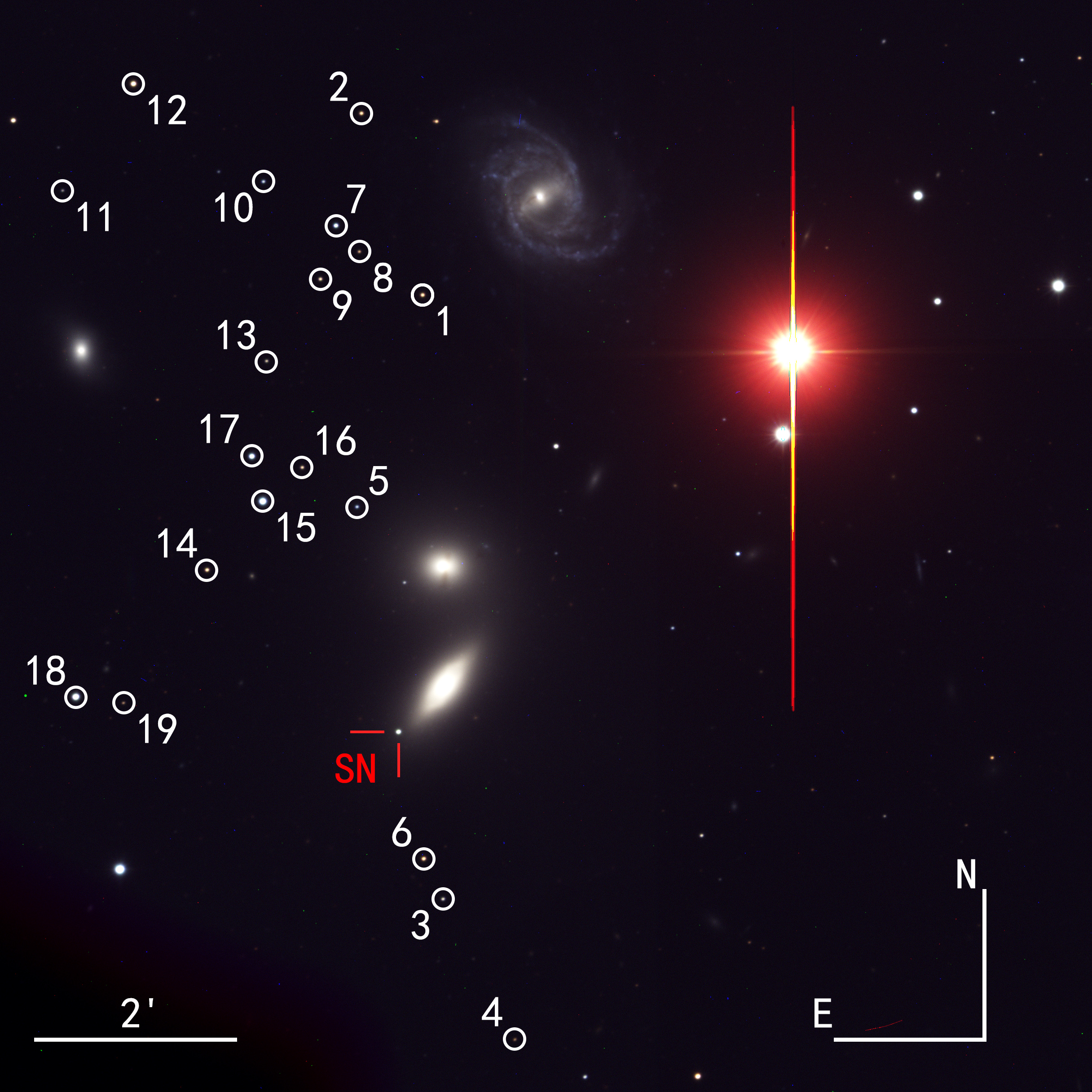}
\vspace{0.0cm}
\caption{An $RVB$-band color image of SN~2019ein (shown by the red crosshairs) in NGC~5353, taken with the AZT-22 1.5~m telescope on MJD 58654.70 ($\sim 35$~days after $B$ maximum). A scale bar is shown in the bottom-left corner. The standard stars listed in Table~\ref{tabstd} are marked. North is up and east is to the left.}
\label{figVimg} \vspace{-0.0cm}
\end{figure*}

SN~2019ein was first discovered at 18.19~mag on MJD~58604.47 (2019 May 1.47; UT dates are used throughout this paper) by the ATLAS survey \citep{2019TNSTR.678....1T} in their {\it cyan} band. Later, \cite{2019ATel12720....1I} detected the transient in $R$-band images taken by the SAO 1~m telescope on MJD~58604.44 (2019 May 1.44) and reported a nondetection on MJD~58602.75 (2019 Apr. 29.75), implying a strict constraint on its explosion time. SN~2019ein is located at J2000 coordinates $\alpha = 13^\mathrm h53^\mathrm m29^\mathrm s.110$ and $\delta = +40^\circ 16' 31''.33$, which is $27''.5$ east and $27''.9$ south of the nucleus of its host NGC~5353, a nearby lenticular galaxy with a heliocentric redshift $z=0.007755$. Figure~\ref{figVimg} shows the $BVR$-band image of SN~2019ein and NGC~5353.

\subsection {Photometric Data}
\label{secobsphoto}

Our follow-up photometric observations were collected with the Tsinghua-NAOC 80~cm telescope \citep[TNT; ][]{2008ApJ...675..626W,2012RAA....12.1585H} at Xinglong Observatory of NAOC and the AZT-22 1.5~m telescope at Maidanak Observatory \citep{2018NatAs...2..349E}. The TNT multiband photometry was obtained in the Johnson-Cousins $BV$ and Sloan $gri$ bands. The AZT observations began on 2019 June 20 (MJD~58654), lasting for $\sim 2$ months. We adopted standard \texttt{IRAF}\footnote{IRAF (Image Reduction and Analysis Facility) is distributed by the National Optical Astronomy Observatories (NOAO), which are operated by the Association of Universities for Research in Astronomy (AURA), Inc., under cooperative agreement with the National Science Foundation.} routines to reduce the CCD images, including processes of bias and flat-field corrections and cosmic-ray removal. The TNT color terms were taken from \cite{2012RAA....12.1585H}, and the extinction coefficients at the site were measured by observing \cite{1992AJ....104..340L} standards during the nights. The magnitudes are calibrated using 19 nearby reference stars from the SDSS catalog \citep{2000AJ....120.1579Y,2006AJ....131.2332G}; see Table~\ref{tabstd}. For calibration, the SDSS $ugriz$ magnitudes of the reference stars are converted to standard $UBVRI$ magnitudes\footnote{http://classic.sdss.org/dr4/algorithms/sdssUBVRITransform.html\#Lupton2005}.

As the AZT observations covered the late-phase evolution of SN~2019ein, when galaxy contamination became relatively important, we applied the template-subtraction technique to the AZT images to obtain better photometry. The template images for AZT photometry were taken on 2021 June 22, more than 26 months after the discovery of the SN. The methodology of template subtraction follows \texttt{Zrutyphot} (Mo et al., in prep.).

We also included the publicly available $gr$-band forced photometry from the Zwicky Transient Facility \citep[ZTF; ][]{2019PASP..131a8003M}\footnote{https://lasair.roe.ac.uk/object/ZTF19aatlmbo/} and $co$-band ({\it cyan}, {\it orange}) forced photometry from the Asteroid Terrestrial-impact Last Alert System \citep[ATLAS; ][]{2018PASP..130f4505T,2020PASP..132h5002S}. These photomeric results were obtained after subtraction of the corresponding template images. The ZTF $g$-band data extend to $t \approx 270$~days after $B$-band maximum, greatly helping constrain the late-time evolution of SN~2019ein.

SN~2019ein was also observed by the Las Cumbres Observatory \citep[LCO; ][]{2013PASP..125.1031B} and the Ultraviolet/Optical Telescope mounted on the {\it Swift} satellite \citep[UVOT; ][]{2005SSRv..120...95R}; the data were published by \cite{2020ApJ...897..159P}. We included the early-time LCO and {\it Swift} UVOT data (taken before $\sim 40$~days past $B$ maximum) in our analysis. Owing to the absence of template subtraction, the late-time data from the above two sources are noticeably flattened by the host-galaxy flux. 

\begin{figure*}
\center
\includegraphics[angle=0,width=0.7\textwidth]{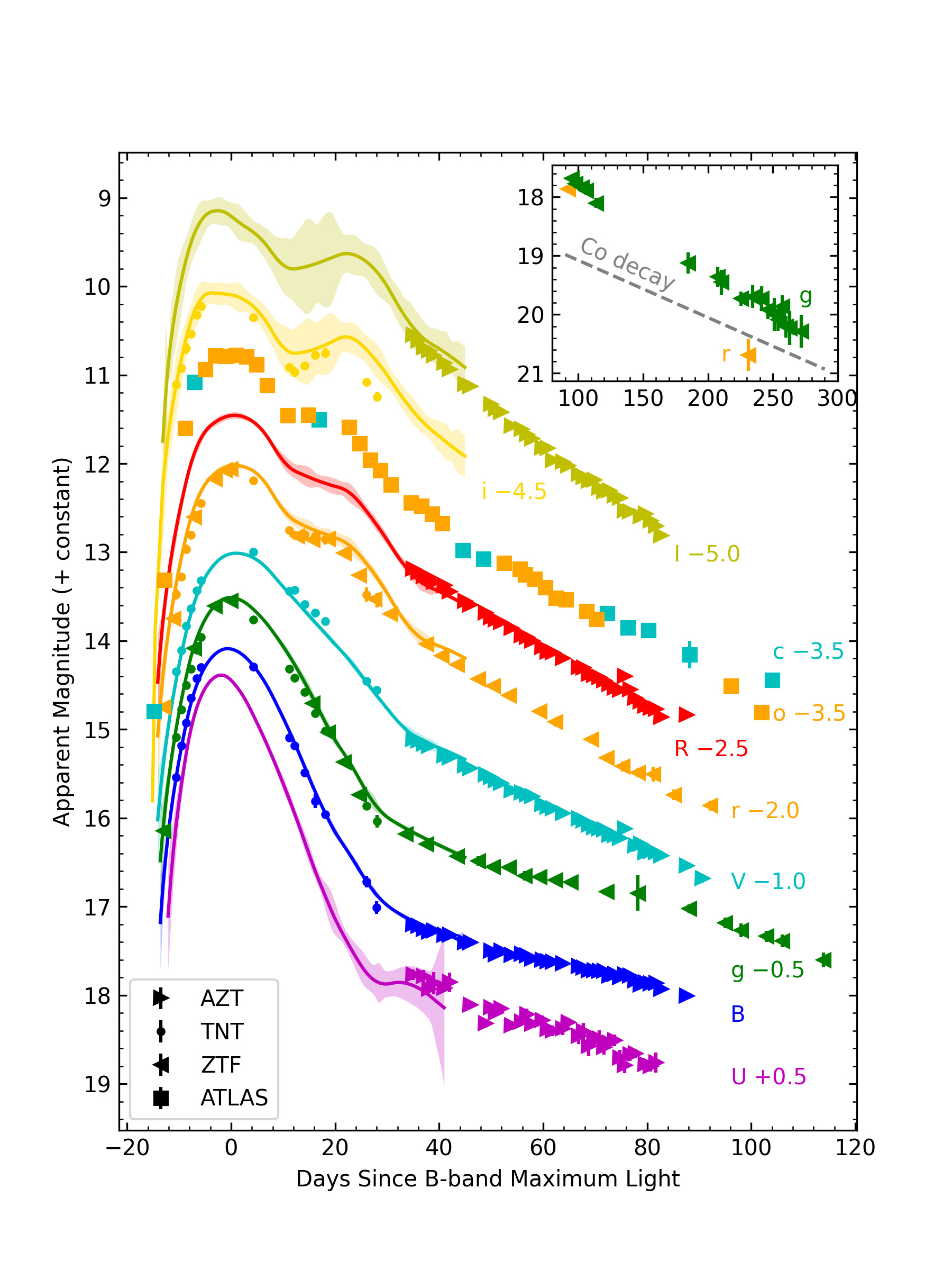}
\vspace{0.0cm}
\caption{The near-infrared, optical, and ultraviolet light curves of SN~2019ein, along with \texttt{SALT2} fits (solid lines). Data from different sources are shown with different symbols. Days are relative to $B$ maximum light on 2019-05-16.36 (MJD~58619.36).}
\label{figlc} \vspace{-0.0cm}
\end{figure*}

The multiband light curves of SN~2019ein, spanning from $\sim 4$ days to $\sim 290$ days after the explosion, are displayed in Figure~\ref{figlc}. The best-fit light curves generated by \texttt{SALT2} \citep{2007A&A...466...11G} are overplotted. 

\subsection {Spectroscopic Data}
\label{secobsspec}

Table~\ref{tabspec} presents a journal of our spectroscopic observations of SN~2019ein. The first spectrum was taken with the 2.16~m telescope at Xinglong Observatory (XLT) of NAOC on 2019 May 3, about 2 days after the discovery. We obtained 6 spectra in total with the XLT BFOSC. We also obtained 8 spectra with the YFOSC mounted on the Lijiang 2.4~m telescope \citep[LJT; ][]{2015RAA....15..918F} of Yunnan Observatories (YNAO), and 2 spectra with the 3.5~m telescope of Apache Point Observatory (APO). The last spectrum was in the nebular phase, taken $\sim 330$~days after maximum light with the Low-Resolution Imaging Spectrometer (LRIS) mounted on the Keck~I 10~m telescope \citep[Keck; ][]{1995PASP..107..375O}.

We performed standard \texttt{IRAF} routines to reduce the spectra obtained with the LJT, XLT, and APO 3.5~m telescope. Telluric lines were removed from the spectra and fluxes were calibrated with standard stars observed on the same night at similar airmasses. The spectra were further corrected for continuum atmospheric extinction using the extinction curves of local observatories. For the Keck/LRIS observation, we follow standard procedures (e.g., \citealt{2012MNRAS.425.1789S}) to extract and calibrate the 1D spectrum from the CCD data. These procedures include flat-fielding, cosmic-ray removal, optimal extraction \citep{1986PASP...98..609H}, sky subtraction, removal of telluric absorption, and flux calibration.

The spectral evolution of SN~2019ein is shown in Figure~\ref{figspec}, where a classification spectrum \citep{2019TNSCR.701....1B}, taken $\sim 2$ days after explosion \citep{2020ApJ...897..159P} using the FLOYDS spectrograph on the 2~m telescope of LCO at Haleakal\=a, is included for comparison.

\section{PHOTOMETRIC RESULTS}
\label{secphoto}

\subsection{Light Curves and Photometric Properties}
\label{secphotolc}

Multiband light curves of SN~2019ein are shown in Figure~\ref{figlc}, covering  phases from two weeks before to over 80 days after $B$-band maximum light. The overall shape of the light curve is quite normal among SNe~Ia, characterized by a secondary shoulder in the $i$ and $r$ bands. The light-curve-fitting tools \texttt{SALT2} \citep{2010A&A...523A...7G,2014A&A...568A..22B} and \texttt{SNooPy} \citep{2011AJ....141...19B,2014ApJ...789...32B} were used to fit the multiband light curve.

The $B$-band peak magnitude and the corresponding epoch derived from \texttt{SNooPy} are $m^{\rm max}_{B}=14.08 \pm 0.02~\mathrm{mag}$ and $T_{\rm Bmax} = \mathrm{MJD}~58619.29 \pm 0.07$ (respectively), while the $B$-band magnitude decline within 15 days after peak is estimated to be $\Delta m_{15} (B) = 1.32 \pm 0.06~\mathrm{mag}$. From the \texttt{SALT2} fit, we derive the maximum-light parameters as $m^{\rm max}_{B}=14.06\pm 0.02~\mathrm{mag}$ and $T_{Bmax}=\mathrm{MJD}~58619.36\pm0.02$. According to \cite{2007A&A...466...11G}, the light-curve-shape parameter $x_1$ of SALT2 can be converted into $\Delta m_{15}(B) = 1.35 \pm 0.01~\mathrm{mag}$. The results given by \texttt{SNooPy} and \texttt{SALT2} fits are consistent within the quoted uncertainties, which are also in good agreement with the estimates by \cite{2020ApJ...893..143K} ($\Delta m_{15}(B) = 1.36 \pm 0.02$~mag) and \cite{2020ApJ...897..159P} ($\Delta m_{15}(B) = 1.40 \pm 0.004$~mag). We adopt the \texttt{SALT2} value throughout the paper. The photometric parameters and distance modulus derived/adopted in this paper are listed in Table~\ref{tablcfit}, together with those used in two previous work \citep{2020ApJ...893..143K,2020ApJ...897..159P}.

\begin{figure*}
\center
\includegraphics[angle=0,width=1\textwidth]{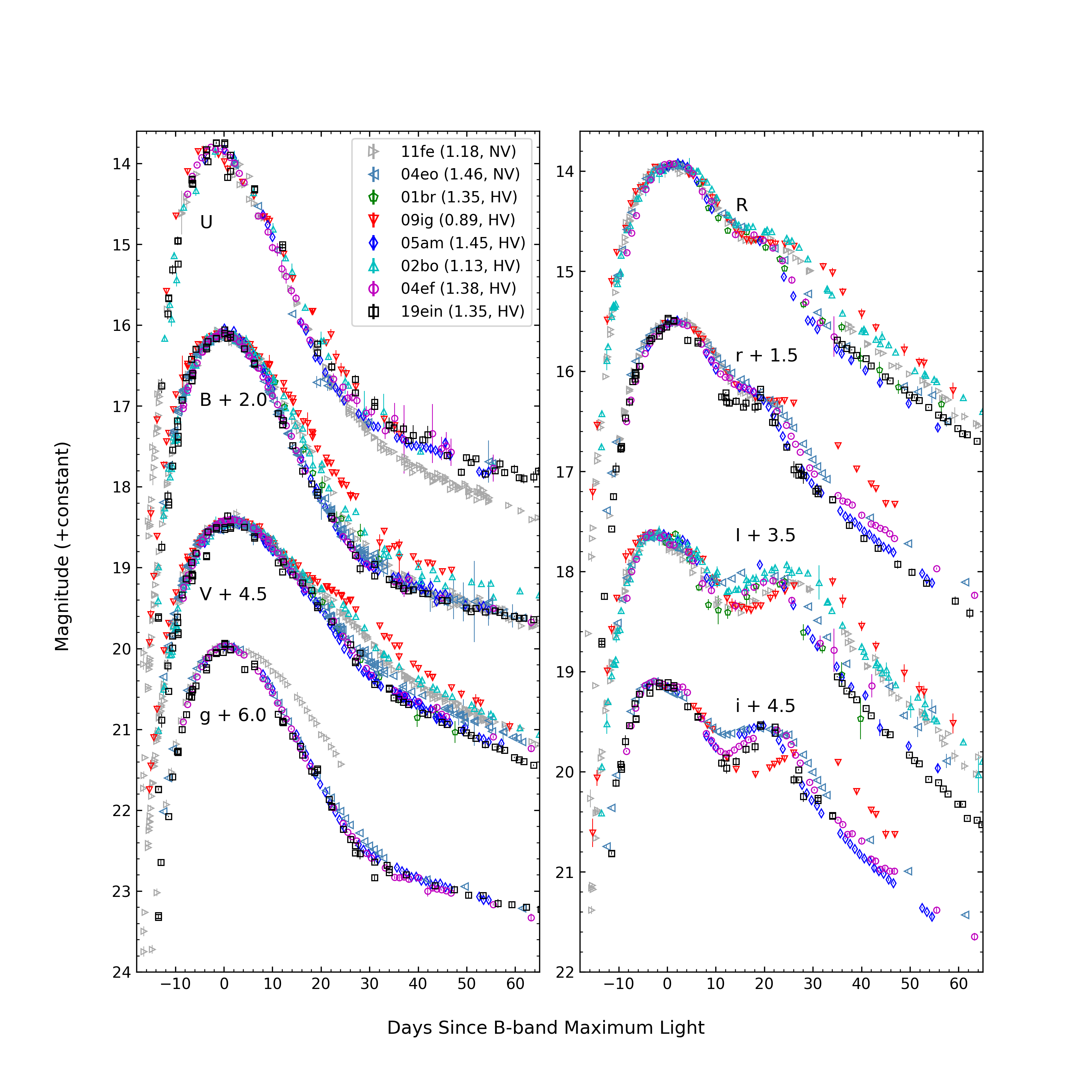}
\vspace{0.0cm}
\caption{Comparison of the multiband light curves of SN~2019ein (black squares) with other well-observed SNe~Ia, including five high-velocity objects (colored markers) and two normal-velocity objects (SN~2011fe and SN~2004eo; gray markers). The light curves of the comparison SNe are normalized to match the peak magnitudes of SN~2019ein. The post-peak decline rate ($\Delta m_{15}(B)$) and spectroscopic subclass (HV/NV) of each SN~Ia are shown in the legend. See Table~\ref{tabcmp} for references of the comparison SNe~Ia.}
\label{figlccmp} \vspace{-0.0cm}
\end{figure*}

In Figure~\ref{figlccmp}, the $UBVg$ and $RrIi$ light curves of SN~2019ein are compared with those of some well-observed SNe~Ia. The comparison sample contain three HV SNe~Ia with similar $\Delta m_{15}(B)$, including SN~2001br ($\Delta m_{15}(B)=1.35\pm 0.05$~mag), SN~2004ef ($\Delta m_{15}(B)=1.38\pm 0.05$~mag), and SN~2005am ($\Delta m_{15}(B)=1.45\pm0.07$~mag), and the other two HV objects with similar velocity near maximum light, including SN~2002bo ($\Delta m_{15}(B)=1.13$~mag) and SN~2009ig ($\Delta m_{15}(B)=0.89$~mag). Two NV~SNe~Ia, SN~2004eo ($\Delta m_{15}(B)=1.46$~mag) and SN~2011fe ($\Delta m_{15}(B)=1.18$~mag), are also included in the comparison. The detailed parameters and references of the above comparison SNe~Ia are given in Table~\ref{tabcmp}. 

The overall morphology of multiband light curves of SN~2019ein shows a close resemblance to that of HV SNe Ia with larger $\Delta m_{15}(B)$, such as SNe~2001br, 2004ef, and 2005am. As suggested by \cite{2008ApJ...675..626W,2019ApJ...882..120W}, the HV~SNe~Ia like 2002bo and 2009ig do exhibit brighter light curve tails (i.e., at $t \approx 1$--3 months after maximum light) in both $U$ and $B$ bands in comparison with their NV counterparts such as SN~2011fe, while this discrepancy tends to become smaller at larger $\Delta m_{15}(B)$ (see Fig. 4 in \cite{2019ApJ...882..120W}), as seen in SN~2019ein and other comparison SNe Ia with similar decline rates. In addition, the $r$-band shoulder and $i$-band secondary peak of SN~2019ein are more distinguishable from the main bulk of the light curves in comparison with the NV counterparts such as SN~2004eo, probably due to the faster decline of its first peak.

\subsection{Reddening and Distance}
\label{secphotored}

The Galactic reddening towards SN~2019ein is estimated to be $E(B-V)_{\rm MW} \approx 0.010$~mag \citep{2011ApJ...737..103S}, corresponding to $A_V=0.031$~mag using the standard extinction coefficient $R_V=3.1$ \citep{1989ApJ...345..245C}. After removing Galactic extinction, we compare the $B-V$ color curve with the Lira-Phillips relation \citep{1999AJ....118.1766P} and deduce a color excess of $E(B-V)=0.024\pm 0.050$~mag. This value is smaller than that derived by \cite{2020ApJ...893..143K} ($E(B-V)_{\rm host}=0.09 \pm 0.02~\mathrm{mag}$). Note that our photometry is based on template subtraction, which allows more accurate estimation of the host-galaxy reddening with late-time $B-V$ color curve. Moreover, our result is more consistent with the fact that no host-galaxy Na~I~D lines are detected in the spectra of SN~2019ein. Additionally, it has been suggested that HV~SNe~Ia are likely located in environments with $R_V$ lower than the standard value of $\sim 3.1$. For HV~SNe~Ia, \cite{2009ApJ...699L.139W} obtained a value of $R_V = 1.55 \pm 0.06$, with which for host-galaxy extinction correction the dispersion in peak luminosity of SNe~Ia can be further minimized. Here we adopt this lower $R_V$ to correct host-galaxy extinction for SN 2019ein.

Since the redshift of the host galaxy NGC~5353 is relatively small, its peculiar velocity may have significant impact on the distance estimated from Hubble's law. We checked the redshift-independent distances for NGC~5353 on the NASA/IPAC Extragalactic Database (NED)\footnote{The NASA/IPAC Extragalactic Database (NED) is funded by the National Aeronautics and Space Administration (NASA) and operated by the California Institute of Technology.}. There are multiple distance-modulus ($\mu$) measurements from the Faber-Jackson or Tully-Fisher methods. However, these results are highly uncertain, ranging from $\mu=31.91$~mag to $32.87$~mag. A most recent study by \cite{2021ApJS..255...21J} estimated $\mu = 32.711 \pm 0.076$~mag through measurements of surface brightness fluctuations, and we adopt it for our analysis. With these distance modulus and extinction values, the $B$-band maximum fitted by SALT2 could be converted into an absolute magnitude of $M_{\rm max}(B)=-18.71\pm 0.15~\mathrm{mag}$.

In Figure~\ref{figdm15} we plot SN 2019ein on $\Delta m_{15}(B)$-$M_{\rm max}(B)$ diagram along with some comparison samples. SN 2019ein is noticeably dimmer, although not as dim as 02es-likes or 91bg-likes, than that predicted by Phillips relation, sitting at the marginal region of Pantheon samples, which were used as normal SNe Ia for cosmological studies \citep{2018ApJ...859..101S}.

\subsection{Color Curves}
\label{secphotocolor}

\begin{figure*}
\center
\includegraphics[angle=0,width=1\textwidth]{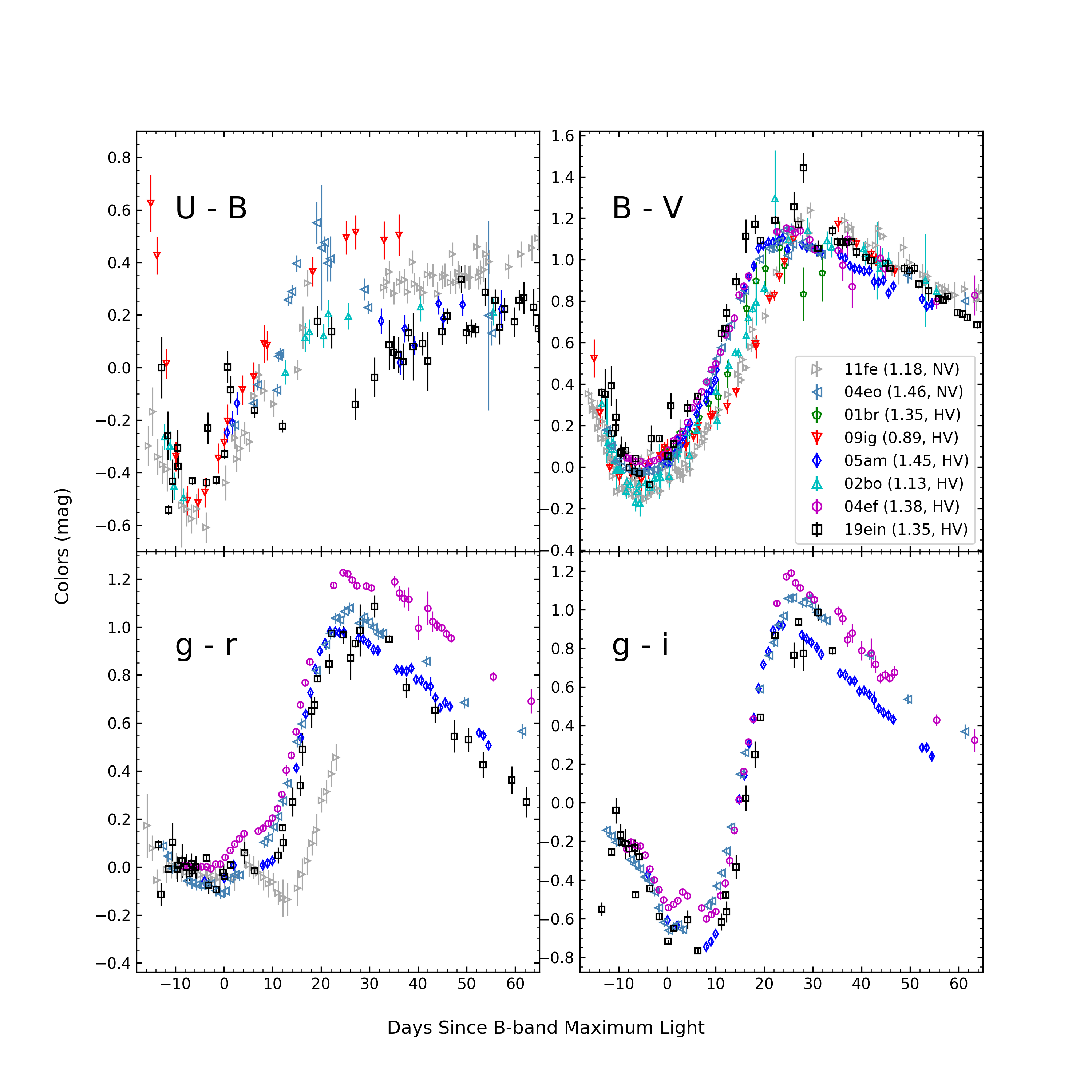}
\vspace{0.0cm}
\caption{Comparison of the $U-B$, $B-V$, $g-r$, and $g-i$ colors of SN~2019ein with those of some well-observed SNe~Ia. All of the color curves are corrected for their Galactic and host-galaxy reddenings. The comparison sample of SNe~Ia and the corresponding symbols are the same as in Figure~\ref{figlccmp}. The data sources are cited in Table~\ref{tabcmp}.}
\label{figcolorcmp} \vspace{-0.0cm}
\end{figure*}

\begin{figure*}
\center
\includegraphics[angle=0,width=1\textwidth]{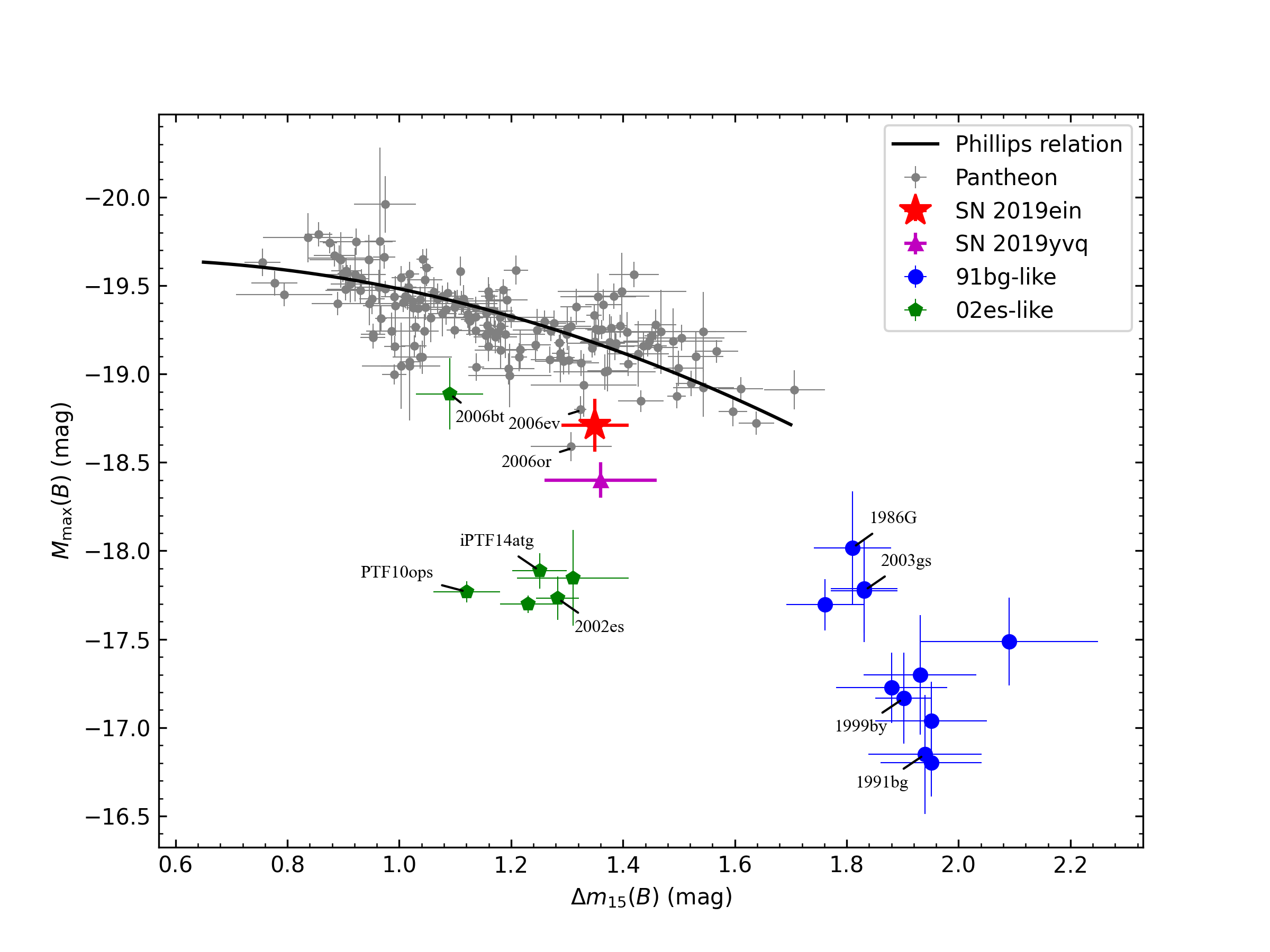}
\vspace{0.0cm}
\caption{Absolute $B$-band maximum magnitudes of a sample of SNe~Ia vs. their corresponding $\Delta m_{15}(B)$ values. Normal SNe~Ia from Pantheon samples \citep{2018ApJ...859..101S} are shown as gray dots. The black solid curve represents the best-fit Lira-Phillips relation \citep{1999AJ....118.1766P}. The SN~1991bg-like and SN~2002es-like samples are taken from \protect\cite{2017hsn..book..317T} and shown as blue and green dots, respectively. The red star and purple triangle represent SN~2019ein (this paper) and SN~2019yvq \citep{2021ApJ...919..142B}, respectively.}
\label{figdm15} \vspace{-0.0cm}
\end{figure*}

\begin{figure*}
\center
\includegraphics[angle=0,width=0.8\textwidth]{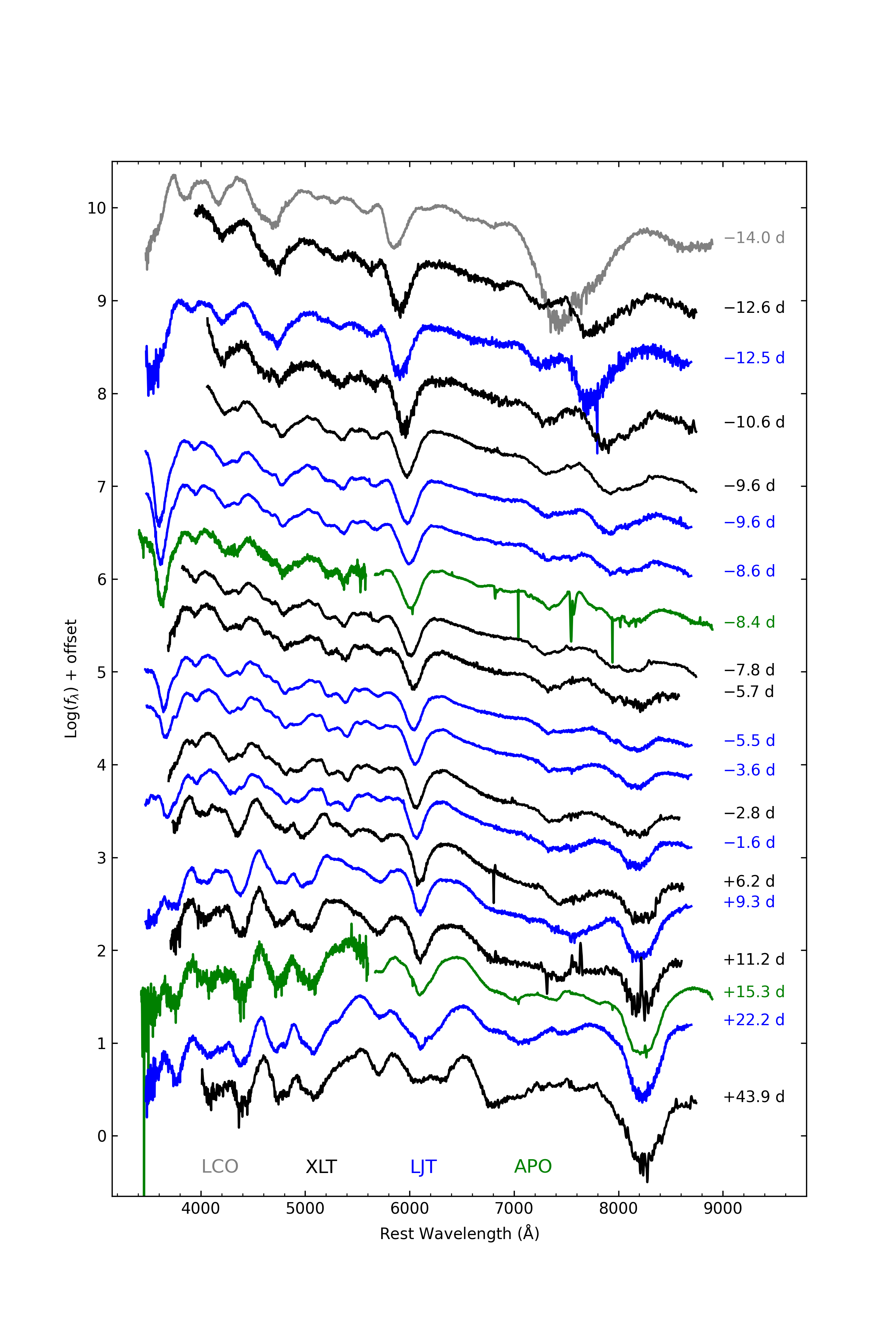}
\vspace{0.0cm}
\caption{Spectral evolution of SN~2019ein, spanning the phases from $-12.6$~days to +43.9~days relative to $B$-band maximum brightness. All spectra have been corrected for the redshift of the host galaxy ($z_{\mathrm{helio}} = 0.007755$). Epochs on the right side of the spectra represent the phases in days with respect to $B$ maximum. Colors indicate the instrument used for observations: black from the XLT, blue from the LJT, and green from the APO 3.5~m telescope. The uppermost spectrum, shown as a gray solid line, is taken from \protect\cite{2020ApJ...897..159P}. Telluric absorption was not removed from the $t = -8.4$~days spectrum.}
\label{figspec} \vspace{-0.0cm}
\end{figure*}

Figure~\ref{figcolorcmp} shows the color evolution of SN~2019ein compared with that of other well-observed SNe~Ia as mentioned in Section~\ref{secphotolc}. All the data are corrected for both Galactic and host-galaxy extinctions. At early times, the $U-B$ and $B-V$ colors of SN~2019ein and the comparison SNe~Ia rapidly evolve blueward and reach their peak values at $t \approx -5$~days relative to $B$-band maximum light, but the HV subclass seems to evolve faster than the NV subclass. After the blue peak, all SNe~Ia evolve redward until they reach the red peak at $t \approx 25$~days. The overall $B-V$ color evolution of SN~2019ein is more similar to that of SNe~2005am and 2004ef. Like the $B-V$ color curve, the $g-r$ and $g-i$ color curves are found to evolve blueward at early times. After $t \approx 10$~days, the $g-r$ and $g-i$ curves show reverse color evolution and become progressively redder until reaching the peak at $t \approx 25$~days after $B$ maximum. A noticeable outlier is the first LCO photometric data point in the $g-i$ color curve, which is very blue with $g-i= -0.55 \pm 0.03$~mag. This is likely caused by the presence of high-velocity Ca~II absorption features, which were visible in the $t\approx -14.0$ day LCO spectrum (see Section~\ref{secspecevol}).

\subsection{Explosion-Time Estimate from Light Curves}
\label{secphotoexpt}

The photometric observations started at $t \approx 2.6$~days after the explosion. With such early-time coverage of the light curve, we can estimate the explosion time by adopting a homologously expanding ``fireball'' model described by \cite{2020ApJ...893..143K}. In the fireball model, the luminosity $L$ increases with post-explosion time $t$ as $L \propto t^2$. The explosion time is assumed to be the same in all bands.

In our analysis, we adopt a $t_{\rm cutoff}$ parameter, only using photometric data earlier than $t_{\rm cutoff}$ to perform the fitting. We also define a reduced $\chi^2$ as an indicator of the goodness of fit,
\begin{equation}
\chi^2_\nu=\frac 1 {\sum_F{(N_F-1)}-1}{\sum_{F,i}\left(\frac{f_{F,i}-A_F(t-t_0)^2}{\sigma_{f,F,i}}\right)^2}\, ,
\end{equation}
where $F$ denotes different filters, $N_F$ is the number of data points in filter $F$, and $t_{F,i}$ and $f_{F,i}\pm\sigma_{f,F,i}$ respectively represent the time and flux of the $i$-th data point in filter $F$. Also, $t_0$ is the tentative explosion time (more precisely, the time of ``first light''), and $A_F$ is the best-fit coefficient for filter $F$, which could be calculated when $t_0$ is given.

Minimizing the $\chi^2_\nu$ can give the best-fit value of the explosion time. However, we found that the result is consistent for a range of selected $t_{\rm cutoff}$. If we adopt $t_{\rm cutoff}=-7$~days and drop data points later than $-7$~days relative to $B$ maximum, the best-fit explosion time is $t_0= \mathrm{MJD}~58603.20 \pm 0.30$, suggesting a rise time of $16.04\pm 0.30$~days. For $t_{\rm cutoff}=-9$ and $-5$ days, the derived rise time becomes 16.02 and 16.19~days, respectively. We thus adopt $t_{\rm rise} = 16.04 \pm 0.30$~days. The determination of rise time will be further discussed in Section~\ref{secspecexpt}.

\subsection{Bolometric Light Curve and Mass of \texorpdfstring{$^{56}$Ni}{Ni56}}
\label{secphotobolo}

Following the procedure of \cite{2019ApJ...870...12L}, we use \texttt{SNooPy2} to construct the spectral energy distribution (SED) and hence the bolometric light curve based on the $UuBVgRrIi$-band photometry. With the multiband photometry, the maximum bolometric luminosity is estimated to be $L \approx 7.0 \times 10^{42}~\mathrm{erg~s^{-1}}$ for SN~2019ein. Adopting a rise time of $t_{\rm rise} \approx 16.0$~days, we estimate the mass of $^{56}\mathrm{Ni}$ synthesized in the explosion as $M_{\rm Ni} \approx 0.31$~M$_\odot$ according to the Arnett's rule \citep{1982ApJ...253..785A,2005A&A...431..423S}.

We also apply the radiation diffusion model of \cite{1982ApJ...253..785A} \citep[see also][]{2012ApJ...746..121C, 2013ApJ...773...76C} to fit the entire bolometric light curve. The bolometric light curve along with the analytic Arnett model is shown in Figure~\ref{figcmppolin} and Figure~\ref{figcmpshen}. The parameters that need to be fitted include $t_0$, the initial mass of the radioactive nickel $M_{\rm Ni}$, the light-curve timescale $t_{\rm lc}$, and the gamma-ray leaking timescale $t_\gamma$. The best-fit values of the above parameters are $t_0 =$ MJD $58604.6\pm 0.9$, $M_{\rm Ni}=(0.27\pm 0.04)$~M$_\odot$, $t_{\rm lc}=11.66 \pm 2.53$~days, and $t_\gamma=30.26\pm 4.49$~days. The best-fit value of $M_{\rm Ni}$ is slightly lower than that estimated from Arnett's rule. Moreover, the best-fit explosion time is $\sim 1.4$~days later than that estimated from the fireball model, and is even later than the first-detection epoch (MJD 58604.4). This difference might be interpreted as a ``dark phase'' \citep{2013ApJ...769...67P,2016ApJ...826...96P} caused by the location of the radioactive $^{56}$Ni within the ejecta. \cite{2016ApJ...826...96P} predict the length of the dark phase to be $\lesssim 2$~days, consistent with our results.

\section{SPECTROSCOPIC RESULTS}
\label{secspec}

\subsection{Spectral Evolution}
\label{secspecevol}

The spectral evolution of SN~2019ein, spanning from $t \approx -12.6$ to +43.9~days relative to $B$ maximum, is shown in Figure~\ref{figspec}, where the $t \approx -14.0$~day LCO spectrum is overplotted to reveal the rapid evolution. The earliest spectrum is characterized by a deep and broad absorption feature over the 7000--8000~\AA\ region, which could be attributed to the blending of O~I and Ca~II NIR triplet absorption features \citep{2020ApJ...897..159P}. This feature evolved very rapidly, with the O~I absorption and Ca~II NIR triplet being clearly separated in our earliest spectrum taken 1.4~days later, implying that either O~I $\lambda$7774 or the high-velocity feature (HVF) of the Ca~II NIR triplet is very strong and disappears rapidly after the explosion.

\begin{figure*}
\center
\includegraphics[angle=0,width=1\textwidth]{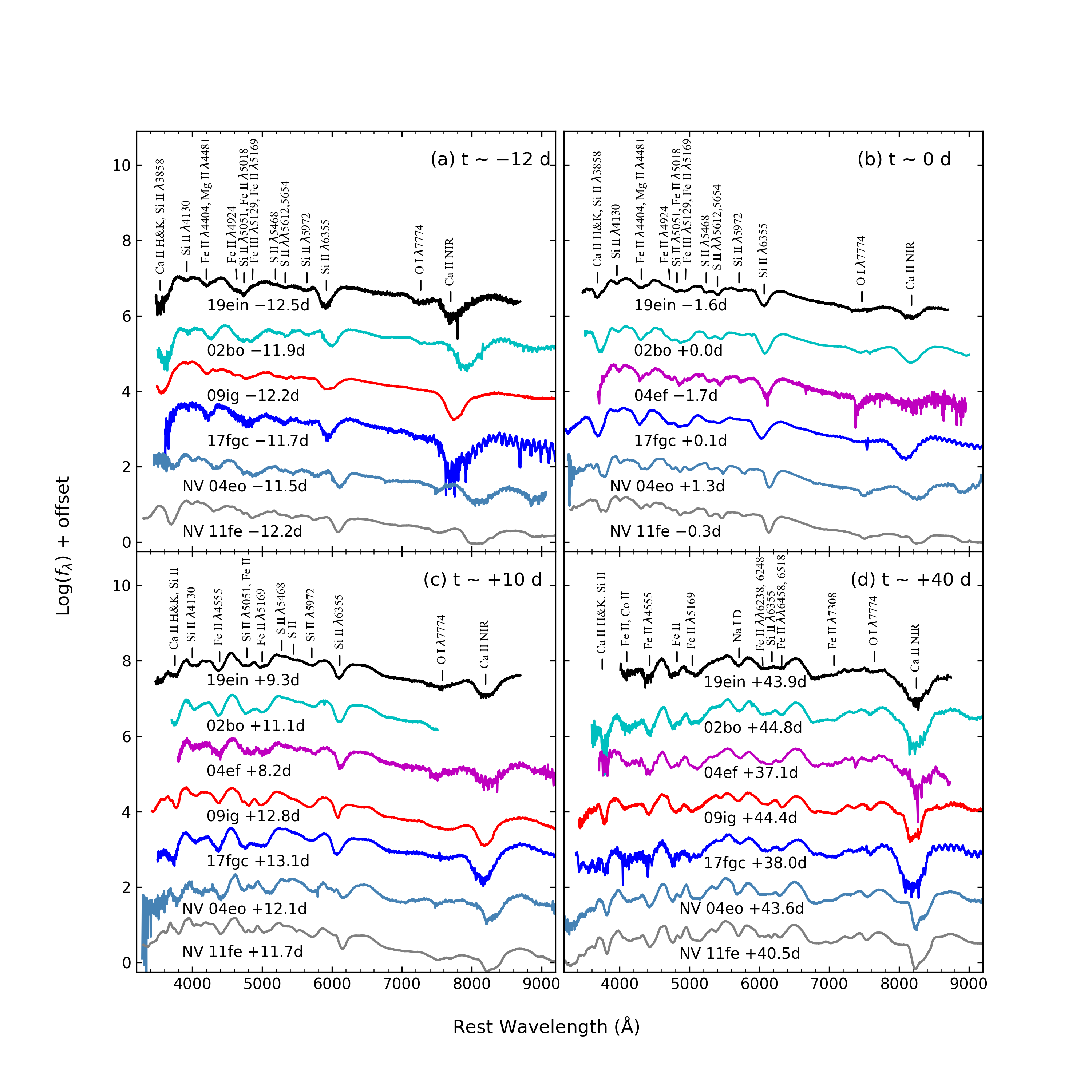}
\vspace{0.0cm}
\caption{Comparison of the spectra of SN~2019ein (at $t \approx -12$, 0, +10, and +40~days after $B$ maximum) with those of other SNe~Ia at similar phases, including HV sample like SNe~2002bo, 2004ef, 2009ig, 2017fgc, and NV sample like SN~2004eo and 2011fe. All spectra have been corrected for the redshift of their host galaxy.}
\label{figspeccmp} \vspace{-0.0cm}
\end{figure*}

Figure~\ref{figspeccmp} shows the spectral comparison between SN~2019ein and several other well-observed SNe~Ia at four selected epochs. The comparison sample includes the HV subclass of SNe~Ia such as SNe 2002bo, 2004ef, 2009ig, and 2017fgc, as well as the NV subclass such as SNe 2004eo and 2011fe. The photometric and spectroscopic parameters along with references for these samples are listed in Table~\ref{tabcmp}, where the relevant parameters of the comparison sample used for later discussions of explosion models are also listed. 

At $t \approx -12$~days (Fig.~\ref{figspeccmp}a), the spectrum of SN~2019ein is characterized by singly ionized lines of intermediate-mass elements (IMEs) such as Si, S, Mg, and Ca. The overall spectral shape at this epoch is similar to that of SN~2002bo, but with significantly higher velocity of Si~II and Ca~II NIR absorption features. The O~I $\lambda$7774 absorption is also present in SN~2019ein, whereas it is absent in the HV~SNe~Ia like SNe~2009ig and 2017fgc. Moreover, the Fe~II/Fe~III absorption feature at $\sim 4800$~\AA\ is found to be deeper than in normal SNe~Ia like SN~2011fe and some HV~SNe~Ia such as SN~2009ig.

Around maximum brightness (see Fig.~\ref{figspeccmp}b), the HVFs of the Ca~II NIR triplet absorption almost disappear in SN~2019ein and the comparison sample, and the double S~II lines at $\sim 5300$~\AA\ (the so-called ``W''-shaped feature) develop well. At this phase, the spectrum of SN~2019ein is quite similar to that of SN~2002bo, except for having less-pronounced absorption features. By $t \approx 10$~days after maximum light (see Fig.~\ref{figspeccmp}c), the broad absorption due to blended Fe~II/Fe~III lines near 5000~\AA\ split into one absorption at $\sim 4800$~\AA\ and another one near 5000~\AA, and the W-shaped profile of the S~II line almost disappeared in SN~2019ein and the comparison sample.

By $t \approx 40$~days (see Fig.~\ref{figspeccmp}d), the spectral profiles of all SNe~Ia become quite homogeneous. During this transition to the early nebular phase, the Si~II $\lambda$6355 absorption is severely affected by nearby Fe~II lines which gradually dominate the spectra. The Si~II $\lambda$5972 absorption is replaced by Na~I~D, and the O~I $\lambda$7774 absorption seems to become narrower and more visible relative to the earlier phases.

\subsection{Ejecta Velocity}
\label{secspecvel}

\begin{figure*}
\center
\includegraphics[angle=0,width=1\textwidth]{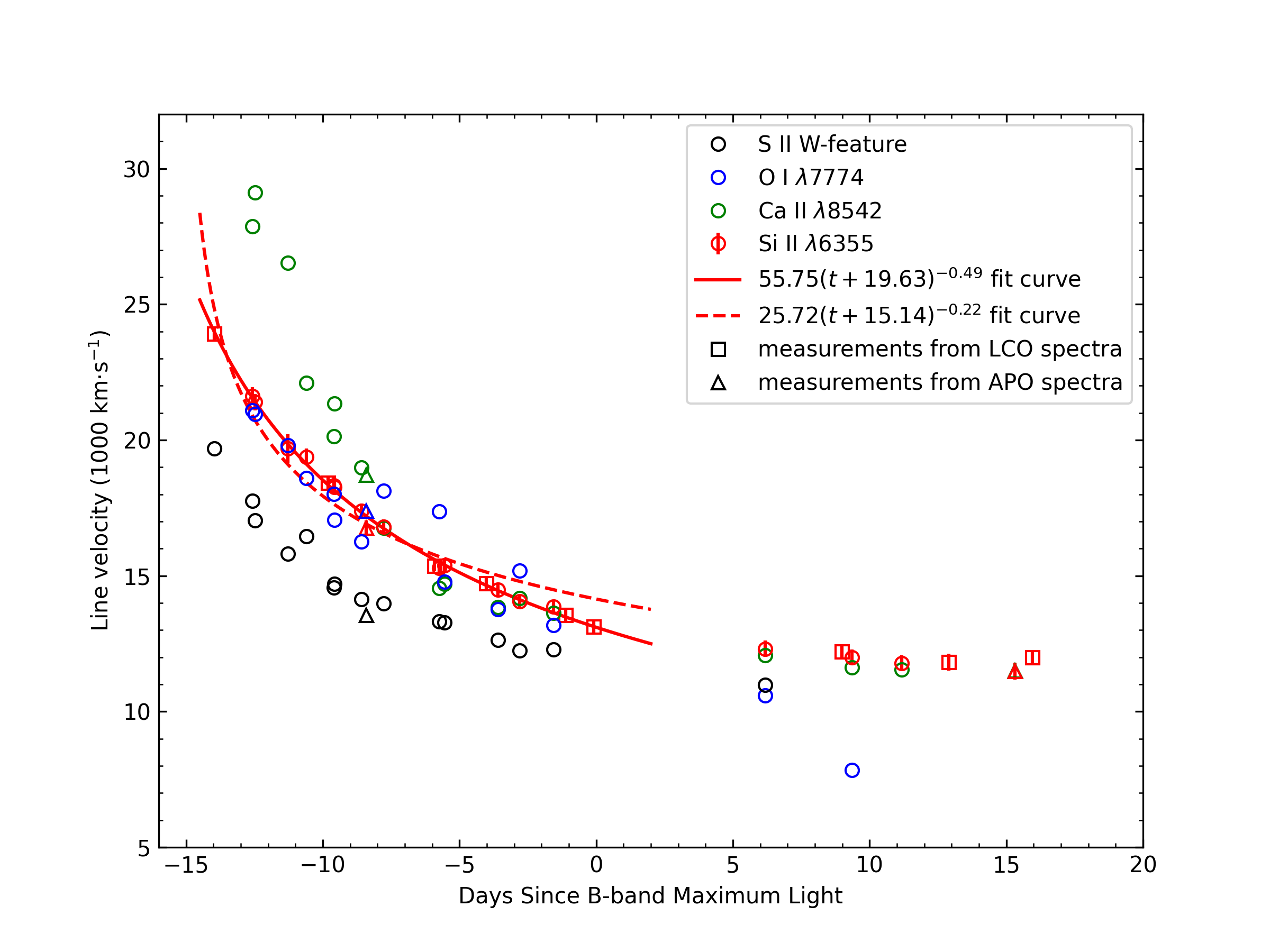}
\vspace{0.0cm}
\caption{Evolution of spectral-line velocities inferred from absorption minima of lines of Si~II $\lambda$6355, Ca~II $\lambda$8542, O~I $\lambda$7774, and the S~II ``W'' feature ($\lambda$5468 and $\lambda\lambda$5612, 5654). The red solid and dashed curves represent the power-law fit to the early-time Si~II velocity evolution, with the corresponding index being the best-fit value $\beta = 0.49$ and a fixed value of $\beta = 0.22$ from \protect\cite{2013ApJ...769...67P}, respectively. }
\label{figspecvel} \vspace{-0.0cm}
\end{figure*}

The velocity evolution of SN~2019ein measured from absorption lines of some IMEs, such as S~II $\lambda\lambda$5460, 5640, O~I $\lambda7774$, Ca~II $\lambda8542$, and Si~II $\lambda6355$, are shown in Figure~\ref{figspecvel}. The velocity of the Ca~II NIR triplet features at $t \approx -14.0$~days is $\sim 29,000~\mathrm{km~s}^{-1}$, much higher than that inferred from Si~II $\lambda 6355$ absorption ($\sim 23,900~\mathrm{km~s}^{-1}$). The early-time Ca~II velocity drops very quickly, and the velocity of Ca~II is comparable to that of Si~II after $t \approx -7$~days. We do not find a significant high-velocity feature (HVF) in the Ca~II NIR triplet of SN~2019ein as seen in some HV~SNe~Ia (e.g., SNe~2012fr and 2017fgc) except for the very first spectrum discussed in Section~\ref{secspecevol}.

At the time of $B$ maximum, the velocity of Si~II $\lambda 6355$ is measured as $13,130 \pm 240~\mathrm{km~s}^{-1}$, which is larger than the typical value ($\lesssim 11,800~\mathrm{km~s}^{-1}$) inferred for normal SNe~Ia. SN~2019ein can thus be put into the HV category of SNe~Ia according to the criteria proposed by \cite{2009ApJ...699L.139W}. After maximum light, the Si~II velocity drops and remains at a high plateau of $\sim 12,000~\mathrm{km~s}^{-1}$. Applying a linear fit to the Si~II velocity evolution from $t \approx 0$~days to $t \approx +14$~days, the velocity gradient is $\dot{v}=105 \pm 20~\mathrm{km~s}^{-1}\mathrm{day}^{-1}$, making SN~2019ein a high-velocity-gradient (HVG) SN~Ia according to the criteria suggested by \cite{2005ApJ...623.1011B}. This velocity gradient is typical compared to other HVG samples \citep{2012AJ....143..126B,2013ApJ...773...53F}. Although the early time velocity of SN~2019ein is extremely high, the unusually rapid decline before the maximum light makes its velocity at and after the maximum light not so extreme.

\begin{figure*}
\center
\includegraphics[angle=0,width=1\textwidth]{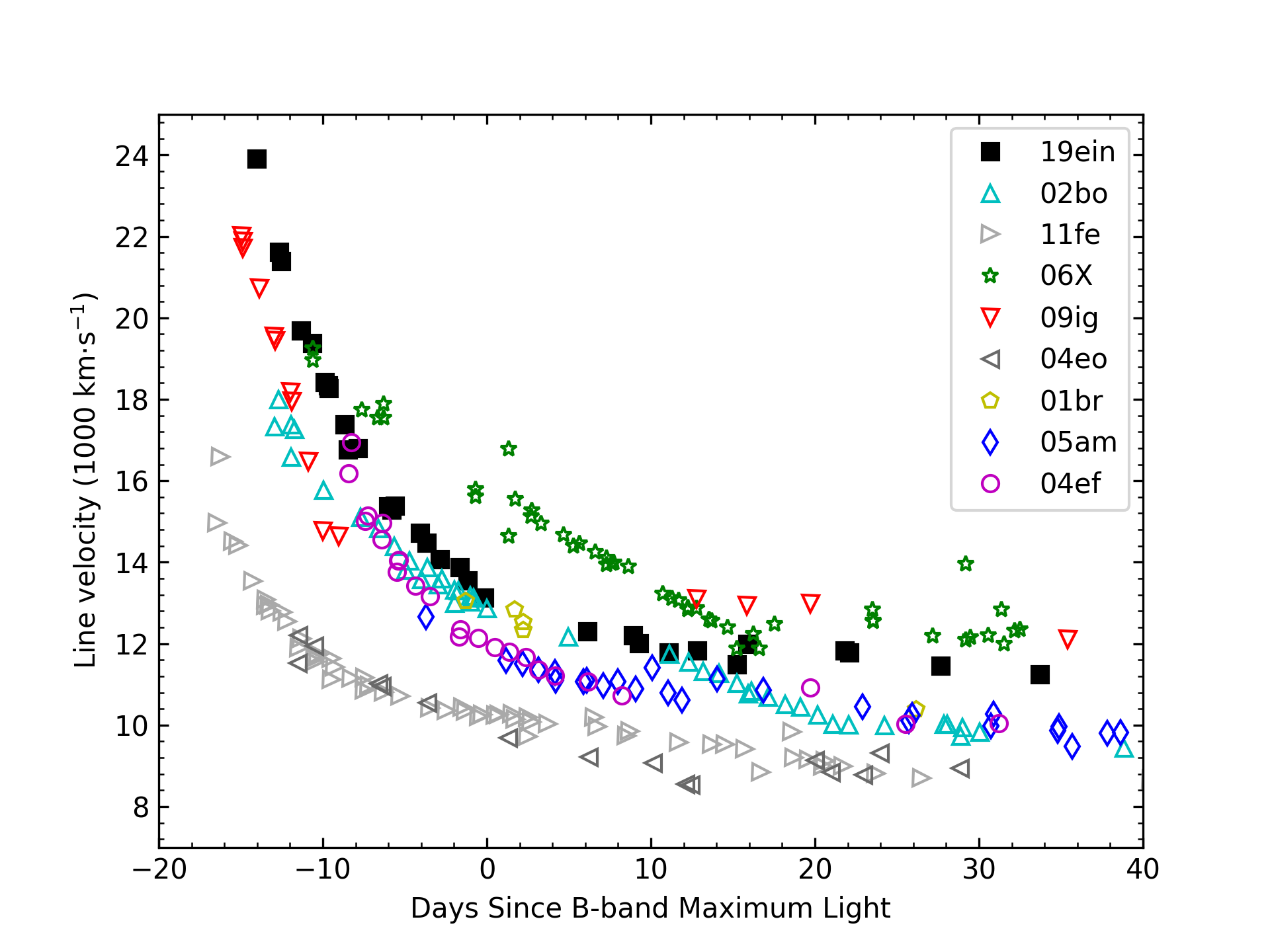}
\vspace{0.0cm}
\caption{Evolution of Si~II $\lambda$6355 velocity obtained for SN~2019ein. The comparison sample includes SNe~2002bo, 2001br, 2004ef, 2005am, 2009ig, 2004eo, and 2011fe, with the first five belonging to the HV subclass and the last two belonging to the NV subclass.}
\label{figSivelcmp} \vspace{-0.0cm}
\end{figure*}

Figure~\ref{figSivelcmp} shows the velocity evolution of Si~II $\lambda 6355$, together with that of some comparison SNe~Ia, including the HV sample like SNe~2002bo, 2006X, 2009ig, 2013gs, and 2017fgc, and two NV objects like SNe~2004eo and 2011fe. The early-time Si~II velocity of SN~2019ein is very high even among the HV~SNe~Ia, while it also shows an unusually rapid decrease before maximum light. The overall velocity evolution of SN~2019ein is similar to that of SN~2009ig, but with relatively higher early-time velocities and lower post-peak values.

\subsection{Explosion-Time Estimate from Si~II lines}
\label{secspecexpt}

\cite{2013ApJ...769...67P} developed a model of ejecta velocity to trace the early-time evolution of an SN explosion. According to their model, the velocity of Si~II lines can be fitted by $v \propto (t-t_0)^\beta$, where $t_0$ is the explosion time and $\beta$ is the power-law index. 

We fit the parameter $t_0$ and $\beta$ by minimizing the reduced goodness-of-fit parameter $\chi_\nu^2$ as well. The best-fit parameter $\beta$ is $\sim -0.5$, which differs significantly from the value $\beta=-0.22$ suggested by \cite{2013ApJ...769...67P}. Figure~\ref{figspecvel} shows the velocity evolution along with the best-fit functions for $\beta=-0.22$ and $\beta=-0.49$. Substantial discrepancy exists between the observed data points and those predicted by using $\beta=-0.22$. The discrepancy may be related to failed model assumptions like spherical symmetry of the ejecta.

Adopting the best-fit $\beta$ value, however, the corresponding rise time is calculated to be 19.6~days, significantly larger than the value derived from the fireball model in Section~\ref{secphotoexpt}. If $\beta$ is fixed at $-0.22$, the deduced explosion time is $t_0 =$ MJD $58604.1$ and the resultant rise time is 15.14~days. This estimate lies between the explosion times derived from the fireball model and the bolometric light-curve fitting, and is $\sim 0.3$~days earlier than the first detection.

\subsection{Nebular-Phase Spectra and Late-Time [Fe~II] Velocity}
\label{secspecneb}

\begin{figure*}
\center
\includegraphics[angle=0,width=1\textwidth]{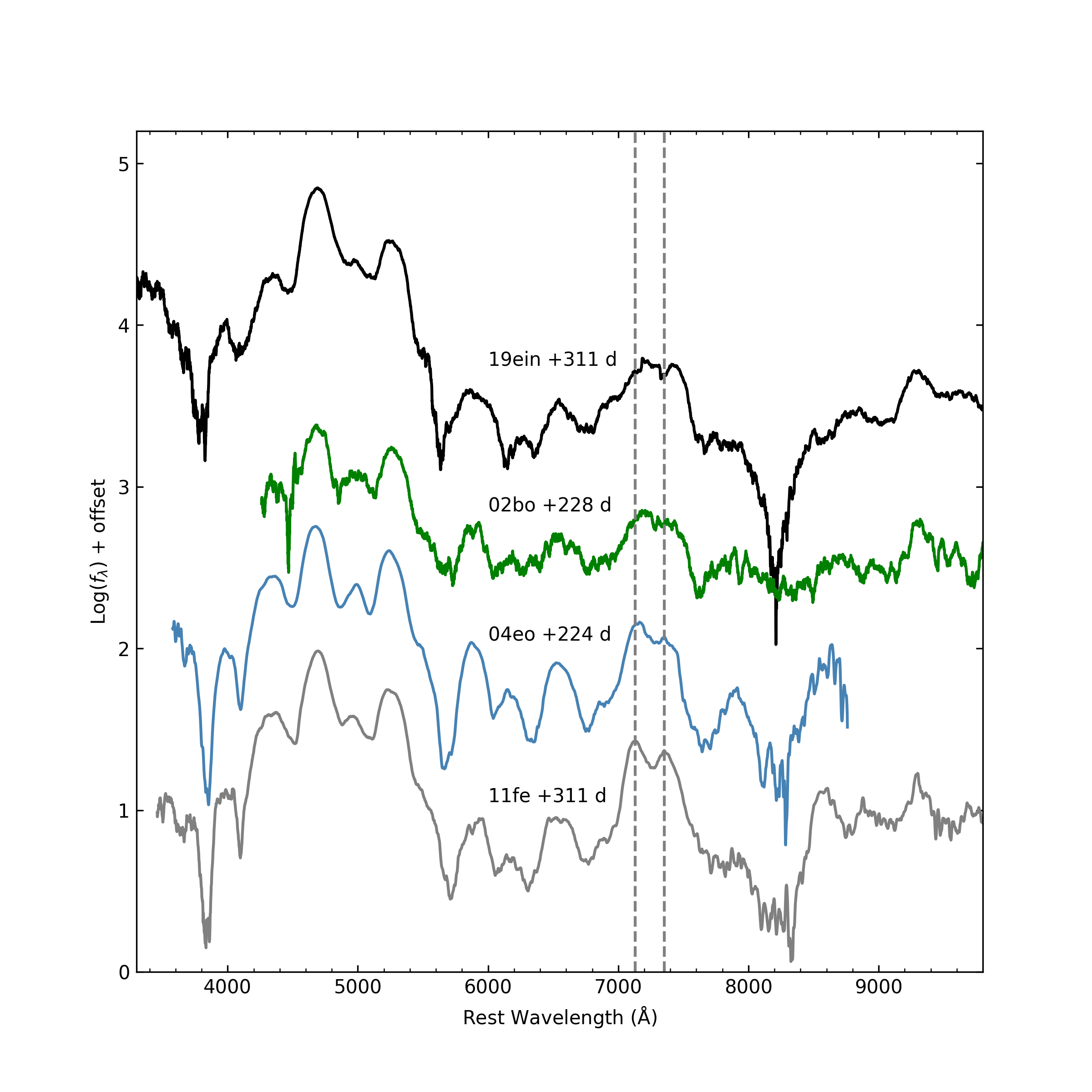}
\vspace{0.0cm}
\caption{Keck-I nebular-phase spectrum of SN~2019ein, compared with those of SNe 2002bo, 2004eo, and 2011fe. All of the spectra are smoothed by a moving-average filter to reduce the noise. The smoothing window is $\sim 60$~\AA\ for SN~2019ein. Two emission peaks of SN~2011fe ([Fe~II] $\lambda$7155 and [Ni~II] $\lambda$7378) are marked by gray dashed lines for comparison.}
\label{fignebcmp} \vspace{-0.0cm}
\end{figure*}

SN~2019ein has a Keck/LRIS spectrum taken at $t \approx 311$~days after maximum light, enabling us to measure the velocity shift of iron-group elements (IGEs) in the nebular phase. Figure~\ref{fignebcmp} compares the nebular spectrum of SN~2019ein with those of the NV subclass (including SNe~2004eo and 2011fe) and the HV subclass (SN~2002bo). Following the method described by \cite{2018MNRAS.477.3567M}, for SN~2019ein we measured the velocity shift of [Fe~II] $\lambda$7155 to be $1125 \pm 688~\mathrm{km~s^{-1}}$ and of [Ni~II] $\lambda$7378 to be $2223 \pm 257~\mathrm{km~s^{-1}}$. The results are consistent with the previous discovery that HV~SNe~Ia tend to have redshifted IGE velocities.

\begin{figure*}
\center
\includegraphics[angle=0,width=1\textwidth]{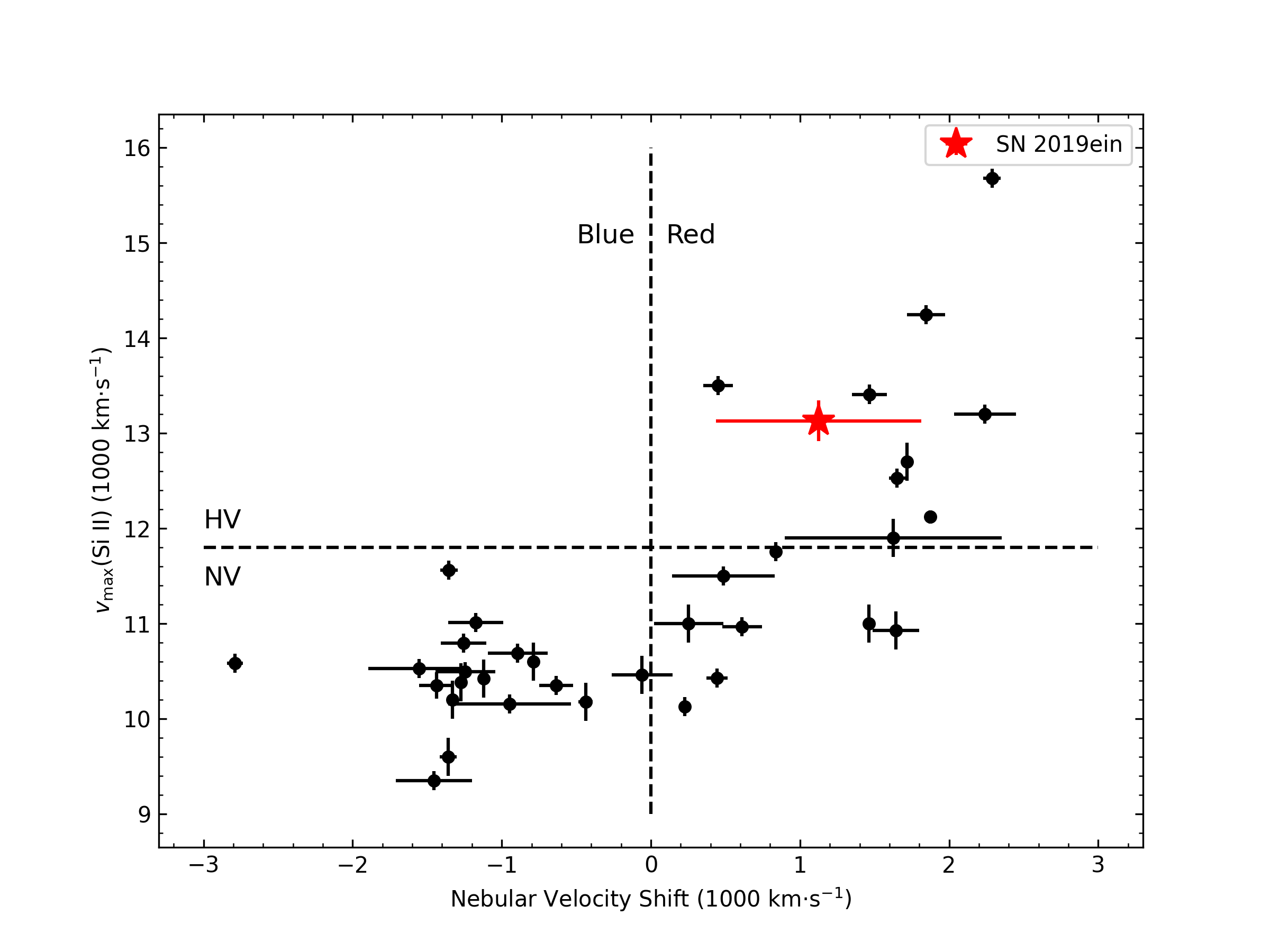}
\vspace{0.0cm}
\caption{The Si~II velocity of SN~2019ein measured from the near-maximum-light spectrum vs. the [Fe~II] velocity shift inferred in the $t \approx 311$~day spectrum, compared with the sample collected by \protect\cite{2021ApJ...906...99L}. The horizontal dashed line marks the division between the HV and NV subclasses, while the vertical dashed line shows no [Fe~II] shift.}
\label{fignebvel} \vspace{-0.0cm}
\end{figure*}

The nebular-phase velocity of the Fe~II emission line and the near-maximum-light velocity of Si~II ($v_{\rm max}$(Si)) measured for a larger sample collected by \cite{2021ApJ...906...99L} are shown in Figure~\ref{fignebvel}, where the corresponding values of SN~2019ein are overplotted for comparison. One can see that all HV~SNe~Ia have redshifted [Fe~II] emission in their nebular spectra, while this feature can be redshifted or blueshifted in the late-time spectra of NV~SNe~Ia.

The discrepancy of late-time emission features of IGEs in spectra of SNe~Ia has been proposed to originate from asymmetric explosions \citep{2010Natur.466...82M}. In such a scenario, the inner ejecta containing the IGEs may experience a velocity kick opposite to the outer ejecta that trace the photospheric velocity. For HV~SNe~Ia, the large early photospheric velocity can be explained as the outer ejecta being asymmetrically moving toward us, while the inner region should move away from us and form redshifted nebular-phase IGE lines. Recently, \cite{2021ApJ...906...99L} found that all HV objects in their 16 SNe~Ia sample have redshifted [Fe~II] velocities, and they proposed that with a geometric/projected effect, a single model of asymmetric explosion (e.g., He-detonation model) may account for the HV and a portion of the NV~SNe~Ia. However, orientation effects of a single explosion mechanism cannot account for the observed preferences of explosion sites of HV~SNe~Ia \citep{2013Sci...340..170W,2015MNRAS.446..354P,2020ApJ...895L...5P}. It is possible that HV~SNe~Ia can be produced by multiple factors in parallel, such as higher metallicity, distinctive explosion mechanism, or asymmetric explosion.

\section{DISCUSSION}
\label{secdis}

\subsection{Carbon Imprint}
\label{secdiscarbon}

The amount and velocity distribution of unburnt materials (C+O) within the expanding ejecta can be useful in distinguishing various explosion mechanisms of SNe~Ia. While oxygen is relatively common in spectra of SNe~Ia, carbon features are present in the spectra of only a small fraction of SNe~Ia. Previous studies show that 20--30\% of SNe~Ia show carbon signatures in their $t < -5$~day (with respect to maximum light) spectra \citep{2011ApJ...743...27T,2012MNRAS.425.1917S}. This ratio could reach $\sim 40$\% in earlier-time spectra ($t < -10$~days; \citealt{2014MNRAS.444.3258M}). Most of the objects with carbon features are NV~SNe~Ia \citep{2011ApJ...732...30P}, while there are either no carbon features or they are very weak and disappear very quickly in HV objects. This tendency might be explained by a viewing-angle effect based on the double-detonation model \citep{2021ApJ...906...99L}, where carbon is more prominent when viewing from the C/O detonation side and becomes hardly detectable when observing from the opposite side.

The delayed-detonation model is another candidate explosion mechanism. The produced total amount of unburnt carbon is suitable, and is only detectable in a portion of events, when the observed line of sight intersects the carbon-rich ``pockets'' in inhomogeneous ejecta \citep{2010ApJ...712..624M}. The deficiency of carbon in the spectra of HV~SNe~Ia may be explained by the outer part of the exploding WDs experiencing more complete burning than the NV counterparts during the delayed detonation. 

\begin{figure*}
\center
\includegraphics[angle=0,width=0.8\textwidth]{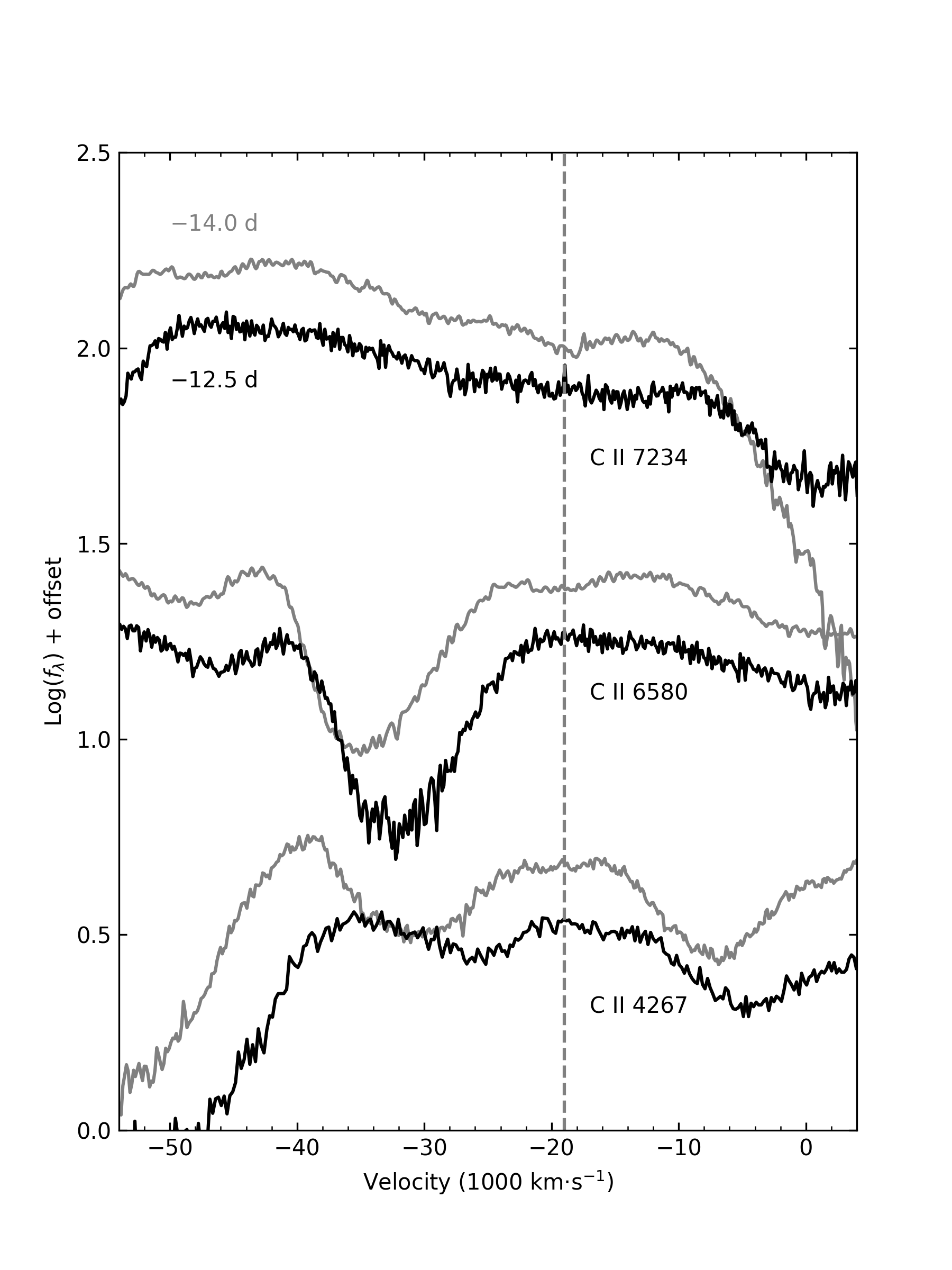}
\vspace{0.0cm}
\caption{Zoomed-in view of early-phase spectra of SN~2019ein around the C~II absorption features (C~II $\lambda$7234, C~II $\lambda$6580, and C~II $\lambda$4267). These carbon imprints are visible in the $t = -14.0$~day spectrum, but all tend to disappear in the $t \approx -12.5$~day spectrum. The vertical gray dashed line marks the carbon velocity inferred from the $t \approx -14.0$~day spectrum.}
\label{figcarbon} \vspace{-0.0cm}
\end{figure*}

In the earliest spectrum of SN~2019ein ($\sim 2.6$~days after explosion), \cite{2020ApJ...897..159P} noticed the presence of a small ``notch'' at $\sim 6200$~\AA, which could be due to C~II $\lambda$6580 at $\sim 20,000$~km~s$^{-1}$. We confirmed this identification and found that there are other two carbon features that can be identified at corresponding velocities --- the ``notch'' at $\sim 6800$~\AA\ caused by C~II $\lambda$7234 and the ``plateau'' at $\sim 4000$~\AA\ due to C~II $\lambda$4267, as shown in Figure~\ref{figcarbon}. We use the code \texttt{SYNAPPS} \citep{2011PASP..123..237T} to generate a synthetic spectrum from a parameterized element distribution in the ejecta and confirm the presence of these two carbon absorption features.

Compared with other NV~SNe~Ia, the carbon imprint disappears very quickly in spectra of SN~2019ein. In the spectrum taken only 0.9~days later ($\sim 3.5$~days after the explosion), the carbon feature is no longer visible. This indicates that the unburnt carbon only exists in the very outer region of the exploding system.

\subsection{Energetics and Progenitor Mass}
\label{secdiserg}

To further examine the explosion mechanism of SN~2019ein, we calculate its explosion energy. The energy of a thermonuclear SN is from runaway nuclear fusion of the progenitor white dwarf. To produce an SN~Ia, the energy released from nuclear fusion ($E_{\rm fusion}$) should be able to unbind the entire WD and account for the kinetic energy ($E_{\rm k}$) of the ejecta. The light curve is powered mainly by the decay of synthesized $^{56}$Ni, not directly by the fusion process. Some of the fusion energy will be released as neutrino luminosity, resulting in only a modest reduction of final kinetic energy by $\sim 2$\% \citep{2015PhRvD..92l4013S}, which will not affect our result. Therefore,
\begin{equation}
E_{\rm k}=E_{\rm fusion}-|E_{\rm binding}|.
\label{eqE}
\end{equation}
This scenario provides a constraint on the initial mass of the WD.

To calculate the binding energy of a WD of a given mass, we adopt the equation of state of degenerate matter \citep{1964ApJ...139.1396C}, and solve the Tolman-Oppenheimer-Volkoff (TOV) equation under the framework of general relativity. The binding energy is the sum of gravitational energy and internal energy. Assuming that the WD is made of elements with the same $\mu_e=A/Z$ parameter (nucleons per electron; e.g., He and C/O), and the electron gas is treated as an ideal degenerate Fermi gas, the energy profile of the WD is irrelevant to the specific abundances of individual elements \citep{2017RAA....17...61M}.

We integrated the equation numerically to get the binding energy as a function of WD mass. The mass and the binding energy first increase with the central density of the WD, reaching a critical point called the Chandrasekhar limit\footnote{This limit is $M_{\rm Ch}=1.42~{\rm M}_{\astrosun}$ and the corresponding binding energy $E_{\rm bind,Ch}=0.55 \times 10^{51}~\mathrm{ergs}$}. Then they turn back and decrease when further increasing the central density.  A WD can no longer support itself when the central pressure passes the turning point. Note that the decreasing branch is physically unstable.

A simple model for nuclear energy generation can be developed following the method described by \cite{2006Natur.443..308H}. Assuming that the WD is composed of equal parts of carbon and oxygen, then it is burnt into a mixture of iron-peak elements and intermediate-mass elements (IMEs), and the energy released will be 
\begin{equation}
E_{\rm fusion}=1.55 \times 10^{51}~\mathrm{ergs} \times (f_{\rm Fe}+0.76f_{\rm IME})\frac{M_{\rm WD}}{{\rm M}_{\astrosun}},
\end{equation}
where $f_{\rm Fe}$ and $f_{\rm IME}$ represent the fraction of iron-peak elements and IMEs produced in the explosion. If $f_{\rm Fe}$ and $f_{\rm IME}$ do not sum to 1 (i.e., $f_{\rm Fe} + f_{\rm IME} = 1-f_{\rm C}$, the remainder ($f_{\rm C}$) denotes the fraction of unburnt material and does not contribute to $E_{\rm fusion}$. The mass of $^{56}$Ni is estimated to be 70\% of the iron-peak elements. We adopt the synthesized $^{56}$Ni mass as $0.3~{\rm M}_{\astrosun}$ for SN~2019ein, which is the average value of our estimation and that suggested by \cite{2020ApJ...897..159P}.

The kinetic energy of the ejecta can be expressed as $E_{\rm k}=v_{\rm ej}^2M_{\rm WD}/2$, where $v_{\rm ej}$ is the kinetic velocity of the ejecta, defined by the relation above. \cite{2006Natur.443..308H} found a good agreement between the kinetic velocity of ejecta and the Si~II velocity measured $\sim 30$~days after maximum light.

\begin{figure*}
\center
\includegraphics[angle=0,width=1\textwidth]{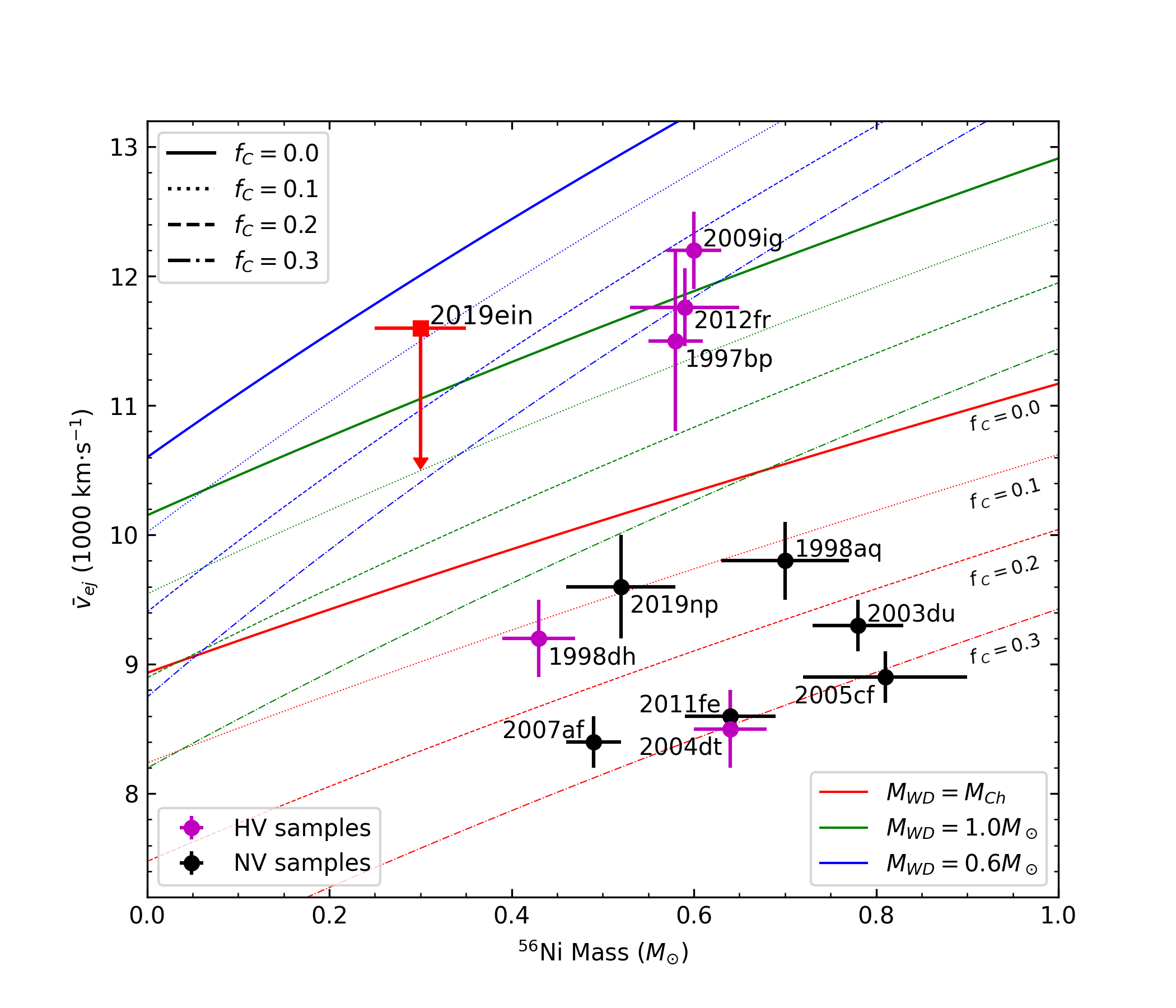}
\vspace{0.0cm}
\caption{Ni mass vs. kinetic ejecta velocity. The solid curves show upper limits of velocity for complete burning of different progenitor WD mass. The dotted, dashed, and dash-dotted curves represent models of incomplete burnings with different $f_{\rm C} > 0$. Normal and high-velocity comparison samples are in black and magenta markers, respectively. The red square denotes SN~2019ein. The vertical downward error bar for SN~2019ein reflects the uncertainty contributed by asymmetry.}
\label{figerg2} \vspace{-0.0cm}
\end{figure*}

Combining all of the models and equations described above, $v_{\rm ej}$ can be evaluated as functions of $M(\rm WD)$, $f_{\rm C}$, and $M(^{56}{\rm Ni})$. In Figure~\ref{figerg2}, we plot the variation of $v_{\rm ej}$ under different models of $M(\rm WD)$ and $f_{\rm C}$, along with the measured values of SN~2019ein and some other well-observed samples of SNe~Ia with accurate distance estimates. The references for these samples can be found in Table~\ref{tabcmp} and their distances are well constrained by recent Cepheid measurements by \citep{2021arXiv211204510R}. The $^{56}$Ni mass of these samples is estimated from their bolometric light curves.

In all of the models with fixed $M(\rm WD)$ and $f_{\rm C}$, more-luminous SNe with more $^{56}$Ni mass tend to expand faster, while $v_{\rm ej}$ tends to decrease with increasing $f_{\rm C}$ (when more unburnt material is left in the explosion). As seen in Figure~\ref{figerg2}, normal SNe~Ia can be well explained with Chandrasekhar-mass models by varying $f_{\rm C}$ from $\sim 0.1$ to $\sim 0.3$. For SN~2019ein, however, the high velocity and low luminosity put it at an extreme position. The Si~II velocity of SN~2019ein stays at 11,500~$\mathrm{km~s}^{-1}$ around 30~days after maximum brightness. If we adopt this velocity as $v_{\rm ej}$, the WD mass is significantly less than 0.8~M$_\odot$, even if we assume complete burning (i.e., $f_{\rm C} = 0$). It is possible that SN~2019ein has some sort of asymmetry, and exhibits a velocity higher than average along our line of sight. For a typical Chandrasekhar-mass delayed-detonation model, the difference in Si~II velocity $\sim 30$~days after maximum light could reach $\sim 1000~\mathrm{km~s}^{-1}$ for different viewing angles \citep{2010Natur.466...82M}. We use a downward arrow in Figure~\ref{figerg2} for SN~2019ein to illustrate this amount of possible asymmetry.

If we adopt a minimum of $\sim 10,500~\mathrm{km~s}^{-1}$ for the kinetic velocity after compensating for the bias due to an asymmetric explosion, the inferred maximum mass for the WD is $\sim 1.22~{\rm M}_\odot$. 
Thus, a Chandrasekhar-mass delayed-detonation model is not favored for SN~2019ein according to the above analysis of explosion energetics. We also note that the HV~SN~Ia samples display very distinct locations in Figure~\ref{figerg2}. Some HV objects, like SN~1998dh and SN~2004dt, have normal or low Si~II velocities 30~days after maximum light, implying a low kinetic velocity for their ejecta, although their velocities at maximum light are very large. Given their close locations to normal SNe~Ia like SN~2019np and SN~2011fe in the $v_{\rm ej} - ^{56}$Ni plot, these two HV~SNe~Ia can be explained by Chandrasekhar-mass models. In contrast, HV objects like SNe~1997bp, 2009ig, 2012fr, and 2019ein may need sub-Chandrasekhar-mass explosions. The above discrepancy indicates that the observed high velocity of SNe~Ia could have different origins and the HV subclass may need further subclassifications.

On the other hand, some effects besides the mass of the progenitor WD may also influence the observed velocity. For example, rapid rotation of the WD can significantly reduce its binding energy, leading to a higher velocity. The actual composition of WD material before explosion will affect the total fusion energy. Thus, the possibility that SN 2019ein originated from a Chandrasekhar-mass WD with peculiar unknown settings cannot be fully ruled out.

\subsection{Comparison with Models}
\label{secdismodel}

\begin{figure*}
\center
\includegraphics[angle=0,width=1\textwidth]{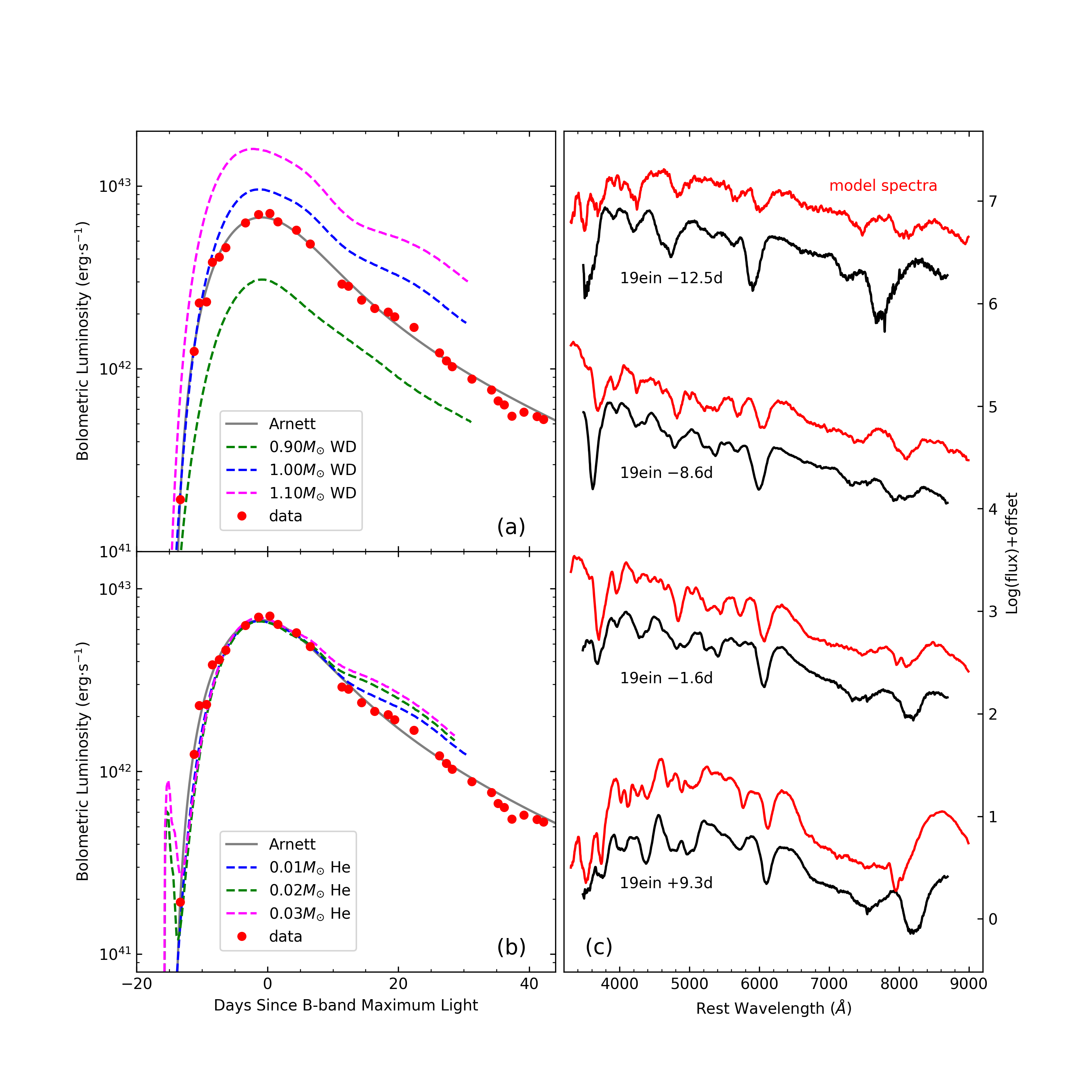}
\vspace{0.0cm}
\caption{Comparison with the double-detonation model of \protect\cite{2019ApJ...873...84P}. (a) Bolometric light curve of SN~2019ein compared with the best-fit Arnett model and double-detonation models having different WD masses and a helium-shell mass of 0.01~M$_\odot$. (b) Bolometric light curve compared with double-detonation models for $M(\rm WD) = 1.00~{\rm M}_\odot$ and a helium-shell mass ranging from 0.01~M$_{\odot}$ to 0.03~M$_{\odot}$; model curves are normalized to match the peak of SN~2019ein. (c) Spectral comparison between SN~2019ein (black) and double-detonation models (red) at different epochs.}
\label{figcmppolin} \vspace{-0.0cm}
\end{figure*}

Carbon/oxygen detonation ignited by detonation of a thin helium layer from accretion has been proposed as a popular mechanism causing sub-Chandrasekhar-mass WDs to explode as SNe~Ia. Theoretically, \cite{2019ApJ...873...84P} analyzed a set of 1D sub-Chandrasekhar-mass WD explosion models which are used to compare with the observed properties of SN~2019ein. Figure~\ref{figcmppolin} shows the comparison of the bolometric light curve and observed spectra of SN~2019ein with those produced by a double-detonation model \citep{2019ApJ...873...84P}. Three model bolometric light curves with helium-shell mass of $M({\rm He}) = 0.01~{\rm M}_{\astrosun}$ and different WD mass are shown in Figure~\ref{figcmppolin}(a). The inferred mass of the WD for SN~2019ein lies between 0.9 and 1.0~M$_{\astrosun}$. Model bolometric light curves with $M({\rm WD}) = 1.0~{\rm M}_{\astrosun}$ and different $M$(He), normalized by peak luminosity, are shown in Figure~\ref{figcmppolin}(b). We note that for $M({\rm He}) > 0.01~{\rm M}_{\astrosun}$, there is a significant early excess in the bolometric light curve, which is due to radioactive heating of the outermost ejecta by heavy elements ($Z>22$) synthesized in the helium shell. As there is no detectable emission excess in the light curves of SN~2019ein, the helium mass is constrained to be $M({\rm He}) \lesssim 0.02~{\rm M}_{\astrosun}$.

In Figure~\ref{figcmppolin}(c), observed spectra of SN~2019ein at different epochs are compared to model spectra with $M({\rm WD}) = 1.0~{\rm M}_{\astrosun}$ and $M({\rm He}) \lesssim 0.01~{\rm M}_{\astrosun}$. The model spectra are fairly similar to the observed spectra of SN~2019ein at $t = -8.6$~days and $t = -1.6$~days. In particular, the velocity and strength of Ca, O, and Si absorption lines in the model spectra match the observed features at these two epochs. A noticeable discrepancy is the strength of Si~II $\lambda$5972, which is too strong in the model spectra, suggesting that the model temperature is lower than the actual SN. For model spectra generated at earlier epochs, the velocities of Ca~II absorption features do not match the observed values. Moreover, the ejecta velocity inferred from model spectra does not evolve, whereas SN~2019ein is characterized by a high velocity gradient.

\begin{figure*}
\center
\includegraphics[angle=0,width=1\textwidth]{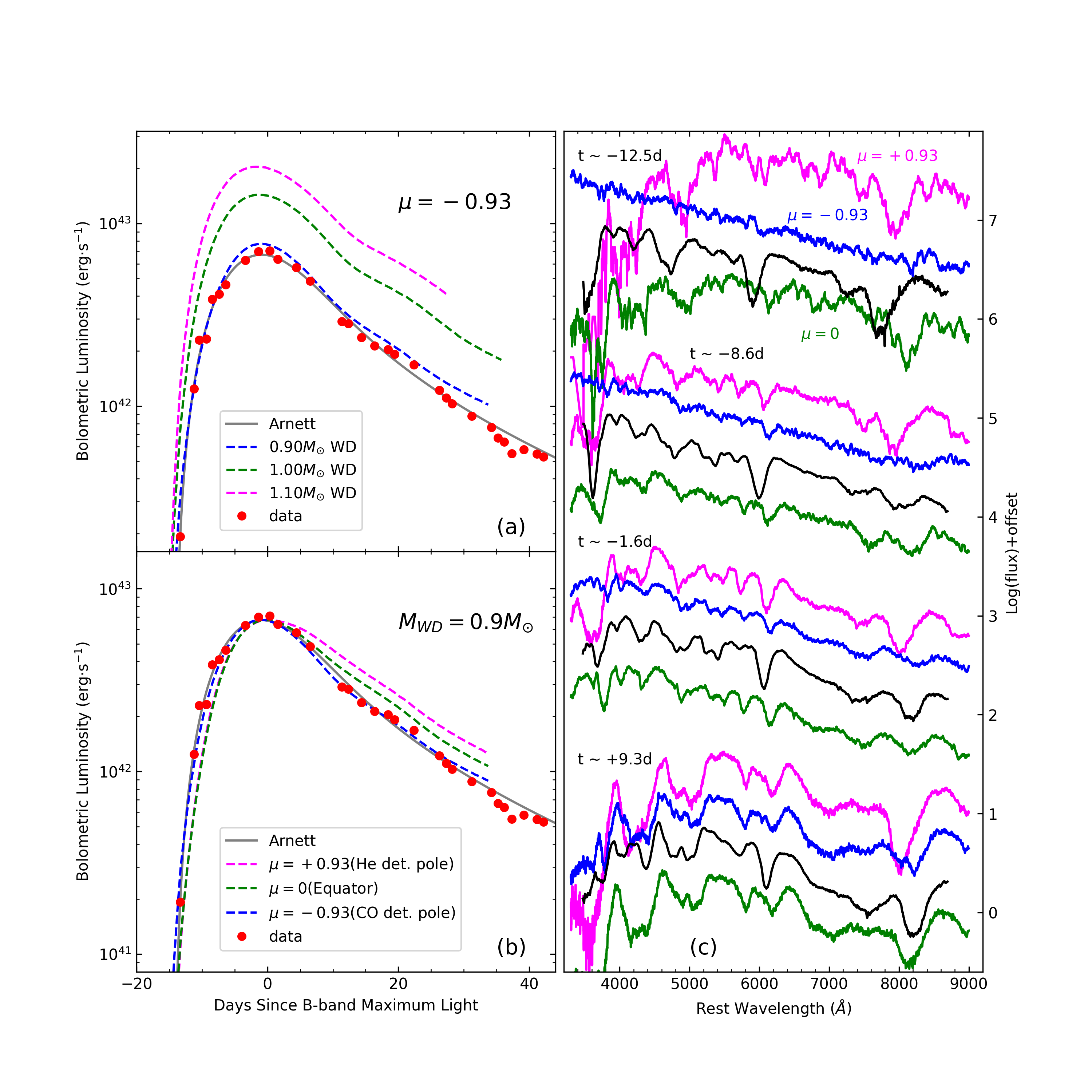}
\vspace{0.0cm}
\caption{Comparisons of the observed bolometric luminosity and spectra with those predicted by double-detonation models \protect\cite{2021ApJ...922...68S}. (a) Bolometric light curve of SN~2019ein, together with the best-fit Arnett model and curves from double-detonation models having different WD masses. (b) Bolometric light curve compared with model curves for different viewing angles (pink for the helium-detonation pole, green for the equator, and blue for the CO detonation pole); all model curves are normalized to match the peak of SN~2019ein. (c) Spectral comparison between SN~2019ein (black) and model spectra at different epochs; colors represent different viewing angles as in panel (b).}
\label{figcmpshen} \vspace{-0.0cm}
\end{figure*}

Since the helium detonation occurs in the outer layers of the WD, the double-detonation model is intrinsically asymmetric. The observed properties should vary with viewing angle --- that is, from the He detonation side, the C/O detonation side, or the equatorial plane. \cite{2019ApJ...878L..38T} and \cite{2021ApJ...922...68S} performed multidimensional radiative transfer calculations of sub-Chandrasekhar-mass WD explosion models. The comparison of SN~2019ein and models of \cite{2021ApJ...922...68S} is shown in Figure~\ref{figcmpshen}. The viewing angle is described by the parameter $\mu=\cos\theta$, where $\theta$ is the angle between the observer and the He detonation pole.

Figure~\ref{figcmpshen}(a) compares the bolometric light curve of SN~2019ein with three models of different $M({\rm WD})$. The model with $M({\rm WD}) = 0.9~{\rm M}_{\astrosun}$ seems to better match the data. Figure~\ref{figcmpshen}(b) shows the model light curve of $M({\rm WD}) = 0.9~{\rm M}_{\astrosun}$ with three different viewing angles, normalized by peak luminosity. We note that the model curve viewed from the C/O detonation side (i.e., $\mu = -0.93$) matches the observed light curve reasonably well, while the light curves viewed from other angles are significantly wider and have a longer rise time than what is observed.

Figure~\ref{figcmpshen}(c) compares the observed spectra of SN~2019ein at different epochs with model spectra as viewed from three different directions. Around maximum light, the overall spectrum is best fit by a model spectrum viewed from the He detonation side (i.e., $\mu = 0.93$), while the O/Ca features better match the $\mu = 0$ model. We noticed that different parts of the observed spectra at different epochs are best fit by model spectra with different viewing angles, and it is hard to choose a definitive viewing angle for SN~2019ein. In addition, the earliest model spectra do not fit the observed spectra very well, especially in the case of $\mu = -0.93$, where the spectrum is almost featureless.

\citet{2020ApJ...897..159P} compared the observed spectra of SN~2019ein with those of a delayed-detonation model, which was used by \cite{2015MNRAS.448.2766B} as the explosion mechanism of fast-expanding objects like SN~2002bo. They found good match between the model and the observed spectra, except for the earliest $t=-14.0$ day spectrum. However, since SN~2019ein is significantly less luminous than SN~2002bo (which has $M_B = -19.41 \pm 0.42$~mag), models that produce less amount of $^{56}$Ni are required.
\cite{2013MNRAS.429.2127B} investigated a sequence of Chandrasekhar-mass delayed-detonation models characterized by different $^{56}$Ni masses. The differences in $^{56}$Ni mass are caused by different time at which the transition from deflagration to detonation occurs. Using a model that generates less $^{56}$Ni will produce a more consistent bolometric peak, but it will also result in an event with more unburnt materials and lower ejecta velocity, like 91bg-like events\citep{1992AJ....104.1543F,1993AJ....105..301L}, which is inconsistent with the unusually high expansion velocity seen in SN 2019ein and the missing signature of significant amount of unburnt C and O in its spectra (see Figure~\ref{figspeccmp}).

\subsection{Comparison with Thick He-Shell Events}
\label{secdisthick}

In contrast to the thin He shell model for SN 2019ein, some peculiar SNe Ia are attributed to double detonation with massive He shells. Such examples include SN 2018byg\citep{2019ApJ...873L..18D} and 2020jgb\citep{2022arXiv220904463L}. They are both characterized by subluminous light curves (with $M_r \sim -18.2$ to $-18.5$ mag), red colors, and spectra with highly suppressed bluer parts and high velocity Ca~II absorption features. Both of the events are well reproduced by the detonation of a massive ($\sim$ 0.15 M$_\odot$) helium shell on a sub-Chandrasekhar mass ($\sim$ 0.8 M$_\odot$) C/O white dwarf.

\begin{figure*}
\center
\includegraphics[angle=0,width=1\textwidth]{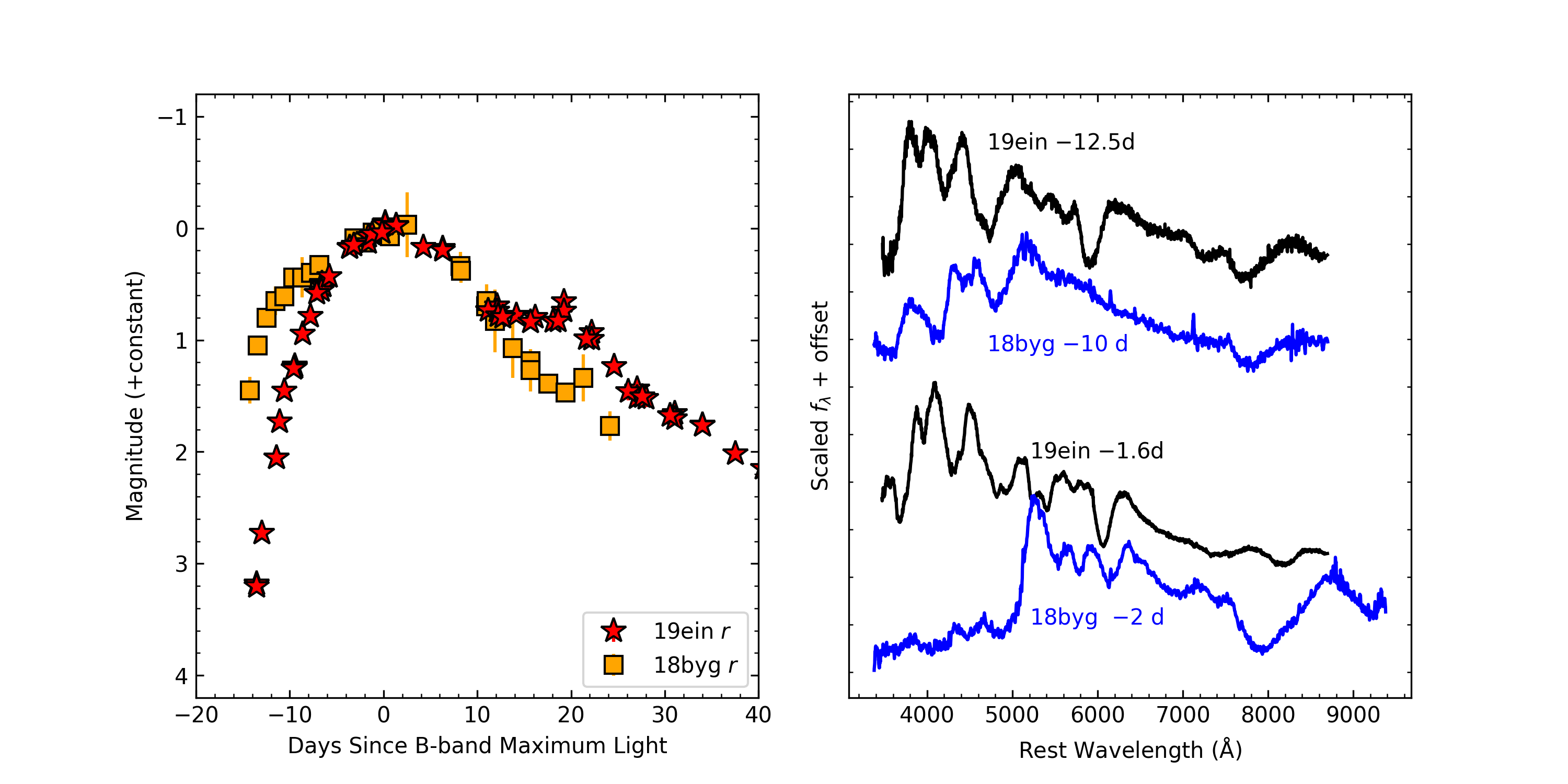}
\vspace{0.0cm}
\caption{
Comparison of the $r$-band light curve (left panel) and spectra (right panel) of SN 2019ein with those of SN 2018byg \protect\citep{2019ApJ...873L..18D}.
}
\label{figcmpbyg} \vspace{-0.0cm}
\end{figure*}

In Figure~\ref{figcmpbyg}, we compared the $r$-band light curve and two-epoch spectra for SN 2019ein and SN 2018byg. The light curve of SN 2018byg shows a strong early excess, which is a signature of radioactive materials formed from detonation of thick He shell. From the earliest spectrum of SN 2018byg taken at $t \approx -10$ days, the velocity of Ca~II NIR triplet is estimated as $\sim 25, 000$ km s$^{-1}$, comparable to that of SN 2019ein at similar phases. However, this velocity is obviously higher than that derived for most SNe Ia, which are found to have photospheric Ca II velocity $\lesssim $ 16,000 km s$^{-1}$ at t$\approx -10$ days\citep{2015ApJS..220...20Z}. In the near-maximum-light spectra, the Ca II velocity remains high in SN~2018byg, while it shows a significant decrease in SN~2019ein. This is probably due to that detonation of massive helium shell produced a large amount of Ca in the outermost layer. The bluer part ($\lesssim 5000$ \AA) of the SN 2018byg spectrum is highly suppressed by the line blanketing effects of Fe group elements, which is also consistent with the expectation that detonation of massive He shell could produce abundant Fe-rich materials in the outer ejecta.

Although SN~2019ein does not show peculiar features caused by detonation of thick He shell as in SN~2018byg, they still share some common properties, like low luminosity and high velocity spectral features. These properties are more consistent with a sub-Chandrasekhar-mass WD explosion, as discussed in Section~\ref{secdiserg}.

\section{CONCLUSION}
\label{secsum}

We have presented extensive optical photometric and spectroscopic observations of SN~2019ein, an SN~Ia with extremely high velocity and rapid velocity evolution at early times. SN~2019ein has an absolute $B$-band peak magnitude of $M(B)=-18.71 \pm 0.15~\mathrm{mag}$ and a post-peak decline rate of $\Delta m_{15}(B) = 1.35 \pm 0.01~\mathrm{mag}$. Using a ``fireball'' model to fit the early-phase light curve, the explosion time is estimated to be $t_0 = \mathrm{MJD}~58603.22$, giving a rise-time estimate of $\sim 16.0$~days. We constructed the bolometric light curve of SN~2019ein from its multiband photometry and fitted it to the radioactive-decay-driven radiation diffusion model \citep{1982ApJ...253..785A}, leading to estimates of the peak luminosity as $L \approx 7.0 \times 10^{42}~\mathrm{erg~s^{-1}}$ and the synthesized nickel mass as $M_{{\rm Ni}} = (0.27 \pm 0.04)~{\rm M}_{\astrosun}$. The luminosity and nickel mass of SN~2019ein are notably lower than those of normal SNe~Ia having similar $\Delta m_{15}(B)$. 

Spectroscopically, SN~2019ein can be put into the group of both HVG and HV~SNe~Ia. The overall spectral evolution of SN~2019ein is similar to that of SN~2002bo except for its much higher ejecta velocity and rapid velocity evolution at early times.

Analysis of the energetics is performed for SN~2019ein based on the synthesized $^{56}$Ni mass and the ejecta velocity, putting strict constraints on the mass of its progenitor WD. The inferred upper limit ($M \lesssim 1.22~{\rm M}_{\astrosun}$) is significantly less than the Chandrasekhar-mass limit, even if we adopt an ejecta velocity ($v_{\rm ej}= 10,500~\mathrm{km~s}^{-1}$) much lower than the Si~II velocity ($\sim 11,600~\mathrm{km~s}^{-1}$) measured several weeks after maximum light. We suggest that SN~2019ein is most likely an asymmetric explosion of a sub-Chandrasekhar WD, and is viewed from the direction along which the ejecta have a larger velocity toward us.

The asymmetric nature of SN~2019ein can also be seen from its nebular-phase spectrum, which we obtained with Keck-I/LRIS 311~days after $B$ maximum. The [Fe~II] and [Ni~II] lines in this spectrum are redshifted by $\sim 1000$~km~s$^{-1}$, which could be interpreted as a ``kick-back effect'' of the inner-core region by the outer ejecta moving toward us with exceedingly high velocity.

We further compared the observed bolometric light curve and spectra of SN~2019ein with those predicted by the double-detonation model. The bolometric light curve and maximum-light spectrum of SN~2019ein are consistent with a $\sim 1.0~{\rm M}_{\astrosun}$ double-detonation model with a minimum amount of helium ($\lesssim 0.02~{\rm M}_{\astrosun}$). Nevertheless, the model spectra at other phases do have discrepancies with the observations, especially in the Ca~II features. 

Spectropolarimetric measurements reported by \cite{2022MNRAS.509.4058P} indicate that SN~2019ein does not exhibit substantial continuum polarization around the maximum light. Explosion models with extreme asymmetry, (e.g., violent merger) are thus disfavored. However, since both the double-detonation and delayed-detonation models are axially symmetric, the possibility that SN~2019ein is observed along the symmetric axis of these explosion model cannot be ruled out.

The unusally high early-time velocities and low luminosity make SN~2019ein a peculiar object even among HV~SNe~Ia, posing multiple challenges for existing models of exploding WDs. The asymmetric explosion mechanism producing SN~2019ein should also account for (at lease some of) the events with different velocities when viewed from different angles. Nevertheless, there are many intrinsic properties of HV~SNe~Ia that do not favor a pure orientation effect, such as their concentrated distribution in the inner region of the host galaxies \citep{2013Sci...340..170W}, the preference of exploding in massive galaxies and metal-rich stellar environments \citep{2015MNRAS.446..354P,2020ApJ...895L...5P}, and the color evolution caused by circumstellar dust \citep{2019ApJ...882..120W}. It is likely that multiple factors, including explosion mechanism, progenitor population, and circumstellar environment, are responsible for the observed diversity of HV~SNe~Ia. Detailed studies of additional HV objects like SN~2019ein are needed to further constrain the explosion physics of SNe~Ia and to improve the precision of their use as distance indicators.

\section*{Acknowledgements}

  This work is supported by the National Science Foundation of China (NSFC grants 12033003 and 11633002), the Scholar Program of Beijing Academy of Science and Technology (DZ:BS202002), and the Tencent Xplorer Prize. Observations at AZT-22 of the Maidanak Observatory were supported by Uzbekistan's Ministry of Innovative Development (grant A-FA-2021-36). The ZTF forced-photometry service was funded under Heising-Simons Foundation grant \#12540303 (PI, M. Graham). A.V.F.'s team received support from the U.C. Berkeley Miller Institute for Basic Research in Science (where A.V.F. was a Miller Senior Fellow), the Christopher R. Redlich Fund, and many individual donors.

  We thank Eddie A. Baron and James M. DerKacy for sharing us two spectra obtained with the Apache Point Observatory 3.5-meter telescope, which is owned and operated by the Astrophysical Research Consortium.  We also thank Abigail Polin, Ken J. Shen, and Dean M. Townsley for sharing us the model data.
  Some of the data presented herein were obtained at the W. M. Keck Observatory, which is operated as a scientific partnership among the California Institute of Technology, the University of California, and NASA; the observatory was made possible by the generous financial support of the W. M. Keck Foundation.
This work makes use of observations from the Las Cumbres Observatory global telescope network, as well as the NASA/IPAC Extragalactic Database (NED), which is funded by NASA and operated by the California Institute of Technology.

  This work has made use of data from the Asteroid Terrestrial-impact Last Alert System (ATLAS) project. The Asteroid Terrestrial-impact Last Alert System (ATLAS) project is primarily funded to search for near-Earth objects through NASA grants NN12AR55G, 80NSSC18K0284, and 80NSSC18K1575; byproducts of the NEO search include images and catalogs from the survey area. This work was partially funded by Kepler/K2 grant J1944/80NSSC19K0112 and HST GO-15889, and by STFC grants ST/T000198/1 and ST/S006109/1. The ATLAS science products have been made possible through the contributions of the University of Hawaii Institute for Astronomy, the Queen's University Belfast, the Space Telescope Science Institute, the South African Astronomical Observatory, and The Millennium Institute of Astrophysics (MAS), Chile.

\section*{Data Availability}

The photometric data underlying this article are available in the article, and the spectroscopic data will be available in Weizmann Interactive Supernova Data Repository (WISeREP) at https://www.wiserep.org/object/12239 .



\bibliographystyle{mnras}
\bibliography{references} 

\begin{thebibliography}{}
\makeatletter
\relax
\def\mn@urlcharsother{\let\do\@makeother \do\$\do\&\do\#\do\^\do\_\do\%\do\~}
\def\mn@doi{\begingroup\mn@urlcharsother \@ifnextchar [ {\mn@doi@}
  {\mn@doi@[]}}
\def\mn@doi@[#1]#2{\def\@tempa{#1}\ifx\@tempa\@empty \href
  {http://dx.doi.org/#2} {doi:#2}\else \href {http://dx.doi.org/#2} {#1}\fi
  \endgroup}
\def\mn@eprint#1#2{\mn@eprint@#1:#2::\@nil}
\def\mn@eprint@arXiv#1{\href {http://arxiv.org/abs/#1} {{\tt arXiv:#1}}}
\def\mn@eprint@dblp#1{\href {http://dblp.uni-trier.de/rec/bibtex/#1.xml}
  {dblp:#1}}
\def\mn@eprint@#1:#2:#3:#4\@nil{\def\@tempa {#1}\def\@tempb {#2}\def\@tempc
  {#3}\ifx \@tempc \@empty \let \@tempc \@tempb \let \@tempb \@tempa \fi \ifx
  \@tempb \@empty \def\@tempb {arXiv}\fi \@ifundefined
  {mn@eprint@\@tempb}{\@tempb:\@tempc}{\expandafter \expandafter \csname
  mn@eprint@\@tempb\endcsname \expandafter{\@tempc}}}

\bibitem[\protect\citeauthoryear{{Altavilla} et~al.,}{{Altavilla}
  et~al.}{2007}]{2007A&A...475..585A}
{Altavilla} G.,  et~al., 2007, \mn@doi [\aap] {10.1051/0004-6361:20077487},
  \href {https://ui.adsabs.harvard.edu/abs/2007A&A...475..585A} {475, 585}

\bibitem[\protect\citeauthoryear{{Anupama}, {Sahu}  \& {Jose}}{{Anupama}
  et~al.}{2005}]{2005A&A...429..667A}
{Anupama} G.~C.,  {Sahu} D.~K.,   {Jose} J.,  2005, \mn@doi [\aap]
  {10.1051/0004-6361:20041687}, \href
  {https://ui.adsabs.harvard.edu/abs/2005A&A...429..667A} {429, 667}

\bibitem[\protect\citeauthoryear{{Arnett}}{{Arnett}}{1982}]{1982ApJ...253..785A}
{Arnett} W.~D.,  1982, \mn@doi [\apj] {10.1086/159681}, \href
  {https://ui.adsabs.harvard.edu/abs/1982ApJ...253..785A} {253, 785}

\bibitem[\protect\citeauthoryear{{Benetti} et~al.,}{{Benetti}
  et~al.}{2004}]{2004MNRAS.348..261B}
{Benetti} S.,  et~al., 2004, \mn@doi [\mnras]
  {10.1111/j.1365-2966.2004.07357.x}, \href
  {https://ui.adsabs.harvard.edu/abs/2004MNRAS.348..261B} {348, 261}

\bibitem[\protect\citeauthoryear{{Benetti} et~al.,}{{Benetti}
  et~al.}{2005}]{2005ApJ...623.1011B}
{Benetti} S.,  et~al., 2005, \mn@doi [\apj] {10.1086/428608}, \href
  {https://ui.adsabs.harvard.edu/abs/2005ApJ...623.1011B} {623, 1011}

\bibitem[\protect\citeauthoryear{{Betoule} et~al.,}{{Betoule}
  et~al.}{2014}]{2014A&A...568A..22B}
{Betoule} M.,  et~al., 2014, \mn@doi [\aap] {10.1051/0004-6361/201423413},
  \href {https://ui.adsabs.harvard.edu/abs/2014A&A...568A..22B} {568, A22}

\bibitem[\protect\citeauthoryear{{Blondin} et~al.,}{{Blondin}
  et~al.}{2012}]{2012AJ....143..126B}
{Blondin} S.,  et~al., 2012, \mn@doi [\aj] {10.1088/0004-6256/143/5/126}, \href
  {https://ui.adsabs.harvard.edu/abs/2012AJ....143..126B} {143, 126}

\bibitem[\protect\citeauthoryear{{Blondin}, {Dessart}, {Hillier}  \&
  {Khokhlov}}{{Blondin} et~al.}{2013}]{2013MNRAS.429.2127B}
{Blondin} S.,  {Dessart} L.,  {Hillier} D.~J.,   {Khokhlov} A.~M.,  2013,
  \mn@doi [\mnras] {10.1093/mnras/sts484}, \href
  {https://ui.adsabs.harvard.edu/abs/2013MNRAS.429.2127B} {429, 2127}

\bibitem[\protect\citeauthoryear{{Blondin}, {Dessart}  \& {Hillier}}{{Blondin}
  et~al.}{2015}]{2015MNRAS.448.2766B}
{Blondin} S.,  {Dessart} L.,   {Hillier} D.~J.,  2015, \mn@doi [\mnras]
  {10.1093/mnras/stv188}, \href
  {https://ui.adsabs.harvard.edu/abs/2015MNRAS.448.2766B} {448, 2766}

\bibitem[\protect\citeauthoryear{{Bloom} et~al.,}{{Bloom}
  et~al.}{2012}]{2012ApJ...744L..17B}
{Bloom} J.~S.,  et~al., 2012, \mn@doi [\apjl] {10.1088/2041-8205/744/2/L17},
  \href {https://ui.adsabs.harvard.edu/abs/2012ApJ...744L..17B} {744, L17}

\bibitem[\protect\citeauthoryear{{Branch}, {Fisher}  \& {Nugent}}{{Branch}
  et~al.}{1993}]{1993AJ....106.2383B}
{Branch} D.,  {Fisher} A.,   {Nugent} P.,  1993, \mn@doi [\aj]
  {10.1086/116810}, \href
  {https://ui.adsabs.harvard.edu/abs/1993AJ....106.2383B} {106, 2383}

\bibitem[\protect\citeauthoryear{{Branch} et~al.,}{{Branch}
  et~al.}{2003}]{2003AJ....126.1489B}
{Branch} D.,  et~al., 2003, \mn@doi [\aj] {10.1086/377016}, \href
  {https://ui.adsabs.harvard.edu/abs/2003AJ....126.1489B} {126, 1489}

\bibitem[\protect\citeauthoryear{{Branch} et~al.,}{{Branch}
  et~al.}{2006}]{2006PASP..118..560B}
{Branch} D.,  et~al., 2006, \mn@doi [\pasp] {10.1086/502778}, \href
  {https://ui.adsabs.harvard.edu/abs/2006PASP..118..560B} {118, 560}

\bibitem[\protect\citeauthoryear{{Brown}, {Dawson}, {Harris}, {Olmstead},
  {Milne}  \& {Roming}}{{Brown} et~al.}{2012}]{2012ApJ...749...18B}
{Brown} P.~J.,  {Dawson} K.~S.,  {Harris} D.~W.,  {Olmstead} M.,  {Milne} P.,
  {Roming} P. W.~A.,  2012, \mn@doi [\apj] {10.1088/0004-637X/749/1/18}, \href
  {https://ui.adsabs.harvard.edu/abs/2012ApJ...749...18B} {749, 18}

\bibitem[\protect\citeauthoryear{{Brown} et~al.,}{{Brown}
  et~al.}{2013}]{2013PASP..125.1031B}
{Brown} T.~M.,  et~al., 2013, \mn@doi [\pasp] {10.1086/673168}, \href
  {https://ui.adsabs.harvard.edu/abs/2013PASP..125.1031B} {125, 1031}

\bibitem[\protect\citeauthoryear{{Brown}, {Breeveld}, {Holland}, {Kuin}  \&
  {Pritchard}}{{Brown} et~al.}{2014}]{2014Ap&SS.354...89B}
{Brown} P.~J.,  {Breeveld} A.~A.,  {Holland} S.,  {Kuin} P.,   {Pritchard} T.,
  2014, \mn@doi [\apss] {10.1007/s10509-014-2059-8}, \href
  {https://ui.adsabs.harvard.edu/abs/2014Ap&SS.354...89B} {354, 89}

\bibitem[\protect\citeauthoryear{{Burke}, {Arcavi}, {Howell}, {Hiramatsu},
  {McCully}  \& {Valenti}}{{Burke} et~al.}{2019}]{2019TNSCR.701....1B}
{Burke} J.,  {Arcavi} I.,  {Howell} D.~A.,  {Hiramatsu} D.,  {McCully} C.,
  {Valenti} S.,  2019, Transient Name Server Classification Report, \href
  {https://ui.adsabs.harvard.edu/abs/2019TNSCR.701....1B} {2019-701, 1}

\bibitem[\protect\citeauthoryear{{Burke} et~al.,}{{Burke}
  et~al.}{2021}]{2021ApJ...919..142B}
{Burke} J.,  et~al., 2021, \mn@doi [\apj] {10.3847/1538-4357/ac126b}, \href
  {https://ui.adsabs.harvard.edu/abs/2021ApJ...919..142B} {919, 142}

\bibitem[\protect\citeauthoryear{{Burns} et~al.,}{{Burns}
  et~al.}{2011}]{2011AJ....141...19B}
{Burns} C.~R.,  et~al., 2011, \mn@doi [\aj] {10.1088/0004-6256/141/1/19}, \href
  {https://ui.adsabs.harvard.edu/abs/2011AJ....141...19B} {141, 19}

\bibitem[\protect\citeauthoryear{{Burns} et~al.,}{{Burns}
  et~al.}{2014}]{2014ApJ...789...32B}
{Burns} C.~R.,  et~al., 2014, \mn@doi [\apj] {10.1088/0004-637X/789/1/32},
  \href {https://ui.adsabs.harvard.edu/abs/2014ApJ...789...32B} {789, 32}

\bibitem[\protect\citeauthoryear{{Cardelli}, {Clayton}  \& {Mathis}}{{Cardelli}
  et~al.}{1989}]{1989ApJ...345..245C}
{Cardelli} J.~A.,  {Clayton} G.~C.,   {Mathis} J.~S.,  1989, \mn@doi [\apj]
  {10.1086/167900}, \href
  {https://ui.adsabs.harvard.edu/abs/1989ApJ...345..245C} {345, 245}

\bibitem[\protect\citeauthoryear{{Chandrasekhar} \& {Tooper}}{{Chandrasekhar}
  \& {Tooper}}{1964}]{1964ApJ...139.1396C}
{Chandrasekhar} S.,  {Tooper} R.~F.,  1964, \mn@doi [\apj] {10.1086/147883},
  \href {https://ui.adsabs.harvard.edu/abs/1964ApJ...139.1396C} {139, 1396}

\bibitem[\protect\citeauthoryear{{Chatzopoulos}, {Wheeler}  \&
  {Vinko}}{{Chatzopoulos} et~al.}{2012}]{2012ApJ...746..121C}
{Chatzopoulos} E.,  {Wheeler} J.~C.,   {Vinko} J.,  2012, \mn@doi [\apj]
  {10.1088/0004-637X/746/2/121}, \href
  {https://ui.adsabs.harvard.edu/abs/2012ApJ...746..121C} {746, 121}

\bibitem[\protect\citeauthoryear{{Chatzopoulos}, {Wheeler}, {Vinko}, {Horvath}
  \& {Nagy}}{{Chatzopoulos} et~al.}{2013}]{2013ApJ...773...76C}
{Chatzopoulos} E.,  {Wheeler} J.~C.,  {Vinko} J.,  {Horvath} Z.~L.,   {Nagy}
  A.,  2013, \mn@doi [\apj] {10.1088/0004-637X/773/1/76}, \href
  {https://ui.adsabs.harvard.edu/abs/2013ApJ...773...76C} {773, 76}

\bibitem[\protect\citeauthoryear{{Childress} et~al.,}{{Childress}
  et~al.}{2013}]{2013ApJ...770...29C}
{Childress} M.~J.,  et~al., 2013, \mn@doi [\apj] {10.1088/0004-637X/770/1/29},
  \href {https://ui.adsabs.harvard.edu/abs/2013ApJ...770...29C} {770, 29}

\bibitem[\protect\citeauthoryear{{Contreras} et~al.,}{{Contreras}
  et~al.}{2010}]{2010AJ....139..519C}
{Contreras} C.,  et~al., 2010, \mn@doi [\aj] {10.1088/0004-6256/139/2/519},
  \href {https://ui.adsabs.harvard.edu/abs/2010AJ....139..519C} {139, 519}

\bibitem[\protect\citeauthoryear{{De} et~al.,}{{De}
  et~al.}{2019}]{2019ApJ...873L..18D}
{De} K.,  et~al., 2019, \mn@doi [\apjl] {10.3847/2041-8213/ab0aec}, \href
  {https://ui.adsabs.harvard.edu/abs/2019ApJ...873L..18D} {873, L18}

\bibitem[\protect\citeauthoryear{{Ehgamberdiev}}{{Ehgamberdiev}}{2018}]{2018NatAs...2..349E}
{Ehgamberdiev} S.,  2018, \mn@doi [Nature Astronomy]
  {10.1038/s41550-018-0459-3}, \href
  {https://ui.adsabs.harvard.edu/abs/2018NatAs...2..349E} {2, 349}

\bibitem[\protect\citeauthoryear{{Fan}, {Bai}, {Zhang}, {Wang}, {Chang}, {Xin}
  \& {Zhang}}{{Fan} et~al.}{2015}]{2015RAA....15..918F}
{Fan} Y.-F.,  {Bai} J.-M.,  {Zhang} J.-J.,  {Wang} C.-J.,  {Chang} L.,  {Xin}
  Y.-X.,   {Zhang} R.-L.,  2015, \mn@doi [Research in Astronomy and
  Astrophysics] {10.1088/1674-4527/15/6/014}, \href
  {https://ui.adsabs.harvard.edu/abs/2015RAA....15..918F} {15, 918}

\bibitem[\protect\citeauthoryear{{Filippenko} et~al.,}{{Filippenko}
  et~al.}{1992a}]{1992AJ....104.1543F}
{Filippenko} A.~V.,  et~al., 1992a, \mn@doi [\aj] {10.1086/116339}, \href
  {https://ui.adsabs.harvard.edu/abs/1992AJ....104.1543F} {104, 1543}

\bibitem[\protect\citeauthoryear{{Filippenko} et~al.,}{{Filippenko}
  et~al.}{1992b}]{1992ApJ...384L..15F}
{Filippenko} A.~V.,  et~al., 1992b, \mn@doi [\apjl] {10.1086/186252}, \href
  {https://ui.adsabs.harvard.edu/abs/1992ApJ...384L..15F} {384, L15}

\bibitem[\protect\citeauthoryear{{Folatelli} et~al.,}{{Folatelli}
  et~al.}{2013}]{2013ApJ...773...53F}
{Folatelli} G.,  et~al., 2013, \mn@doi [\apj] {10.1088/0004-637X/773/1/53},
  \href {https://ui.adsabs.harvard.edu/abs/2013ApJ...773...53F} {773, 53}

\bibitem[\protect\citeauthoryear{{Foley} et~al.,}{{Foley}
  et~al.}{2012}]{2012ApJ...744...38F}
{Foley} R.~J.,  et~al., 2012, \mn@doi [\apj] {10.1088/0004-637X/744/1/38},
  \href {https://ui.adsabs.harvard.edu/abs/2012ApJ...744...38F} {744, 38}

\bibitem[\protect\citeauthoryear{{Ganeshalingam} et~al.,}{{Ganeshalingam}
  et~al.}{2010}]{2010ApJS..190..418G}
{Ganeshalingam} M.,  et~al., 2010, \mn@doi [\apjs]
  {10.1088/0067-0049/190/2/418}, \href
  {https://ui.adsabs.harvard.edu/abs/2010ApJS..190..418G} {190, 418}

\bibitem[\protect\citeauthoryear{{Ganeshalingam} et~al.,}{{Ganeshalingam}
  et~al.}{2012}]{2012ApJ...751..142G}
{Ganeshalingam} M.,  et~al., 2012, \mn@doi [\apj]
  {10.1088/0004-637X/751/2/142}, \href
  {https://ui.adsabs.harvard.edu/abs/2012ApJ...751..142G} {751, 142}

\bibitem[\protect\citeauthoryear{{Graham} et~al.,}{{Graham}
  et~al.}{2017}]{2017MNRAS.472.3437G}
{Graham} M.~L.,  et~al., 2017, \mn@doi [\mnras] {10.1093/mnras/stx2224}, \href
  {https://ui.adsabs.harvard.edu/abs/2017MNRAS.472.3437G} {472, 3437}

\bibitem[\protect\citeauthoryear{{Gunn} et~al.,}{{Gunn}
  et~al.}{2006}]{2006AJ....131.2332G}
{Gunn} J.~E.,  et~al., 2006, \mn@doi [\aj] {10.1086/500975}, \href
  {https://ui.adsabs.harvard.edu/abs/2006AJ....131.2332G} {131, 2332}

\bibitem[\protect\citeauthoryear{{Guy} et~al.,}{{Guy}
  et~al.}{2007}]{2007A&A...466...11G}
{Guy} J.,  et~al., 2007, \mn@doi [\aap] {10.1051/0004-6361:20066930}, \href
  {https://ui.adsabs.harvard.edu/abs/2007A&A...466...11G} {466, 11}

\bibitem[\protect\citeauthoryear{{Guy} et~al.,}{{Guy}
  et~al.}{2010}]{2010A&A...523A...7G}
{Guy} J.,  et~al., 2010, \mn@doi [\aap] {10.1051/0004-6361/201014468}, \href
  {https://ui.adsabs.harvard.edu/abs/2010A&A...523A...7G} {523, A7}

\bibitem[\protect\citeauthoryear{{Hicken} et~al.,}{{Hicken}
  et~al.}{2009}]{2009ApJ...700..331H}
{Hicken} M.,  et~al., 2009, \mn@doi [\apj] {10.1088/0004-637X/700/1/331}, \href
  {https://ui.adsabs.harvard.edu/abs/2009ApJ...700..331H} {700, 331}

\bibitem[\protect\citeauthoryear{{Hicken} et~al.,}{{Hicken}
  et~al.}{2012}]{2012ApJS..200...12H}
{Hicken} M.,  et~al., 2012, \mn@doi [\apjs] {10.1088/0067-0049/200/2/12}, \href
  {https://ui.adsabs.harvard.edu/abs/2012ApJS..200...12H} {200, 12}

\bibitem[\protect\citeauthoryear{{Hoeflich}, {Khokhlov}, {Wheeler}, {Phillips},
  {Suntzeff}  \& {Hamuy}}{{Hoeflich} et~al.}{1996}]{1996ApJ...472L..81H}
{Hoeflich} P.,  {Khokhlov} A.,  {Wheeler} J.~C.,  {Phillips} M.~M.,  {Suntzeff}
  N.~B.,   {Hamuy} M.,  1996, \mn@doi [\apjl] {10.1086/310363}, \href
  {https://ui.adsabs.harvard.edu/abs/1996ApJ...472L..81H} {472, L81}

\bibitem[\protect\citeauthoryear{{Horne}}{{Horne}}{1986}]{1986PASP...98..609H}
{Horne} K.,  1986, \mn@doi [\pasp] {10.1086/131801}, \href
  {https://ui.adsabs.harvard.edu/abs/1986PASP...98..609H} {98, 609}

\bibitem[\protect\citeauthoryear{{Howell}}{{Howell}}{2011}]{2011NatCo...2..350H}
{Howell} D.~A.,  2011, \mn@doi [Nature Communications] {10.1038/ncomms1344},
  \href {https://ui.adsabs.harvard.edu/abs/2011NatCo...2..350H} {2, 350}

\bibitem[\protect\citeauthoryear{{Howell} et~al.,}{{Howell}
  et~al.}{2006}]{2006Natur.443..308H}
{Howell} D.~A.,  et~al., 2006, \mn@doi [\nat] {10.1038/nature05103}, \href
  {https://ui.adsabs.harvard.edu/abs/2006Natur.443..308H} {443, 308}

\bibitem[\protect\citeauthoryear{{Hoyle} \& {Fowler}}{{Hoyle} \&
  {Fowler}}{1960}]{1960ApJ...132..565H}
{Hoyle} F.,  {Fowler} W.~A.,  1960, \mn@doi [\apj] {10.1086/146963}, \href
  {https://ui.adsabs.harvard.edu/abs/1960ApJ...132..565H} {132, 565}

\bibitem[\protect\citeauthoryear{{Huang}, {Li}, {Wang}, {Shang}, {Zhang}, {Hu},
  {Qiu}  \& {Jiang}}{{Huang} et~al.}{2012}]{2012RAA....12.1585H}
{Huang} F.,  {Li} J.-Z.,  {Wang} X.-F.,  {Shang} R.-C.,  {Zhang} T.-M.,  {Hu}
  J.-Y.,  {Qiu} Y.-L.,   {Jiang} X.-J.,  2012, \mn@doi [Research in Astronomy
  and Astrophysics] {10.1088/1674-4527/12/11/012}, \href
  {https://ui.adsabs.harvard.edu/abs/2012RAA....12.1585H} {12, 1585}

\bibitem[\protect\citeauthoryear{{Iben} \& {Tutukov}}{{Iben} \&
  {Tutukov}}{1984}]{1984ApJS...54..335I}
{Iben} I. J.,  {Tutukov} A.~V.,  1984, \mn@doi [\apjs] {10.1086/190932}, \href
  {https://ui.adsabs.harvard.edu/abs/1984ApJS...54..335I} {54, 335}

\bibitem[\protect\citeauthoryear{{Im}, {Lim}, {Paek}, {Choi}  \& {Sung}}{{Im}
  et~al.}{2019}]{2019ATel12720....1I}
{Im} M.,  {Lim} G.,  {Paek} G. S.~H.,  {Choi} C.,   {Sung} H.-I.,  2019, The
  Astronomer's Telegram, \href
  {https://ui.adsabs.harvard.edu/abs/2019ATel12720....1I} {12720, 1}

\bibitem[\protect\citeauthoryear{{Jensen} et~al.,}{{Jensen}
  et~al.}{2021}]{2021ApJS..255...21J}
{Jensen} J.~B.,  et~al., 2021, \mn@doi [\apjs] {10.3847/1538-4365/ac01e7},
  \href {https://ui.adsabs.harvard.edu/abs/2021ApJS..255...21J} {255, 21}

\bibitem[\protect\citeauthoryear{{Jha} et~al.,}{{Jha}
  et~al.}{2006}]{2006AJ....131..527J}
{Jha} S.,  et~al., 2006, \mn@doi [\aj] {10.1086/497989}, \href
  {https://ui.adsabs.harvard.edu/abs/2006AJ....131..527J} {131, 527}

\bibitem[\protect\citeauthoryear{{Jha}, {Riess}  \& {Kirshner}}{{Jha}
  et~al.}{2007}]{2007ApJ...659..122J}
{Jha} S.,  {Riess} A.~G.,   {Kirshner} R.~P.,  2007, \mn@doi [\apj]
  {10.1086/512054}, \href
  {https://ui.adsabs.harvard.edu/abs/2007ApJ...659..122J} {659, 122}

\bibitem[\protect\citeauthoryear{{Kawabata} et~al.,}{{Kawabata}
  et~al.}{2020}]{2020ApJ...893..143K}
{Kawabata} M.,  et~al., 2020, \mn@doi [\apj] {10.3847/1538-4357/ab8236}, \href
  {https://ui.adsabs.harvard.edu/abs/2020ApJ...893..143K} {893, 143}

\bibitem[\protect\citeauthoryear{{Khokhlov}}{{Khokhlov}}{1991}]{1991A&A...245..114K}
{Khokhlov} A.~M.,  1991, \aap, \href
  {https://ui.adsabs.harvard.edu/abs/1991A&A...245..114K} {245, 114}

\bibitem[\protect\citeauthoryear{{Krisciunas} et~al.,}{{Krisciunas}
  et~al.}{2004}]{2004AJ....128.3034K}
{Krisciunas} K.,  et~al., 2004, \mn@doi [\aj] {10.1086/425629}, \href
  {https://ui.adsabs.harvard.edu/abs/2004AJ....128.3034K} {128, 3034}

\bibitem[\protect\citeauthoryear{{Landolt}}{{Landolt}}{1992}]{1992AJ....104..340L}
{Landolt} A.~U.,  1992, \mn@doi [\aj] {10.1086/116242}, \href
  {https://ui.adsabs.harvard.edu/abs/1992AJ....104..340L} {104, 340}

\bibitem[\protect\citeauthoryear{{Leibundgut} et~al.,}{{Leibundgut}
  et~al.}{1993}]{1993AJ....105..301L}
{Leibundgut} B.,  et~al., 1993, \mn@doi [\aj] {10.1086/116427}, \href
  {https://ui.adsabs.harvard.edu/abs/1993AJ....105..301L} {105, 301}

\bibitem[\protect\citeauthoryear{{Li} et~al.,}{{Li}
  et~al.}{2019}]{2019ApJ...870...12L}
{Li} W.,  et~al., 2019, \mn@doi [\apj] {10.3847/1538-4357/aaec74}, \href
  {https://ui.adsabs.harvard.edu/abs/2019ApJ...870...12L} {870, 12}

\bibitem[\protect\citeauthoryear{{Li} et~al.,}{{Li}
  et~al.}{2021}]{2021ApJ...906...99L}
{Li} W.,  et~al., 2021, \mn@doi [\apj] {10.3847/1538-4357/abc9b5}, \href
  {https://ui.adsabs.harvard.edu/abs/2021ApJ...906...99L} {906, 99}

\bibitem[\protect\citeauthoryear{{Liu} et~al.,}{{Liu}
  et~al.}{2022}]{2022arXiv220904463L}
{Liu} C.,  et~al., 2022, arXiv e-prints, \href
  {https://ui.adsabs.harvard.edu/abs/2022arXiv220904463L} {p. arXiv:2209.04463}

\bibitem[\protect\citeauthoryear{{Livio} \& {Mazzali}}{{Livio} \&
  {Mazzali}}{2018}]{2018PhR...736....1L}
{Livio} M.,  {Mazzali} P.,  2018, \mn@doi [\physrep]
  {10.1016/j.physrep.2018.02.002}, \href
  {https://ui.adsabs.harvard.edu/abs/2018PhR...736....1L} {736, 1}

\bibitem[\protect\citeauthoryear{{Livne}}{{Livne}}{1990}]{1990ApJ...354L..53L}
{Livne} E.,  1990, \mn@doi [\apjl] {10.1086/185721}, \href
  {https://ui.adsabs.harvard.edu/abs/1990ApJ...354L..53L} {354, L53}

\bibitem[\protect\citeauthoryear{{Maeda} et~al.,}{{Maeda}
  et~al.}{2010a}]{2010Natur.466...82M}
{Maeda} K.,  et~al., 2010a, \mn@doi [\nat] {10.1038/nature09122}, \href
  {https://ui.adsabs.harvard.edu/abs/2010Natur.466...82M} {466, 82}

\bibitem[\protect\citeauthoryear{{Maeda}, {R{\"o}pke}, {Fink}, {Hillebrandt},
  {Travaglio}  \& {Thielemann}}{{Maeda} et~al.}{2010b}]{2010ApJ...712..624M}
{Maeda} K.,  {R{\"o}pke} F.~K.,  {Fink} M.,  {Hillebrandt} W.,  {Travaglio} C.,
    {Thielemann} F.~K.,  2010b, \mn@doi [\apj] {10.1088/0004-637X/712/1/624},
  \href {https://ui.adsabs.harvard.edu/abs/2010ApJ...712..624M} {712, 624}

\bibitem[\protect\citeauthoryear{{Maguire} et~al.,}{{Maguire}
  et~al.}{2014}]{2014MNRAS.444.3258M}
{Maguire} K.,  et~al., 2014, \mn@doi [\mnras] {10.1093/mnras/stu1607}, \href
  {https://ui.adsabs.harvard.edu/abs/2014MNRAS.444.3258M} {444, 3258}

\bibitem[\protect\citeauthoryear{{Maguire} et~al.,}{{Maguire}
  et~al.}{2018}]{2018MNRAS.477.3567M}
{Maguire} K.,  et~al., 2018, \mn@doi [\mnras] {10.1093/mnras/sty820}, \href
  {https://ui.adsabs.harvard.edu/abs/2018MNRAS.477.3567M} {477, 3567}

\bibitem[\protect\citeauthoryear{{Maoz}, {Mannucci}  \& {Nelemans}}{{Maoz}
  et~al.}{2014}]{2014ARA&A..52..107M}
{Maoz} D.,  {Mannucci} F.,   {Nelemans} G.,  2014, \mn@doi [\araa]
  {10.1146/annurev-astro-082812-141031}, \href
  {https://ui.adsabs.harvard.edu/abs/2014ARA&A..52..107M} {52, 107}

\bibitem[\protect\citeauthoryear{{Masci} et~al.,}{{Masci}
  et~al.}{2019}]{2019PASP..131a8003M}
{Masci} F.~J.,  et~al., 2019, \mn@doi [\pasp] {10.1088/1538-3873/aae8ac}, \href
  {https://ui.adsabs.harvard.edu/abs/2019PASP..131a8003M} {131, 018003}

\bibitem[\protect\citeauthoryear{{Matheson} et~al.,}{{Matheson}
  et~al.}{2008}]{2008AJ....135.1598M}
{Matheson} T.,  et~al., 2008, \mn@doi [\aj] {10.1088/0004-6256/135/4/1598},
  \href {https://ui.adsabs.harvard.edu/abs/2008AJ....135.1598M} {135, 1598}

\bibitem[\protect\citeauthoryear{{Mathew} \& {Nandy}}{{Mathew} \&
  {Nandy}}{2017}]{2017RAA....17...61M}
{Mathew} A.,  {Nandy} M.~K.,  2017, \mn@doi [Research in Astronomy and
  Astrophysics] {10.1088/1674-4527/17/6/61}, \href
  {https://ui.adsabs.harvard.edu/abs/2017RAA....17...61M} {17, 061}

\bibitem[\protect\citeauthoryear{{Mazzali} et~al.,}{{Mazzali}
  et~al.}{2014}]{2014MNRAS.439.1959M}
{Mazzali} P.~A.,  et~al., 2014, \mn@doi [\mnras] {10.1093/mnras/stu077}, \href
  {https://ui.adsabs.harvard.edu/abs/2014MNRAS.439.1959M} {439, 1959}

\bibitem[\protect\citeauthoryear{{Mazzali} et~al.,}{{Mazzali}
  et~al.}{2015}]{2015MNRAS.450.2631M}
{Mazzali} P.~A.,  et~al., 2015, \mn@doi [\mnras] {10.1093/mnras/stv761}, \href
  {https://ui.adsabs.harvard.edu/abs/2015MNRAS.450.2631M} {450, 2631}

\bibitem[\protect\citeauthoryear{{Nomoto}}{{Nomoto}}{1982a}]{1982ApJ...253..798N}
{Nomoto} K.,  1982a, \mn@doi [\apj] {10.1086/159682}, \href
  {https://ui.adsabs.harvard.edu/abs/1982ApJ...253..798N} {253, 798}

\bibitem[\protect\citeauthoryear{{Nomoto}}{{Nomoto}}{1982b}]{1982ApJ...257..780N}
{Nomoto} K.,  1982b, \mn@doi [\apj] {10.1086/160031}, \href
  {https://ui.adsabs.harvard.edu/abs/1982ApJ...257..780N} {257, 780}

\bibitem[\protect\citeauthoryear{{Nomoto}, {Thielemann}  \& {Yokoi}}{{Nomoto}
  et~al.}{1984}]{1984ApJ...286..644N}
{Nomoto} K.,  {Thielemann} F.~K.,   {Yokoi} K.,  1984, \mn@doi [\apj]
  {10.1086/162639}, \href
  {https://ui.adsabs.harvard.edu/abs/1984ApJ...286..644N} {286, 644}

\bibitem[\protect\citeauthoryear{{Nugent} et~al.,}{{Nugent}
  et~al.}{2011}]{2011Natur.480..344N}
{Nugent} P.~E.,  et~al., 2011, \mn@doi [\nat] {10.1038/nature10644}, \href
  {https://ui.adsabs.harvard.edu/abs/2011Natur.480..344N} {480, 344}

\bibitem[\protect\citeauthoryear{{Oke} et~al.,}{{Oke}
  et~al.}{1995}]{1995PASP..107..375O}
{Oke} J.~B.,  et~al., 1995, \mn@doi [\pasp] {10.1086/133562}, \href
  {https://ui.adsabs.harvard.edu/abs/1995PASP..107..375O} {107, 375}

\bibitem[\protect\citeauthoryear{{Pan}}{{Pan}}{2020}]{2020ApJ...895L...5P}
{Pan} Y.-C.,  2020, \mn@doi [\apjl] {10.3847/2041-8213/ab8e47}, \href
  {https://ui.adsabs.harvard.edu/abs/2020ApJ...895L...5P} {895, L5}

\bibitem[\protect\citeauthoryear{{Pan}, {Sullivan}, {Maguire}, {Gal-Yam},
  {Hook}, {Howell}, {Nugent}  \& {Mazzali}}{{Pan}
  et~al.}{2015}]{2015MNRAS.446..354P}
{Pan} Y.~C.,  {Sullivan} M.,  {Maguire} K.,  {Gal-Yam} A.,  {Hook} I.~M.,
  {Howell} D.~A.,  {Nugent} P.~E.,   {Mazzali} P.~A.,  2015, \mn@doi [\mnras]
  {10.1093/mnras/stu2121}, \href
  {https://ui.adsabs.harvard.edu/abs/2015MNRAS.446..354P} {446, 354}

\bibitem[\protect\citeauthoryear{{Parrent} et~al.,}{{Parrent}
  et~al.}{2011}]{2011ApJ...732...30P}
{Parrent} J.~T.,  et~al., 2011, \mn@doi [\apj] {10.1088/0004-637X/732/1/30},
  \href {https://ui.adsabs.harvard.edu/abs/2011ApJ...732...30P} {732, 30}

\bibitem[\protect\citeauthoryear{{Parrent} et~al.,}{{Parrent}
  et~al.}{2012}]{2012ApJ...752L..26P}
{Parrent} J.~T.,  et~al., 2012, \mn@doi [\apjl] {10.1088/2041-8205/752/2/L26},
  \href {https://ui.adsabs.harvard.edu/abs/2012ApJ...752L..26P} {752, L26}

\bibitem[\protect\citeauthoryear{{Parrent}, {Friesen}  \&
  {Parthasarathy}}{{Parrent} et~al.}{2014}]{2014Ap&SS.351....1P}
{Parrent} J.,  {Friesen} B.,   {Parthasarathy} M.,  2014, \mn@doi [\apss]
  {10.1007/s10509-014-1830-1}, \href
  {https://ui.adsabs.harvard.edu/abs/2014Ap&SS.351....1P} {351, 1}

\bibitem[\protect\citeauthoryear{{Pastorello} et~al.,}{{Pastorello}
  et~al.}{2007}]{2007MNRAS.377.1531P}
{Pastorello} A.,  et~al., 2007, \mn@doi [\mnras]
  {10.1111/j.1365-2966.2007.11700.x}, \href
  {https://ui.adsabs.harvard.edu/abs/2007MNRAS.377.1531P} {377, 1531}

\bibitem[\protect\citeauthoryear{{Patra} et~al.,}{{Patra}
  et~al.}{2022}]{2022MNRAS.509.4058P}
{Patra} K.~C.,  et~al., 2022, \mn@doi [\mnras] {10.1093/mnras/stab3136}, \href
  {https://ui.adsabs.harvard.edu/abs/2022MNRAS.509.4058P} {509, 4058}

\bibitem[\protect\citeauthoryear{{Pellegrino} et~al.,}{{Pellegrino}
  et~al.}{2020}]{2020ApJ...897..159P}
{Pellegrino} C.,  et~al., 2020, \mn@doi [\apj] {10.3847/1538-4357/ab8e3f},
  \href {https://ui.adsabs.harvard.edu/abs/2020ApJ...897..159P} {897, 159}

\bibitem[\protect\citeauthoryear{{Pereira} et~al.,}{{Pereira}
  et~al.}{2013}]{2013A&A...554A..27P}
{Pereira} R.,  et~al., 2013, \mn@doi [\aap] {10.1051/0004-6361/201221008},
  \href {https://ui.adsabs.harvard.edu/abs/2013A&A...554A..27P} {554, A27}

\bibitem[\protect\citeauthoryear{{Perlmutter} et~al.,}{{Perlmutter}
  et~al.}{1999}]{1999ApJ...517..565P}
{Perlmutter} S.,  et~al., 1999, \mn@doi [\apj] {10.1086/307221}, \href
  {https://ui.adsabs.harvard.edu/abs/1999ApJ...517..565P} {517, 565}

\bibitem[\protect\citeauthoryear{{Phillips}}{{Phillips}}{1993}]{1993ApJ...413L.105P}
{Phillips} M.~M.,  1993, \mn@doi [\apjl] {10.1086/186970}, \href
  {https://ui.adsabs.harvard.edu/abs/1993ApJ...413L.105P} {413, L105}

\bibitem[\protect\citeauthoryear{{Phillips}, {Wells}, {Suntzeff}, {Hamuy},
  {Leibundgut}, {Kirshner}  \& {Foltz}}{{Phillips}
  et~al.}{1992}]{1992AJ....103.1632P}
{Phillips} M.~M.,  {Wells} L.~A.,  {Suntzeff} N.~B.,  {Hamuy} M.,  {Leibundgut}
  B.,  {Kirshner} R.~P.,   {Foltz} C.~B.,  1992, \mn@doi [\aj]
  {10.1086/116177}, \href
  {https://ui.adsabs.harvard.edu/abs/1992AJ....103.1632P} {103, 1632}

\bibitem[\protect\citeauthoryear{{Phillips}, {Lira}, {Suntzeff}, {Schommer},
  {Hamuy}  \& {Maza}}{{Phillips} et~al.}{1999}]{1999AJ....118.1766P}
{Phillips} M.~M.,  {Lira} P.,  {Suntzeff} N.~B.,  {Schommer} R.~A.,  {Hamuy}
  M.,   {Maza} J.,  1999, \mn@doi [\aj] {10.1086/301032}, \href
  {https://ui.adsabs.harvard.edu/abs/1999AJ....118.1766P} {118, 1766}

\bibitem[\protect\citeauthoryear{{Piro} \& {Morozova}}{{Piro} \&
  {Morozova}}{2016}]{2016ApJ...826...96P}
{Piro} A.~L.,  {Morozova} V.~S.,  2016, \mn@doi [\apj]
  {10.3847/0004-637X/826/1/96}, \href
  {https://ui.adsabs.harvard.edu/abs/2016ApJ...826...96P} {826, 96}

\bibitem[\protect\citeauthoryear{{Piro} \& {Nakar}}{{Piro} \&
  {Nakar}}{2013}]{2013ApJ...769...67P}
{Piro} A.~L.,  {Nakar} E.,  2013, \mn@doi [\apj] {10.1088/0004-637X/769/1/67},
  \href {https://ui.adsabs.harvard.edu/abs/2013ApJ...769...67P} {769, 67}

\bibitem[\protect\citeauthoryear{{Polin}, {Nugent}  \& {Kasen}}{{Polin}
  et~al.}{2019}]{2019ApJ...873...84P}
{Polin} A.,  {Nugent} P.,   {Kasen} D.,  2019, \mn@doi [\apj]
  {10.3847/1538-4357/aafb6a}, \href
  {https://ui.adsabs.harvard.edu/abs/2019ApJ...873...84P} {873, 84}

\bibitem[\protect\citeauthoryear{{Richmond} \& {Smith}}{{Richmond} \&
  {Smith}}{2012}]{2012JAVSO..40..872R}
{Richmond} M.~W.,  {Smith} H.~A.,  2012, \jaavso, \href
  {https://ui.adsabs.harvard.edu/abs/2012JAVSO..40..872R} {40, 872}

\bibitem[\protect\citeauthoryear{{Riess} et~al.,}{{Riess}
  et~al.}{1998}]{1998AJ....116.1009R}
{Riess} A.~G.,  et~al., 1998, \mn@doi [\aj] {10.1086/300499}, \href
  {https://ui.adsabs.harvard.edu/abs/1998AJ....116.1009R} {116, 1009}

\bibitem[\protect\citeauthoryear{{Riess} et~al.,}{{Riess}
  et~al.}{2005}]{2005ApJ...627..579R}
{Riess} A.~G.,  et~al., 2005, \mn@doi [\apj] {10.1086/430497}, \href
  {https://ui.adsabs.harvard.edu/abs/2005ApJ...627..579R} {627, 579}

\bibitem[\protect\citeauthoryear{{Riess}, {Casertano}, {Yuan}, {Macri}  \&
  {Scolnic}}{{Riess} et~al.}{2019}]{2019ApJ...876...85R}
{Riess} A.~G.,  {Casertano} S.,  {Yuan} W.,  {Macri} L.~M.,   {Scolnic} D.,
  2019, \mn@doi [\apj] {10.3847/1538-4357/ab1422}, \href
  {https://ui.adsabs.harvard.edu/abs/2019ApJ...876...85R} {876, 85}

\bibitem[\protect\citeauthoryear{{Riess} et~al.,}{{Riess}
  et~al.}{2021}]{2021arXiv211204510R}
{Riess} A.~G.,  et~al., 2021, arXiv e-prints, \href
  {https://ui.adsabs.harvard.edu/abs/2021arXiv211204510R} {p. arXiv:2112.04510}

\bibitem[\protect\citeauthoryear{{Roming} et~al.,}{{Roming}
  et~al.}{2005}]{2005SSRv..120...95R}
{Roming} P. W.~A.,  et~al., 2005, \mn@doi [\ssr] {10.1007/s11214-005-5095-4},
  \href {https://ui.adsabs.harvard.edu/abs/2005SSRv..120...95R} {120, 95}

\bibitem[\protect\citeauthoryear{{Sai} et~al.,}{{Sai}
  et~al.}{2022}]{2022MNRAS.tmp.1454S}
{Sai} H.,  et~al., 2022, \mn@doi [\mnras] {10.1093/mnras/stac1525}, \href
  {https://ui.adsabs.harvard.edu/abs/2022MNRAS.tmp.1454S} {}

\bibitem[\protect\citeauthoryear{{Schlafly} \& {Finkbeiner}}{{Schlafly} \&
  {Finkbeiner}}{2011}]{2011ApJ...737..103S}
{Schlafly} E.~F.,  {Finkbeiner} D.~P.,  2011, \mn@doi [\apj]
  {10.1088/0004-637X/737/2/103}, \href
  {https://ui.adsabs.harvard.edu/abs/2011ApJ...737..103S} {737, 103}

\bibitem[\protect\citeauthoryear{{Schmidt} et~al.,}{{Schmidt}
  et~al.}{1998}]{1998ApJ...507...46S}
{Schmidt} B.~P.,  et~al., 1998, \mn@doi [\apj] {10.1086/306308}, \href
  {https://ui.adsabs.harvard.edu/abs/1998ApJ...507...46S} {507, 46}

\bibitem[\protect\citeauthoryear{{Scolnic} et~al.,}{{Scolnic}
  et~al.}{2018}]{2018ApJ...859..101S}
{Scolnic} D.~M.,  et~al., 2018, \mn@doi [\apj] {10.3847/1538-4357/aab9bb},
  \href {https://ui.adsabs.harvard.edu/abs/2018ApJ...859..101S} {859, 101}

\bibitem[\protect\citeauthoryear{{Seitenzahl}, {Herzog}, {Ruiter}, {Marquardt},
  {Ohlmann}  \& {R{\"o}pke}}{{Seitenzahl} et~al.}{2015}]{2015PhRvD..92l4013S}
{Seitenzahl} I.~R.,  {Herzog} M.,  {Ruiter} A.~J.,  {Marquardt} K.,  {Ohlmann}
  S.~T.,   {R{\"o}pke} F.~K.,  2015, \mn@doi [\prd]
  {10.1103/PhysRevD.92.124013}, \href
  {https://ui.adsabs.harvard.edu/abs/2015PhRvD..92l4013S} {92, 124013}

\bibitem[\protect\citeauthoryear{{Shen}, {Boos}, {Townsley}  \& {Kasen}}{{Shen}
  et~al.}{2021}]{2021ApJ...922...68S}
{Shen} K.~J.,  {Boos} S.~J.,  {Townsley} D.~M.,   {Kasen} D.,  2021, \mn@doi
  [\apj] {10.3847/1538-4357/ac2304}, \href
  {https://ui.adsabs.harvard.edu/abs/2021ApJ...922...68S} {922, 68}

\bibitem[\protect\citeauthoryear{{Silverman} \& {Filippenko}}{{Silverman} \&
  {Filippenko}}{2012}]{2012MNRAS.425.1917S}
{Silverman} J.~M.,  {Filippenko} A.~V.,  2012, \mn@doi [\mnras]
  {10.1111/j.1365-2966.2012.21276.x}, \href
  {https://ui.adsabs.harvard.edu/abs/2012MNRAS.425.1917S} {425, 1917}

\bibitem[\protect\citeauthoryear{{Silverman} et~al.,}{{Silverman}
  et~al.}{2012}]{2012MNRAS.425.1789S}
{Silverman} J.~M.,  et~al., 2012, \mn@doi [\mnras]
  {10.1111/j.1365-2966.2012.21270.x}, \href
  {https://ui.adsabs.harvard.edu/abs/2012MNRAS.425.1789S} {425, 1789}

\bibitem[\protect\citeauthoryear{{Simon} et~al.,}{{Simon}
  et~al.}{2007}]{2007ApJ...671L..25S}
{Simon} J.~D.,  et~al., 2007, \mn@doi [\apjl] {10.1086/524707}, \href
  {https://ui.adsabs.harvard.edu/abs/2007ApJ...671L..25S} {671, L25}

\bibitem[\protect\citeauthoryear{{Smith} et~al.,}{{Smith}
  et~al.}{2020}]{2020PASP..132h5002S}
{Smith} K.~W.,  et~al., 2020, \mn@doi [\pasp] {10.1088/1538-3873/ab936e}, \href
  {https://ui.adsabs.harvard.edu/abs/2020PASP..132h5002S} {132, 085002}

\bibitem[\protect\citeauthoryear{{Stahl} et~al.,}{{Stahl}
  et~al.}{2019}]{2019MNRAS.490.3882S}
{Stahl} B.~E.,  et~al., 2019, \mn@doi [\mnras] {10.1093/mnras/stz2742}, \href
  {https://ui.adsabs.harvard.edu/abs/2019MNRAS.490.3882S} {490, 3882}

\bibitem[\protect\citeauthoryear{{Stahl} et~al.,}{{Stahl}
  et~al.}{2020}]{2020MNRAS.492.4325S}
{Stahl} B.~E.,  et~al., 2020, \mn@doi [\mnras] {10.1093/mnras/staa102}, \href
  {https://ui.adsabs.harvard.edu/abs/2020MNRAS.492.4325S} {492, 4325}

\bibitem[\protect\citeauthoryear{{Stanishev} et~al.,}{{Stanishev}
  et~al.}{2007}]{2007A&A...469..645S}
{Stanishev} V.,  et~al., 2007, \mn@doi [\aap] {10.1051/0004-6361:20066020},
  \href {https://ui.adsabs.harvard.edu/abs/2007A&A...469..645S} {469, 645}

\bibitem[\protect\citeauthoryear{{Stritzinger} \& {Leibundgut}}{{Stritzinger}
  \& {Leibundgut}}{2005}]{2005A&A...431..423S}
{Stritzinger} M.,  {Leibundgut} B.,  2005, \mn@doi [\aap]
  {10.1051/0004-6361:20041630}, \href
  {https://ui.adsabs.harvard.edu/abs/2005A&A...431..423S} {431, 423}

\bibitem[\protect\citeauthoryear{{Taubenberger}}{{Taubenberger}}{2017}]{2017hsn..book..317T}
{Taubenberger} S.,  2017, in {Alsabti} A.~W.,  {Murdin} P.,  eds, , Handbook of
  Supernovae.
p.~317, \mn@doi{10.1007/978-3-319-21846-5\_37}

\bibitem[\protect\citeauthoryear{{Thomas}, {Nugent}  \& {Meza}}{{Thomas}
  et~al.}{2011a}]{2011PASP..123..237T}
{Thomas} R.~C.,  {Nugent} P.~E.,   {Meza} J.~C.,  2011a, \mn@doi [\pasp]
  {10.1086/658673}, \href
  {https://ui.adsabs.harvard.edu/abs/2011PASP..123..237T} {123, 237}

\bibitem[\protect\citeauthoryear{{Thomas} et~al.,}{{Thomas}
  et~al.}{2011b}]{2011ApJ...743...27T}
{Thomas} R.~C.,  et~al., 2011b, \mn@doi [\apj] {10.1088/0004-637X/743/1/27},
  \href {https://ui.adsabs.harvard.edu/abs/2011ApJ...743...27T} {743, 27}

\bibitem[\protect\citeauthoryear{{Tonry} et~al.,}{{Tonry}
  et~al.}{2018}]{2018PASP..130f4505T}
{Tonry} J.~L.,  et~al., 2018, \mn@doi [\pasp] {10.1088/1538-3873/aabadf}, \href
  {https://ui.adsabs.harvard.edu/abs/2018PASP..130f4505T} {130, 064505}

\bibitem[\protect\citeauthoryear{{Tonry} et~al.,}{{Tonry}
  et~al.}{2019}]{2019TNSTR.678....1T}
{Tonry} J.,  et~al., 2019, Transient Name Server Discovery Report, \href
  {https://ui.adsabs.harvard.edu/abs/2019TNSTR.678....1T} {2019-678, 1}

\bibitem[\protect\citeauthoryear{{Townsley}, {Miles}, {Shen}  \&
  {Kasen}}{{Townsley} et~al.}{2019}]{2019ApJ...878L..38T}
{Townsley} D.~M.,  {Miles} B.~J.,  {Shen} K.~J.,   {Kasen} D.,  2019, \mn@doi
  [\apjl] {10.3847/2041-8213/ab27cd}, \href
  {https://ui.adsabs.harvard.edu/abs/2019ApJ...878L..38T} {878, L38}

\bibitem[\protect\citeauthoryear{{Tsvetkov}, {Shugarov}, {Volkov}, {Goranskij},
  {Pavlyuk}, {Katysheva}, {Barsukova}  \& {Valeev}}{{Tsvetkov}
  et~al.}{2013}]{2013CoSka..43...94T}
{Tsvetkov} D.~Y.,  {Shugarov} S.~Y.,  {Volkov} I.~M.,  {Goranskij} V.~P.,
  {Pavlyuk} N.~N.,  {Katysheva} N.~A.,  {Barsukova} E.~A.,   {Valeev} A.~F.,
  2013, Contributions of the Astronomical Observatory Skalnate Pleso, \href
  {https://ui.adsabs.harvard.edu/abs/2013CoSka..43...94T} {43, 94}

\bibitem[\protect\citeauthoryear{{Wang}, {Wang}, {Zhou}, {Lou}  \& {Li}}{{Wang}
  et~al.}{2005}]{2005ApJ...620L..87W}
{Wang} X.,  {Wang} L.,  {Zhou} X.,  {Lou} Y.-Q.,   {Li} Z.,  2005, \mn@doi
  [\apjl] {10.1086/428774}, \href
  {https://ui.adsabs.harvard.edu/abs/2005ApJ...620L..87W} {620, L87}

\bibitem[\protect\citeauthoryear{{Wang} et~al.,}{{Wang}
  et~al.}{2008}]{2008ApJ...675..626W}
{Wang} X.,  et~al., 2008, \mn@doi [\apj] {10.1086/526413}, \href
  {https://ui.adsabs.harvard.edu/abs/2008ApJ...675..626W} {675, 626}

\bibitem[\protect\citeauthoryear{{Wang}, {Li}  \& {Filippenko}}{{Wang}
  et~al.}{2009a}]{2009AAS...21331204W}
{Wang} X.,  {Li} W.,   {Filippenko} A.~V.,  2009a, in American Astronomical
  Society Meeting Abstracts \#213. p. 312.04

\bibitem[\protect\citeauthoryear{{Wang} et~al.,}{{Wang}
  et~al.}{2009b}]{2009ApJ...699L.139W}
{Wang} X.,  et~al., 2009b, \mn@doi [\apjl] {10.1088/0004-637X/699/2/L139},
  \href {https://ui.adsabs.harvard.edu/abs/2009ApJ...699L.139W} {699, L139}

\bibitem[\protect\citeauthoryear{{Wang}, {Wang}, {Filippenko}, {Zhang}  \&
  {Zhao}}{{Wang} et~al.}{2013}]{2013Sci...340..170W}
{Wang} X.,  {Wang} L.,  {Filippenko} A.~V.,  {Zhang} T.,   {Zhao} X.,  2013,
  \mn@doi [Science] {10.1126/science.1231502}, \href
  {https://ui.adsabs.harvard.edu/abs/2013Sci...340..170W} {340, 170}

\bibitem[\protect\citeauthoryear{{Wang}, {Chen}, {Wang}, {Hu}, {Xi}, {Yang},
  {Zhao}  \& {Li}}{{Wang} et~al.}{2019}]{2019ApJ...882..120W}
{Wang} X.,  {Chen} J.,  {Wang} L.,  {Hu} M.,  {Xi} G.,  {Yang} Y.,  {Zhao} X.,
   {Li} W.,  2019, \mn@doi [\apj] {10.3847/1538-4357/ab26b5}, \href
  {https://ui.adsabs.harvard.edu/abs/2019ApJ...882..120W} {882, 120}

\bibitem[\protect\citeauthoryear{{Whelan} \& {Iben}}{{Whelan} \&
  {Iben}}{1973}]{1973ApJ...186.1007W}
{Whelan} J.,  {Iben} Icko J.,  1973, \mn@doi [\apj] {10.1086/152565}, \href
  {https://ui.adsabs.harvard.edu/abs/1973ApJ...186.1007W} {186, 1007}

\bibitem[\protect\citeauthoryear{{White} et~al.,}{{White}
  et~al.}{2015}]{2015ApJ...799...52W}
{White} C.~J.,  et~al., 2015, \mn@doi [\apj] {10.1088/0004-637X/799/1/52},
  \href {https://ui.adsabs.harvard.edu/abs/2015ApJ...799...52W} {799, 52}

\bibitem[\protect\citeauthoryear{{Woosley} \& {Weaver}}{{Woosley} \&
  {Weaver}}{1994}]{1994ApJ...423..371W}
{Woosley} S.~E.,  {Weaver} T.~A.,  1994, \mn@doi [\apj] {10.1086/173813}, \href
  {https://ui.adsabs.harvard.edu/abs/1994ApJ...423..371W} {423, 371}

\bibitem[\protect\citeauthoryear{{Yaron} \& {Gal-Yam}}{{Yaron} \&
  {Gal-Yam}}{2012}]{2012PASP..124..668Y}
{Yaron} O.,  {Gal-Yam} A.,  2012, \mn@doi [\pasp] {10.1086/666656}, \href
  {https://ui.adsabs.harvard.edu/abs/2012PASP..124..668Y} {124, 668}

\bibitem[\protect\citeauthoryear{{York} et~al.,}{{York}
  et~al.}{2000}]{2000AJ....120.1579Y}
{York} D.~G.,  et~al., 2000, \mn@doi [\aj] {10.1086/301513}, \href
  {https://ui.adsabs.harvard.edu/abs/2000AJ....120.1579Y} {120, 1579}

\bibitem[\protect\citeauthoryear{{Zeng} et~al.,}{{Zeng}
  et~al.}{2021}]{2021ApJ...919...49Z}
{Zeng} X.,  et~al., 2021, \mn@doi [\apj] {10.3847/1538-4357/ac0e9c}, \href
  {https://ui.adsabs.harvard.edu/abs/2021ApJ...919...49Z} {919, 49}

\bibitem[\protect\citeauthoryear{{Zhang}, {Wang}, {Bai}, {Zhang}, {Wang},
  {Liu}, {Zhao}  \& {Chen}}{{Zhang} et~al.}{2014}]{2014AJ....148....1Z}
{Zhang} J.-J.,  {Wang} X.-F.,  {Bai} J.-M.,  {Zhang} T.-M.,  {Wang} B.,  {Liu}
  Z.-W.,  {Zhao} X.-L.,   {Chen} J.-C.,  2014, \mn@doi [\aj]
  {10.1088/0004-6256/148/1/1}, \href
  {https://ui.adsabs.harvard.edu/abs/2014AJ....148....1Z} {148, 1}

\bibitem[\protect\citeauthoryear{{Zhang} et~al.,}{{Zhang}
  et~al.}{2016}]{2016ApJ...820...67Z}
{Zhang} K.,  et~al., 2016, \mn@doi [\apj] {10.3847/0004-637X/820/1/67}, \href
  {https://ui.adsabs.harvard.edu/abs/2016ApJ...820...67Z} {820, 67}

\bibitem[\protect\citeauthoryear{{Zhao} et~al.,}{{Zhao}
  et~al.}{2015}]{2015ApJS..220...20Z}
{Zhao} X.,  et~al., 2015, \mn@doi [\apjs] {10.1088/0067-0049/220/1/20}, \href
  {https://ui.adsabs.harvard.edu/abs/2015ApJS..220...20Z} {220, 20}

\makeatother
\end{thebibliography}

\clearpage

\begin{table*}
\caption{Photometric Standards in the SN~2019ein Field\label{tabstd}}
\begin{threeparttable}[b]
\begin{tabular}{r c c c c c c c}
\hline
   Num. & $\alpha(J2000)$ & $\delta$(J2000) & $u$ (mag) & $g$ (mag) & $r$ (mag) & $i$ (mag) & $z$ (mag)  \\
\hline
1 & $13^{\rm h}53^{\rm m}27^{\rm s}.757$ & $+40^{\circ}20'52''.112$ & 20.593(059) & 18.040(017) & 16.628(018) & 15.571(027) & 15.080(019)\\
2 & $13^{\rm h}53^{\rm m}30^{\rm s}.900$ & $+40^{\circ}22'40''.724$ & 20.755(065) & 18.185(017) & 16.750(018) & 15.945(027) & 15.523(019)\\
3 & $13^{\rm h}53^{\rm m}26^{\rm s}.842$ & $+40^{\circ}14'51''.720$ & 20.369(050) & 18.430(025) & 17.632(028) & 17.285(020) & 17.189(025)\\
4 & $13^{\rm h}53^{\rm m}23^{\rm s}.152$ & $+40^{\circ}13'27''.638$ & 22.389(167) & 19.836(026) & 18.547(026) & 17.896(013) & 17.528(025)\\
5 & $13^{\rm h}53^{\rm m}31^{\rm s}.221$ & $+40^{\circ}18'45''.738$ & 18.995(024) & 18.191(017) & 17.926(019) & 17.815(028) & 17.846(029)\\
6 & $13^{\rm h}53^{\rm m}27^{\rm s}.801$ & $+40^{\circ}15'16''.085$ & 20.400(045) & 18.015(024) & 16.596(027) & 15.308(013) & 14.708(021)\\
7 & $13^{\rm h}53^{\rm m}32^{\rm s}.258$ & $+40^{\circ}21'33''.746$ & 18.041(016) & 16.762(017) & 16.282(018) & 16.106(027) & 16.045(020)\\
8 & $13^{\rm h}53^{\rm m}31^{\rm s}.044$ & $+40^{\circ}21'18''.162$ & 22.132(198) & 19.245(019) & 17.984(019) & 17.463(027) & 17.232(022)\\
9 & $13^{\rm h}53^{\rm m}33^{\rm s}.091$ & $+40^{\circ}21'02''.092$ & 20.614(061) & 18.268(017) & 17.159(018) & 16.723(027) & 16.488(020)\\
10 & $13^{\rm h}53^{\rm m}36^{\rm s}.042$ & $+40^{\circ}22'00''.098$ & 19.205(025) & 18.191(017) & 17.824(019) & 17.683(027) & 17.678(025)\\
11 & $13^{\rm h}53^{\rm m}46^{\rm s}.505$ & $+40^{\circ}21'55''.192$ & 20.946(075) & 19.285(019) & 18.668(020) & 18.430(028) & 18.316(032)\\
12 & $13^{\rm h}53^{\rm m}42^{\rm s}.809$ & $+40^{\circ}22'58''.426$ & 19.216(025) & 16.562(017) & 15.163(018) & 14.522(027) & 14.211(019)\\
13 & $13^{\rm h}53^{\rm m}35^{\rm s}.902$ & $+40^{\circ}20'12''.635$ & 20.900(069) & 18.784(018) & 17.952(019) & 17.562(027) & 17.398(023)\\
14 & $13^{\rm h}53^{\rm m}39^{\rm s}.056$ & $+40^{\circ}18'08''.503$ & 20.648(062) & 18.012(017) & 16.663(018) & 16.032(027) & 15.724(019)\\
15 & $13^{\rm h}53^{\rm m}36^{\rm s}.146$ & $+40^{\circ}18'49''.057$ & 15.684(011) & 14.550(016) & 17.295(086) & 17.505(086) & 14.001(019)\\
16 & $13^{\rm h}53^{\rm m}34^{\rm s}.056$ & $+40^{\circ}19'09''.160$ & 20.868(074) & 18.481(017) & 17.389(018) & 16.912(027) & 16.665(020)\\
17 & $13^{\rm h}53^{\rm m}36^{\rm s}.688$ & $+40^{\circ}19'15''.960$ & 17.043(013) & 15.763(016) & 15.299(018) & 15.111(027) & 15.076(019)\\
18 & $13^{\rm h}53^{\rm m}45^{\rm s}.929$ & $+40^{\circ}16'53''.166$ & 16.556(012) & 14.921(016) & 14.343(018) & 14.913(002) & 14.022(019)\\
19 & $13^{\rm h}53^{\rm m}43^{\rm s}.459$ & $+40^{\circ}16'49''.519$ & 21.958(170) & 19.675(020) & 18.491(019) & 18.014(027) & 17.681(025)\\
\hline
\end{tabular}
\begin{tablenotes}
\item[]{Uncertainties, in units of 0.001~mag, are 1$\sigma$.}
\item[]{See Figure~\ref{figVimg} for a finder chart of SN~2019ein and the comparison stars.}
\end{tablenotes}
\end{threeparttable}
\end{table*}

\clearpage
\begin{table*}
\caption{Comparison Objects Used in This Paper\label{tabcmp}}
\begin{threeparttable}[b]
\begin{tabular}{l l l l l l l l l}
\hline
  SN Ia & Type & $B_{\mathrm{max}}$ & $\Delta m_{15}(B)$ & $z_{\mathrm{helio}}$ & $\mu$\tnote{*} & $E(B-V)_{\mathrm{gal}}$ & $E(B-V)_{\mathrm{host}}$ & References\tnote{c}  \\
    &    & (mag) & (mag) & & (mag) & (mag) & (mag) & \\
\hline
2019ein & HV & $-18.71_{\pm 0.15}$ & 1.35$_{\pm 0.01}$ & 0.007755 & $32.711_{\pm 0.076}$   & 0.010 & 0.024 & This paper\\
1997bp  & HV & $-19.43_{\pm 0.29}$ & 1.0 $_{\pm 0.15}$ & 0.008309 & $32.599^*_{\pm 0.205}$ & 0.038 & 0.172 & 17, 19, 20\\
1998dh  & HV & $-19.09_{\pm 0.18}$ & 1.20$_{\pm 0.03}$ & 0.00897  & $32.500^*_{\pm 0.119}$ & 0.058 & 0.095 & 17$\sim$20, 30\\
2001br  & HV & $-18.93$\tnote{a} & 1.35$_{\pm 0.05}$ & 0.020628 & 34.63\tnote{a} & 0.056 & 0.25  & 19, 20, 30\\
2002bo  & HV & $-19.41_{\pm 0.42}$ & 1.13$_{\pm 0.05}$ & 0.00424  & 31.67                  & 0.022 & 0.43  & 2, 14, 20, 22, 30\\
2004dt  & HV & $-19.26$\tnote{a} & 1.11$_{\pm 0.05}$  & 0.01973  & 34.47\tnote{a} & 0.023 & 0.042 & 16, 18, 21, 30\\
2004ef  & HV & $-18.99$\tnote{a} & 1.38$_{\pm 0.05}$ & 0.030985 & 35.48\tnote{a} & 0.047 & 0.18  & 18$\sim$21, 23, 30\\
2005am  & HV & $-19.05$\tnote{b} & 1.45$_{\pm 0.07}$ & 0.007899 & 32.67\tnote{b} & 0.047 & 0.05  & 19$\sim$23, 30\\
2006X   & HV & $-19.10_{\pm 0.20}$ & 1.31$_{\pm 0.05}$ & 0.00524  & $30.91_{\pm 0.14}$     & 0.023 & 1.42  & 7\\
2009ig  & HV & $-19.22_{\pm 0.12}$ & 0.89$_{\pm 0.02}$ & 0.00877  & $32.563^*_{\pm 0.074}$ & 0.027 & 0     & 9, 25$\sim$30\\
2012fr  & HV & $-19.48_{\pm 0.11}$ & 0.80$_{\pm 0.01}$ & 0.0054   & $31.378^*_{\pm 0.056}$ & 0.018 & $\le$ 0.015 & 10, 24\\
2017fgc & HV & $-19.32_{\pm 0.13}$ & 1.05$_{\pm 0.07}$ & 0.007722 & $32.39_{\pm 0.06} $    & 0.029 & 0.17  & 12\\
1998aq  & NV & $-19.47_{\pm 0.11}$ & 1.14$_{\pm 0.05}$ & 0.003699 & $31.722^*_{\pm 0.071}$ & 0.012 & 0     & 1, 15\\
2003du  & NV & $-19.32_{\pm 0.11}$ & 1.04$_{\pm 0.04}$ & 0.006381 & $32.848^*_{\pm 0.067}$ & 0.009 & 0     & 3, 4\\
2004eo  & NV & $-19.08_{\pm 0.10}$ & 1.46$_{\pm 0.04}$ & 0.01571  & $34.12_{\pm 0.10}$     & 0.108 & 0     & 6, 18, 21, 23, 30\\
2005cf  & NV & $-19.28_{\pm 0.15}$ & 1.07$_{\pm 0.03}$ & 0.006461 & $32.363^*_{\pm 0.120}$ & 0.093 & 0.09  & 8\\
2007af  & NV & $-18.97_{\pm 0.09}$ & 1.15$_{\pm 0.03}$ & 0.005464 & $31.772^*_{\pm 0.052}$ & 0.034 & 0.11  & 5, 18$\sim$22, 30, 31\\
2011fe  & NV & $-19.40_{\pm 0.12}$ & 1.18$_{\pm 0.03}$ & 0.000804 & $29.178^*_{\pm 0.041}$ & 0.008 & 0.014 & 11, 27$\sim$40\\
2019np  & NV & $-19.13_{\pm 0.11}$ & 1.04$_{\pm 0.07}$ & 0.00452  & $32.331^*_{\pm 0.076}$ & 0.017 & 0.11  & 13\\
\hline
\end{tabular}
\begin{tablenotes}
\item[*]{Distance moduli from the recent Cepheid measurements of \cite{2021arXiv211204510R} are tagged with asterisks.}
\item[a]{Estimated from the cosmological redshift and H$_0 = 73.04~\mathrm{km~s}^{-1}~\mathrm{Mpc}^{-1}$ \citep{2021arXiv211204510R}. }
\item[b]{Estimated by fitting to SN~Ia templates using \texttt{SNooPy} \citep{2011AJ....141...19B,2014ApJ...789...32B}.}
\item[c]{References: 1, \cite{2003AJ....126.1489B}; 2, \cite{2004MNRAS.348..261B}; 3, \cite{2005A\string&A...429..667A}; 4, \cite{2007A\string&A...469..645S}; 5, \cite{2007ApJ...671L..25S}; 6, \cite{2007MNRAS.377.1531P}; 7, \cite{2008ApJ...675..626W}; 8, \cite{2009AAS...21331204W}; 9, \cite{2012ApJ...744...38F}; 10, \cite{2013ApJ...770...29C}; 11, \cite{2016ApJ...820...67Z}; 12, \cite{2021ApJ...919...49Z}; 13, \cite{2022MNRAS.tmp.1454S}; 14, \cite{2004AJ....128.3034K}; 15, \cite{2005ApJ...627..579R}; 16, \cite{2007A\string&A...475..585A}; 17, \cite{2006AJ....131..527J}; 18, \cite{2010ApJS..190..418G}; 19, \cite{2008AJ....135.1598M}; 20, \cite{2012AJ....143..126B}; 21, \cite{2010AJ....139..519C}; 22, \cite{2009ApJ...700..331H}; 23, \cite{2013ApJ...773...53F}; 24, \cite{2014AJ....148....1Z}; 25, \cite{2012ApJS..200...12H}; 26, \cite{2012ApJ...749...18B}; 27, \cite{2020MNRAS.492.4325S}; 28, \cite{2012PASP..124..668Y}; 29, \cite{2019MNRAS.490.3882S}; 30, \cite{2012MNRAS.425.1789S}; 31, \cite{2014Ap\string&SS.354...89B}; 32, \cite{2012ApJ...752L..26P}; 33, \cite{2012JAVSO..40..872R}; 34, \cite{2011Natur.480..344N}; 35, \cite{2014MNRAS.439.1959M}; 36, \cite{2014MNRAS.444.3258M}; 37, \cite{2015MNRAS.450.2631M}; 38, \cite{2017MNRAS.472.3437G}; 39, \cite{2013A\string&A...554A..27P}; 40, \cite{2013CoSka..43...94T}. }
\end{tablenotes}
\end{threeparttable}
\end{table*}

\clearpage
\begin{table}
\caption{Summary of Photometric Parameters for SN 2019ein\label{tablcfit}}
\begin{threeparttable}[b]
\begin{tabular}{l c c c}
\hline
   Parameter  & \cite{2020ApJ...893..143K} & \cite{2020ApJ...897..159P} & This work\\
\hline
   $t_{\rm max}(B)$ (MJD)    & 58618.24$_{\pm 0.07}$  & 58619.45$_{\pm 0.031}$  & 58619.36$_{\pm 0.02}$\tnote{a}  \\
   $t_{\rm exp}$ (MJD)       & 58602.87$_{\pm 0.55}$  & 58602.73                & 58603.20$_{\pm 0.30}$  \\
   $B_{\rm max}$ (mag)       &    13.99$_{\pm 0.03}$  & 14.05$_{\pm 0.05}$\tnote{c} &   14.056$_{\pm 0.016}$\tnote{a} \\
   $\Delta m_{15}(B)$ (mag)  &     1.36$_{\pm 0.02}$  &    1.400$_{\pm 0.004}$  &    1.348$_{\pm 0.005}$\tnote{a} \\
   $c$                       &      -                 &    0.003$_{\pm 0.0174}$ &    0.039$_{\pm 0.015}$\tnote{a} \\
   $x_0$                     &      -                 &    0.044$_{\pm 0.0007}$ &   0.0429$_{\pm 0.0006}$\tnote{a}\\
   $x_1$                     &      -                 &   $-1.678_{\pm 0.026}$  &   $-1.416_{\pm 0.023}$\tnote{a} \\
   $\mu$ (mag)               &    32.95$_{\pm 0.12}$  &    32.59$_{\pm 0.11}$   &   32.711$_{\pm 0.076}$\tnote{b} \\
   $E(B-V)_{\rm host}$ (mag) &     0.09$_{\pm 0.02}$  &     0.09$_{\pm 0.02}$   &    0.024$_{\pm 0.050}$ \\
   $M_{\rm max}(B)$ (mag)    &   $-19.14_{\pm 0.10}$  &   $-18.81_{\pm 0.059}$  &   $-18.71_{\pm 0.15}$ \\
   $M(\rm Ni)$ (M$_\odot$)   &      -                 &      0.33               &     0.27$_{\pm 0.04}$  \\
\hline
\end{tabular}
\begin{tablenotes}
\item[a] {From \texttt{SALT2} fits \citep{2007A&A...466...11G}.}
\item[b] {From \cite{2021ApJS..255...21J}.}
\item[c] {Inferred from $M_{\rm max}(B)$ along with adopted distant modulus and extinction.}
\end{tablenotes}
\end{threeparttable}
\end{table}

\begin{table*}
\caption{SN~2019ein Photometry from TNT\label{tablctnt}}
\begin{threeparttable}[b]
\begin{tabular}{c c c c c c}
\hline
   MJD & $B$ (mag) & $V$ (mag) & $g$ (mag) & $r$ (mag) & $i$ (mag) \\
\hline
58608.74 & 15.54(04) & 15.35(02) & 15.59(06) & 15.48(07) & 15.61(05)\\
58609.77 & 15.19(04) & 15.11(02) & 15.28(04) & 15.28(05) & 15.43(05)\\
58610.74 & 14.93(04) & 14.84(02) & 15.00(05) & 14.97(06) & 15.20(07)\\
58611.60 & 14.65(04) & 14.64(02) & 14.82(04) & 14.81(05) & 15.04(02)\\
58612.73 & 14.43(04) & 14.44(02) & 14.60(04) & 14.58(03) & 14.82(03)\\
58613.59 & 14.30(04) & 14.32(02) & 14.46(03) & 14.45(05) & 14.72(05)\\
58623.69 & 14.29(04) & 14.00(02) & 14.26(03) & 14.19(04) & 14.85(04)\\
58630.69 & 15.09(04) & 14.44(02) & 14.81(01) & 14.76(04) & 15.41(05)\\
58631.72 & 15.18(04) & 14.43(03) & 14.92(01) & 14.80(04) & 15.46(06)\\
58633.69 & 15.49(04) & 14.59(02) & 15.08(05) & 14.80(04) & 15.40(05)\\
58635.72 & 15.81(08) & 14.69(03) & 15.32(05) & 14.82(05) & 15.28(05)\\
58637.66 & 15.96(05) & 14.78(03) & 15.52(04) & 14.86(06) & 15.25(06)\\
58645.71 & 16.72(07) & 15.45(03) & 16.36(05) & 15.48(09) & 15.58(05)\\
58647.66 & 17.01(07) & 15.56(03) & 16.54(08) & 15.54(09) & 15.75(06)\\
\hline
\end{tabular}
\begin{tablenotes}
\item[] {Uncertainties, in units of 0.01~mag, are 1$\sigma$.}
\end{tablenotes}
\end{threeparttable}
\end{table*}

\begin{table*}
\caption{SN~2019ein Photometry from AZT\label{tablcazt}}
\begin{threeparttable}[b]
\begin{tabular}{c c c c c c}
\hline
   MJD & $U$ (mag) & $B$ (mag) & $V$ (mag) & $R$ (mag) & $I$ (mag) \\
\hline
58654.70 & 17.27(06) & 17.20(02) & 16.11(01) & 15.69(02) & 15.55(01)\\
58655.72 & - & 17.22(03) & 16.13(03) & 15.73(02) & 15.62(02)\\
58656.70 & 17.28(07) & 17.25(02) & 16.16(02) & 15.77(02) & 15.69(01)\\
58657.68 & 17.43(04) & 17.29(01) & 16.19(02) & 15.79(01) & 15.73(01)\\
58658.72 & 17.36(13) & 17.27(03) & - & 15.84(02) & 15.78(01)\\
58660.70 & 17.42(05) & 17.32(02) & 16.30(01) & 15.87(01) & 15.87(01)\\
58661.73 & 17.36(11) & 17.32(03) & 16.32(04) & 15.95(02) & 15.94(01)\\
58664.69 & - & 17.40(02) & 16.41(01) & 16.06(03) & 16.11(01)\\
58665.68 & 17.61(03) & 17.41(01) & 16.44(01) & 16.09(01) & 16.13(01)\\
58668.71 & 17.82(05) & - & 16.51(02) & 16.19(02) & -\\
58669.68 & 17.64(04) & 17.50(03) & 16.54(02) & 16.24(02) & 16.33(01)\\
58670.69 & 17.70(03) & 17.54(02) & 16.57(02) & 16.26(02) & 16.39(02)\\
58671.68 & 17.65(03) & 17.50(02) & 16.61(01) & 16.29(01) & 16.42(01)\\
58673.70 & 17.84(06) & 17.55(02) & 16.69(05) & 16.36(01) & 16.58(01)\\
58675.68 & 17.80(04) & 17.54(01) & 16.72(01) & 16.44(01) & 16.61(01)\\
58676.69 & 17.71(07) & 17.55(02) & 16.74(01) & 16.46(01) & 16.67(01)\\
58677.69 & 17.82(05) & 17.59(01) & 16.76(02) & 16.50(01) & 16.72(01)\\
58679.68 & 17.78(06) & 17.60(02) & 16.85(01) & 16.57(01) & 16.82(01)\\
58680.69 & 17.89(04) & 17.62(02) & 16.88(01) & 16.62(01) & 16.83(01)\\
58681.72 & 17.90(06) & 17.63(01) & 16.90(01) & 16.63(01) & 16.97(01)\\
58683.70 & 17.88(07) & 17.64(02) & 16.95(02) & 16.70(01) & 16.98(01)\\
58684.71 & 17.81(06) & - & - & - & 17.03(02)\\
58686.70 & 17.96(10) & 17.67(03) & 17.01(02) & 16.80(02) & 17.12(02)\\
58687.70 & 17.91(10) & 17.69(03) & 17.03(02) & 16.80(02) & 17.16(02)\\
58688.70 & 18.08(11) & 17.72(02) & 17.08(01) & 16.87(02) & 17.20(02)\\
58689.69 & 17.99(08) & 17.72(03) & 17.11(01) & 16.89(02) & 17.18(02)\\
58690.70 & 17.99(10) & 17.72(02) & 17.12(03) & 16.91(02) & 17.28(02)\\
58691.69 & 18.09(09) & 17.73(03) & 17.13(02) & 16.95(02) & 17.32(02)\\
58692.67 & 18.04(06) & 17.78(02) & 17.19(01) & 16.99(02) & 17.30(02)\\
58693.67 & 18.01(07) & 17.76(02) & 17.20(02) & 17.03(02) & 17.37(02)\\
58694.69 & 18.21(09) & 17.80(02) & 17.23(03) & 17.06(02) & 17.39(02)\\
58695.67 & 18.29(09) & 17.77(02) & 17.12(02) & 16.90(03) & 17.53(02)\\
58696.70 & 18.17(07) & 17.78(03) & - & 17.05(02) & 17.54(02)\\
58697.68 & 18.16(06) & 17.84(02) & 17.31(02) & 17.14(02) & -\\
58698.67 & - & 17.88(02) & 17.29(02) & 17.19(02) & 17.60(02)\\
58699.67 & 18.28(06) & 17.86(02) & 17.39(01) & 17.24(02) & 17.57(02)\\
58700.67 & 18.30(06) & 17.87(03) & 17.38(02) & 17.26(02) & 17.64(02)\\
58701.66 & 18.26(12) & 17.86(02) & 17.39(01) & 17.27(02) & 17.71(02)\\
58702.66 & - & 17.93(02) & 17.43(02) & 17.36(02) & 17.81(02)\\
58707.67 & - & 18.01(03) & 17.54(03) & 17.34(04) & -\\
58710.68 & - & - & 17.68(05) & - & -\\
\hline
\end{tabular}
\begin{tablenotes}
\item[] {Uncertainties, in units of 0.01~mag, are 1$\sigma$.}
\end{tablenotes}
\end{threeparttable}
\end{table*}

\clearpage
\begin{table*}
\caption{SN~2019ein Photometry from ZTF\label{tablcztf}}
\begin{threeparttable}[b]
\begin{tabular}{c c c |c c c |c c c}
\hline
   MJD & $g$ (mag) & $r$ (mag) & MJD & $g$ (mag) & $r$ (mag) & MJD & $g$ (mag) & $r$ (mag)\\
\hline
58606.33 & 16.64(04) & 16.75(04) & 58667.27 & 16.98(06) & 16.43(05) & 58723.16 & 17.83(07) & -\\
58608.26 & - & 15.75(03) & 58670.19 & 17.05(04) & 16.51(04) & 58726.16 & 17.89(08) & -\\
58612.27 & 14.59(03) & 14.60(03) & 58673.19 & 17.06(05) & 16.62(04) & 58734.15 & 18.10(09) & -\\
58616.22 & 14.11(03) & 14.17(03) & 58676.30 & 17.15(06) & - & 58805.53 & 19.12(18) & -\\
58618.22 & - & 14.07(04) & 58679.20 & 17.17(05) & 16.79(05) & 58828.55 & 19.36(18) & -\\
58619.29 & 14.05(02) & 14.06(04) & 58682.21 & 17.20(04) & 16.92(06) & 58831.56 & 19.45(22) & -\\
58632.27 & - & 14.82(03) & 58685.19 & 17.22(05) & - & 58846.55 & 19.73(13) & -\\
58635.25 & 15.21(04) & 14.86(03) & 58689.18 & - & 17.11(05) & 58852.57 & - & 20.69(28)\\
58638.21 & 15.53(03) & 14.85(03) & 58692.19 & 17.33(04) & 17.32(04) & 58855.51 & 19.70(20) & -\\
58641.24 & 15.87(04) & 15.01(03) & 58695.21 & - & 17.42(06) & 58862.48 & 19.73(22) & -\\
58644.23 & 16.24(03) & 15.26(03) & 58698.19 & 17.35(20) & 17.49(05) & 58867.46 & 19.92(16) & -\\
58647.24 & - & 15.53(03) & 58701.16 & - & 17.50(09) & 58872.45 & 20.00(28) & -\\
58650.23 & - & 15.70(03) & 58705.16 & - & 17.74(07) & 58875.46 & 20.09(19) & -\\
58653.28 & 16.68(04) & - & 58708.16 & 17.52(06) & - & 58878.51 & 19.86(19) & -\\
58657.23 & 16.80(04) & 16.04(03) & 58712.24 & - & 17.86(06) & 58881.49 & 20.21(19) & -\\
58660.22 & - & 16.17(03) & 58715.16 & 17.68(06) & - & 58884.47 & 20.23(29) & -\\
58663.21 & 16.93(05) & 16.27(04) & 58718.16 & 17.77(08) & - & 58893.47 & 20.28(28) & -\\
\hline
\end{tabular}
\begin{tablenotes}
\item[] {Uncertainties, in units of 0.01~mag, are 1$\sigma$.}
\end{tablenotes}
\end{threeparttable}
\end{table*}

\begin{table*}
\caption{SN~2019ein Photometry from ATLAS\label{tablcatlas}}
\begin{threeparttable}[b]
\begin{tabular}{c c c |c c c |c c c}
\hline
   MJD & filter & mag & MJD & filter & mag & MJD & filter & mag\\
\hline
58604.49 & c & 18.30(05) & 58642.29 & o & 15.09(01) & 58678.30 & o & 16.80(03)\\
58606.50 & o & 16.81(02) & 58644.34 & o & 15.27(01) & 58680.28 & o & 16.90(02)\\
58610.47 & o & 15.10(01) & 58646.33 & o & 15.46(01) & 58682.33 & o & 17.02(03)\\
58612.40 & c & 14.58(01) & 58648.32 & o & 15.58(01) & 58684.28 & o & 17.03(02)\\
58614.40 & o & 14.44(01) & 58650.35 & o & 15.74(01) & 58688.28 & o & 17.16(02)\\
58616.39 & o & 14.28(01) & 58654.32 & o & 15.94(01) & 58690.27 & o & 17.26(02)\\
58618.40 & o & 14.29(01) & 58656.33 & o & 15.98(01) & 58692.29 & c & 17.19(02)\\
58620.37 & o & 14.28(01) & 58658.32 & o & 16.07(08) & 58696.26 & c & 17.35(05)\\
58622.37 & o & 14.29(01) & 58660.30 & o & 16.17(04) & 58700.30 & c & 17.38(03)\\
58624.31 & o & 14.38(01) & 58664.30 & c & 16.48(01) & 58708.31 & c & 17.65(16)\\
58626.38 & o & 14.62(01) & 58668.36 & c & 16.58(01) & 58716.26 & o & 18.01(04)\\
58630.36 & o & 14.95(01) & 58672.31 & o & 16.62(02) & 58722.25 & o & 18.31(07)\\
58634.37 & o & 14.95(01) & 58675.33 & o & 16.69(02) & 58724.25 & c & 17.95(04)\\
58636.41 & c & 15.01(01) & 58676.39 & o & 16.75(05) & 58740.24 & o & 18.77(62)\\
\hline
\end{tabular}
\begin{tablenotes}
\item[] {Uncertainties, in units of 0.01~mag, are 1$\sigma$.}
\end{tablenotes}
\end{threeparttable}
\end{table*}

\begin{table*}
\caption{Spectroscopic Observations of SN~2019ein\label{tabspec}}
\begin{threeparttable}[b]
\begin{tabular}{c c c c c c c c}
\hline
   UT  & MJD & Epoch\tnote{a} & Telescope & Instrument & Range(\AA) & Resolution(\AA)\\
\hline
   2019-05-03 & 58606.7 & $-12.6$ & XLT & BFOSC(G4) & 3980--8810 & 2.8\\
   2019-05-03 & 58606.8 & $-12.5$ & LJT & YFOSC(G3) & 3500--8760 & 2.8\\
   2019-05-05 & 58608.7 & $-10.6$ & XLT & BFOSC(G4) & 4100--8810 & 2.8\\
   2019-05-06 & 58609.7 & $-9.6$ & XLT & BFOSC(G4) & 4100--8810 & 2.8\\
   2019-05-06 & 58609.7 & $-9.6$ & LJT & YFOSC(G3) & 3500--8760 & 2.8\\
   2019-05-07 & 58610.7 & $-8.6$ & LJT & YFOSC(G3) & 3500--8760 & 2.8\\
   2019-05-07 & 58610.9 & $-8.4$ & APO 3.5 m & DIS & 3440--9160 & 2.3\\
   2019-05-08 & 58611.5 & $-7.8$ & XLT & BFOSC(G4) & 3850--8810 & 2.8\\
   2019-05-10 & 58613.6 & $-5.7$ & XLT & BFOSC(G4) & 3720--8640 & 2.8\\
   2019-05-10 & 58613.8 & $-5.5$ & LJT & YFOSC(G3) & 3500--8760 & 2.8\\
   2019-05-12 & 58615.7 & $-3.6$ & LJT & YFOSC(G3) & 3500--8760 & 2.8\\
   2019-05-13 & 58616.5 & $-2.8$ & XLT & OMR & 3720--8650 & 2.8\\
   2019-05-14 & 58617.8 & $-1.6$ & LJT & YFOSC(G3) & 3500--8760 & 2.8\\
   2019-05-22 & 58625.6 & 6.2 & XLT & OMR & 3760--8690 & 2.8\\
   2019-05-25 & 58628.8 & 9.3 & LJT & YFOSC(G3) & 3500--8760 & 2.8\\
   2019-05-27 & 58630.6 & 11.2 & XLT & OMR & 3740--8670 & 2.8\\
   2019-05-31 & 58634.8 & 15.3 & APO 3.5 m & DIS & 3450--9050 & 2.3\\
   2019-06-07 & 58641.7 & 22.2 & LJT & YFOSC(G3) & 3500--8760 & 2.8\\
   2019-06-29 & 58663.6 & 43.9 & XLT & BFOSC(G4) & 4040--8810 & 2.8\\
   2020-03-24 & 58932.6 & 310.8 & Keck-I & LRIS & 3147--10,279 & 0.6\\
\hline
\end{tabular}
\begin{tablenotes}
\item[a]{Days relative to $B$-band maximum brightness on 2019-05-16.36 (MJD 58619.36).}
\end{tablenotes}
\end{threeparttable}
\end{table*}

\clearpage

\clearpage

\clearpage

\clearpage

\clearpage

\clearpage

\clearpage

\clearpage

\clearpage

\clearpage

\clearpage

\clearpage

\clearpage

\clearpage

\clearpage

\clearpage

\clearpage

\bsp	
\label{lastpage}
\end{document}